\documentclass[onecolumn,prd,floats,aps,amsmath,amssymb,nofootinbib, superscriptaddress,preprintnumbers,10pt]{revtex4-2}
\usepackage{comment}
\usepackage[utf8]{inputenc}
\usepackage{mathtools}
\usepackage{bbm}
\usepackage{cancel} 
\usepackage{bm}
\usepackage{blindtext}
\usepackage{xcolor}
\usepackage{hyperref}
\usepackage{graphicx}
\usepackage{amsthm}
\usepackage{bm}
\usepackage{verbatim}
\usepackage{amsmath}
\usepackage{siunitx}
\usepackage{amssymb}
\usepackage{slashed}
\usepackage{tikz}
\usetikzlibrary{positioning, shapes, arrows.meta, decorations.pathmorphing}

\newcommand{\Romatre}{Dipartimento di Matematica e Fisica, Universit\`a  Roma Tre and INFN, Sezione di Roma Tre,\\ Via della Vasca Navale 84, I-00146 Rome, Italy}
\newcommand{\RomatreINFN}{Istituto Nazionale di Fisica Nucleare, Sezione di Roma Tre,\\ Via della Vasca Navale 84, I-00146 Rome, Italy}
\newcommand{\Romadue}{Dipartimento di Fisica and INFN, Universit\`a di Roma ``Tor Vergata",\\ Via della Ricerca Scientifica 1, I-00133 Roma, Italy}
\newcommand{\LaSapienza}{Physics Department and INFN Sezione di Roma La Sapienza,\\ Piazzale Aldo Moro 5, 00185 Roma, Italy}
\newcommand{\soton}{Department of Physics and Astronomy, University of Southampton,\\ Southampton SO17 1BJ, UK}
\newcommand{\be}{\begin{equation}}
\newcommand{\ee}{\end{equation}}
\newcommand{\bs}[1]{\bm{#1}}
\renewcommand{\vec}[1]{\bm{#1}}
\newcommand{\dfour}{d^{\hspace{1.35pt}4}}

\usepackage{hyperref}
\usepackage{physics}
\usepackage{graphicx}

\usepackage{simpler-wick}
\usepackage{comment}
\usetikzlibrary{decorations.markings}
\tikzset{W->-/.style={decoration={
  markings,
  mark=at position 0.5*\pgfdecoratedpathlength+2pt with
  {\draw[-latex] (-2pt,0pt) -- (1pt,0pt);}},postaction={decorate}},
  W-<-/.style={decoration={
  markings,
  mark=at position 0.5*\pgfdecoratedpathlength with
  {\draw[latex-] (-2pt,0pt) -- (1pt,0pt);}},postaction={decorate}}
  }
\newif\ifWickBelow
\WickBelowfalse
\pgfkeys{
  /simplerwick/.cd,
  arrows/.store in=\LstWickArrows,
  arrows={-,-,-,-,-,-,-,-,-},
  arrows/.initial={-,-,-,-,-,-,-,-,-}, 
  below/.code={\WickBelowtrue},
}

\makeatletter
\def\swick@end#1#2{
  \swick@setfalse@#1
  \tikzexternaldisable
  \begin{tikzpicture}[remember picture, baseline=(swick-close#1.base)]
    \node[use as bounding box, inner sep=0pt, outer sep=0pt] (swick-close#1) {$\displaystyle #2$};
  \end{tikzpicture}
  \tikz[remember picture, overlay]
{
\foreach \W@X[count=\W@C] in \LstWickArrows
{\ifnum\W@C=#1
\xdef\myW@style{\W@X}
\fi}
\ifWickBelow
    \draw[\myW@style] ($(swick-open#1.south) + (0, -3pt)$) 
          -- ($(swick-open#1.base) + (0, -\swick@offset) + #1*(0, -\swick@sep)$) 
          -- ($(swick-close#1.base) + (0, -\swick@offset) + #1*(0, -\swick@sep)$) 
          -- ($(swick-close#1.south) + (0, -3pt)$);
\else
    \draw[\myW@style] ($(swick-open#1.north) + (0, 3pt)$) 
          -- ($(swick-open#1.base) + (0, \swick@offset) + #1*(0, \swick@sep)$) 
          -- ($(swick-close#1.base) + (0, \swick@offset) + #1*(0, \swick@sep)$) 
          -- ($(swick-close#1.north) + (0, 3pt)$);
\fi}
  \tikzexternalenable}
\makeatother

\begin{document}
\title{Kaon radiative leptonic decay rates from lattice QCD simulations  at the physical point}
\author{R.\,Di\,Palma}\affiliation{\Romatre}
\author{R.\,Frezzotti}\affiliation{\Romadue} 
\author{G.\,Gagliardi}\affiliation{\Romatre}
\author{V.\,Lubicz}\affiliation{\Romatre} 
\author{G.\,Martinelli}\affiliation{\LaSapienza}
\author{C.T.\,Sachrajda}\affiliation{\soton}
\author{F.\,Sanfilippo}\affiliation{\RomatreINFN}
\author{S.\,Simula}\affiliation{\RomatreINFN}
\author{N.\,Tantalo}\affiliation{\Romadue}
\date{\today}
\begin{abstract}
We present a lattice QCD calculation of the radiative leptonic decay rates of the kaon, improving upon our previous work in Ref.~\cite{Desiderio:2020oej}. Our analysis uses gauge ensembles generated by the Extended Twisted Mass Collaboration (ETMC) with $N_{f} = 2 + 1 + 1$ flavors of Wilson-clover twisted mass fermions. For the first time, we go beyond the electroquenched approximation by including quark-disconnected contributions. Several key improvements have been implemented: (i) the simulations are now performed directly at physical light- and strange-quark masses, (ii) finite-size effects are carefully investigated using lattices with spatial extents ranging from $L \simeq 3.8\,\mathrm{fm}$ to $L \simeq 7.7\,\mathrm{fm}$, and (iii) the continuum extrapolation is based on three lattice spacings in the range $a \in [0.08, 0.058]\,\mathrm{fm}$. As a result of the high-precision determination of the relevant correlation functions, we reduce the uncertainties on both the axial and vector form factors by nearly a factor of two compared to our previous analysis. When compared to experimental measurements in the electron channel ($K^- \to e^- \bar{\nu}_e \gamma$), our results show a tension---at the level of $2.6$ standard deviations---with respect to KLOE data. On the other hand, they are compatible with   measurements from the E36 Collaboration at J-PARC. In the muonic decay channel ($K^- \to \mu^- \bar{\nu}_\mu \gamma$), we confirm the tensions, already observed in our previous study, between lattice QCD predictions and ISTRA+ and OKA data, which are both primarily sensitive to the value of the negative-helicity form factor $F^{-}$.
\end{abstract}

\maketitle

\section{Introduction}
Charged pseudoscalar mesons decaying into light leptons, $P \to \ell\,\nu_\ell (\gamma)$ (with $\ell = e,\mu$), provide important avenues for probing the flavor sector of the Standard Model (SM). 
In the absence of the emission of a hard real photon, these processes are helicity-suppressed by the SM’s $V-A$ structure. 
Radiative leptonic decays, $P \to \ell\,\nu_\ell\,\gamma$, lift this suppression and can reveal new-physics signals in the presence of non-standard currents and/or non-universal lepton couplings. Beyond their role in constraining possible SM extensions, radiative leptonic decays allow for alternative determinations of the Cabibbo--Kobayashi--Maskawa (CKM) matrix elements, complementing the standard purely leptonic channels. Moreover, in the region of hard photon energies, radiative leptonic decays serve as powerful probes of the mesonic structure. In addition to the usual axial decay constant $f_P$, non-perturbative structure-dependent (SD) contributions, parametrized through the vector ($F_V$) and axial ($F_A$) form factors, govern the emission of real photons from quarks. A first-principle calculation of these contributions requires the use of non-perturbative methods, a requirement naturally addressed by lattice QCD + QED. 

For pion and kaon decays, Chiral Perturbation Theory (ChPT) provides analytical predictions 
for the axial and vector form factors up to $O(p^6)$~\cite{Bijnens:1992en, Chen:2008mg}. 
These calculations, however, involve low-energy constants that are only partially constrained 
by experimental data, and in some cases rely on model-dependent assumptions. 
For heavier pseudoscalar mesons ($D, D_s, B$), where ChPT does not apply, the importance 
of lattice QCD calculations becomes even more evident (see also our recent calculation of the $D_{s}\to \ell\nu_{\ell}\gamma$ decay rate~\cite{Frezzotti:2023ygt}).
In this work, we present an improved lattice QCD calculation of the radiative leptonic decay rates of the kaon, updating our previous study in Ref.~\cite{Desiderio:2020oej}. 
Our analysis employs gauge ensembles generated by the Extended Twisted Mass  Collaboration (ETMC) with $N_f = 2+1+1$ flavors of Wilson--Clover twisted mass fermions. With respect to Ref.~\cite{Desiderio:2020oej}, several key improvements have been implemented:
\begin{enumerate}
\item For the first time, we go beyond the so-called electroquenched approximation by including quark-disconnected contributions (see Fig.~\ref{Fig:Diagrams}).
    \item We perform simulations directly at physical light- and strange-quark masses, thereby eliminating the need for chiral extrapolations.
    \item We systematically investigate finite-size effects (FSE) by utilizing ensembles with spatial extents ranging from $L \simeq 3.8$\,fm to $L \simeq 7.7$\,fm.
    \item We carry out a continuum extrapolation using three fine lattice spacings in the range $a \in [0.08,\,0.058]$\,fm.
\end{enumerate}
Although performing the simulations directly at the physical point significantly increases the statistical noise, our high-statistics evaluations of the relevant Euclidean correlation functions, combined with significant improvements of the numerical procedures, allows us to reduce the uncertainties on both $F_{A}$ and $F_{V}$ by nearly a factor of two compared to our previous analysis.
We anticipate here our final result for the $R_{\gamma}$-ratio, defined as
\begin{align}
R_{\gamma} \equiv \frac{ {\rm Br}[K^{-}\to e^{-}\bar{\nu}_{e} \gamma]}{{\rm Br}[K^{-}\to \mu^{-}\bar{\nu}_{\mu}(\gamma)]}~,\qquad E_{\gamma} > 10~{\rm MeV},\qquad  \vert \bs{p}_e \vert > 200~{\rm MeV}\,,
\end{align}
where $E_{\gamma}$ and $\bs{p}_e$ denote, respectively, the photon energy and the electron momentum in the rest frame of the kaon. Combining statistical and systematic uncertainties, we obtain
\begin{align}
\label{eq:Rgamma}
R_{\gamma} = 1.84~(12) \times 10^{-5}~,
\end{align}
to be compared with our previous determination $R_{\gamma}^{\rm ETMC-21} = 1.74~(21) \times 10^{-5}$~\cite{Desiderio:2020oej, Frezzotti:2020bfa}. Our result can be compared with the experimental data from KLOE~\cite{KLOE:2009urs}, which reports  
$R_{\gamma}^{\rm KLOE} = 1.483(67) \times 10^{-5}$,  
and from the E36 Collaboration at J-PARC~\cite{J-PARCE36:2021yvz,J-PARCE36:2022wfk}, which finds  
$R_{\gamma}^{\rm E36} = 1.98(11) \times 10^{-5}$.  
Our estimate differs from the KLOE data by 2.6 standard deviations, while it is consistent at $1\sigma$ level with the E36 measurement. Furthermore, our result turns out to be larger than the ChPT predictions at order $O(p^4)$, namely $R_{\gamma}^{\rm ChPT-p^{4}} = 1.28\,(32) \times 10^{-5}$.  In this paper, we will also compare our results for the decay rate of the muonic channel ($K^- \to \mu^- \bar{\nu}_\mu \gamma$) with those obtained by the E787, ISTRA+ and OKA experimental collaborations, updating the analysis of Ref.~\cite{Frezzotti:2020bfa}. 

The remainder of this paper is organized as follows. 
In Sec.~\ref{sec:definitions}, we define the axial and vector form factors relevant for describing radiative leptonic decays. 
We explain how these form factors can be extracted from Euclidean correlation functions on the lattice in Sec.~\ref{sec:FF_from_ECF}. 
Details about our numerical setup, including the gauge ensembles employed for the present calculation and the strategy adopted for the calculation of both the quark-connected and quark-disconnected contributions, are given in Sec.~\ref{sec:Lattice_setup}. 
We present our numerical results in Sec.~\ref{sec:numerical_results} , highlighting the strategies employed to estimate the form factors and carefully address finite-size (both spatial and temporal) effects as well as discretization effects. 
In Sec.~\ref{sec:exp_comparison}, we compare our predictions with available experimental data, focusing in particular on results from the KLOE experiment, the E36 experiment at J-PARC, E787, ISTRA+ and OKA. 
Finally, Sec.~\ref{sec:conclusions} contains our conclusions.

\section{Definitions of the form factors}
\label{sec:definitions}
\begin{figure}[]
    \centering
\includegraphics[width=1.\columnwidth]{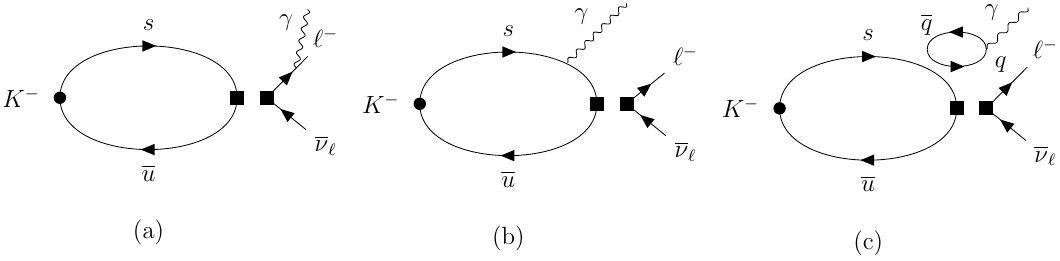}
\caption{Diagrams contributing to the $K^{-} \to \ell^{-}\bar{\nu}_{\ell} \gamma$ at leading order in $\alpha_{\rm em}$. The double square represents the 4-fermion weak operator in the Fermi effective theory. Diagram (a) corresponds to the final-state radiation contribution where the photon is emitted from a lepton leg. Diagrams (b) and (c) correspond respectively to the quark-connected and quark-disconnected contributions to the structure dependent part of the amplitude, with the photon emitted by a quark. In the case of the quark-disconnected contribution the photon is emitted by a sea quark ($q=u,d,s,c$). Similarly to diagram (b), there is also a contribution where the photon is emitted by the valence $\overline{u}$ quark, which it is not shown in the figure.}
\label{Fig:Diagrams}
\end{figure}
In the low energy effective description of the SM (Fermi theory), the relevant diagrams describing the $K^{-} \to \ell^-\overline{\nu}_{\ell}\gamma$ decay at lowest order in the electroweak interactions are given in Fig.~\ref{Fig:Diagrams}. Diagram (a) corresponds to the so-called final-state-radiation contribution and the only non-perturbative input required for its calculation is the decay constant of the kaon, $f_{K}$. Diagram  (b) and diagram (c)  describe instead the non-perturbative quark-photon interaction, and are the main focus of the present work. We refer to the latter two diagrams as the \textit{quark-connected} and \textit{quark-disconnected} contributions, respectively.  While the (dominant) quark-connected diagram has already been calculated in our previous exploratory study~\cite{Desiderio:2020oej}, in this work we go beyond the so-called electroquenched approximation by computing the disconnected contribution for the first time. 

The quark-connected and quark-disconnected contributions correspond to the two distinct Wick contractions of the following hadronic tensor: 
\be
H_{W}^{r, \nu}(E_{\gamma}, \bs{k}, \bs{p})=\epsilon_\mu^r(k)\,H_{W}^{\mu\nu}(E_{\gamma}, \bs{k}, \bs{p})= \epsilon_\mu^r(k) \int \dfour y_{\gamma}\, e^{ik\cdot y_{\gamma}}\, \bra{0} \hat{\mathrm{T}}[\,j_W^\nu(0) j^\mu_\mathrm{em}(y_{\gamma})]\ket{K^-(\bs{p})}~,
\label{Eq:Def_hadronictensor}
\ee
where $\epsilon_{\mu}^r(k)$ is the polarisation vector of the outgoing photon with four-momentum $k=(E_{\gamma}, \bs{k})$, while $\bs{p}$ is the momentum of the decaying kaon. The $\hat{\mathrm{T}}$ operator in Eq.~(\ref{Eq:Def_hadronictensor}) stands for \textit{time-ordered-product}. The electromagnetic current $j_{\rm em}^{\mu}(x)$ and the hadronic weak current $j_{W}^{\nu}$ are given by 
\begin{align}
\label{Eq:Def_em_current}
j^{\mu}_{\mathrm{em}}(x) &=  \sum_{f} ~ j^{\mu}_{f}(x) =  \sum_f e_f~ \overline{q}_f(x) \gamma^{\mu} q_f(x)~,\\[8pt]
\label{Eq:Def_weak_current}
j^{\nu}_W(x) &= j^{\nu}_{V}(x) - j^{\nu}_A(x) = \overline{q}_u(x) (\gamma^{\nu} - \gamma^{\nu} \gamma_5) q_s(x)~,
\end{align}
where $q_{f}(x)$ is the quark-field of the flavour $f$, and we have indicated by $e_f$ its electric charge in units of the charge of the positron. Throughout this paper, whenever the subscript $V$ (or $A$) is used in place of $W$, it indicates that the vector (or axial) component of the weak hadronic current in 
Eq.~(\ref{Eq:Def_weak_current}) is being considered.

Following the definitions given in Ref.~\cite{Desiderio:2020oej},  $H^{\mu\nu}_W(E_{\gamma}, \bs{k}, \bs{p})$ can be decomposed, for a general value of $k^2$, in terms of the kaon decay constant $f_K$ and four scalar structure-dependent (SD) form factors $F_V$, $F_A$, $H_1$ and $H_2$ as 
\be
H^{\mu\nu}_W(E_{\gamma}, \bs{k}, \bs{p}) =  \  H^{\mu\nu}_{\mathrm{pt}}(E_{\gamma}, \bs{k}, \bs{p}) +   H^{\mu\nu}_{\mathrm{SD}}(E_{\gamma}, \bs{k}, \bs{p}), \\[10pt]
\label{Eq:Hmunu_decomposition}
\ee
where
\begin{align}
& H^{\mu \nu}_{\mathrm{pt}}(E_{\gamma}, \bs{k}, \bs{p}) = \ f_{K} \bigg[g^{\mu\nu} + \frac{(2p-k)^\mu(p-k)^\nu}{2p\cdot k - k^2}\bigg], \\[10pt] 
& H^{\mu\nu}_{\mathrm{SD}}(E_{\gamma}, \bs{k}, \bs{p}) =  -i \frac{F_V}{m_K} \varepsilon^{\mu\nu\alpha\beta}k_\alpha p_\beta + \frac{F_A}{m_K}[(p\cdot k -k^2)g^{\mu\nu} - (p-k)^\mu k^\nu]
 \\ \nonumber & \qquad \qquad \qquad \quad +\frac{H_1}{m_K} (k^2 g^{\mu \nu} - k^\mu k^\nu) + \frac{H_2}{m_K} \frac{(p\cdot k - k^2)k^\mu - k^2(p-k)^\mu}{(p-k)^2 - m_K^2} (p-k)^\nu, 
\end{align}
and $m_K$ is the mass of the kaon. The four independent  form factors are functions of two Lorentz invariant, $k^2$ and $p\cdot k$. Notice that the formal separation between the point-like  ($H^{\mu\nu}_{\mathrm{pt}})$ and the structure-dependent  ($H^{\mu\nu}_{\mathrm{SD}}$) contribution in Eq.~(\ref{Eq:Hmunu_decomposition}) is such that the Ward-Identity (WI) satisfied by $H^{\mu\nu}_W(E_{\gamma}, \bs{k}, \bs{p})$ reads 
\be
k_{\mu}H^{\mu\nu}_{\mathrm{pt}}(E_{\gamma}, \bs{k},\bs{p}) = i \bra{0}j^{\nu}_W(0)\ket{K^-(\bs{p})} = f_Kp^\nu, \qquad k_\mu H^{\mu\nu}_{\mathrm{SD}}(E_{\gamma},\bs{k}, \bs{p}) = 0.
\label{Eq:WI}
\ee 
The decomposition of the hadronic tensor  in Eq.~(\ref{Eq:Hmunu_decomposition}) corresponds to the general case in which the emitted photon can be either real or virtual. For real photon emission, the focus of the present work, after setting
\be
k^2=0, \qquad \qquad \qquad \epsilon^r(k) \cdot k=0,
\label{Eq:RF_settings}
\ee
in Eq.~(\ref{Eq:Def_hadronictensor}), one gets
\be
H^{r, \nu}_W(E_{\gamma}, \bs{k}, \bs{p}) = \epsilon^r_\mu(k) \bigg[ -i \frac{F_V}{m_K} \varepsilon^{\mu\nu\alpha\beta}k_\alpha p_\beta + \frac{F_A}{m_K}[(p\cdot k)g^{\mu\nu} - p^\mu k^\nu]  +  f_{K} \bigg(g^{\mu\nu} + \frac{p^\mu p^\nu - p^\mu k^\nu}{p\cdot k }\bigg) \bigg]~,
\label{Eq:RealHmunu}
\ee
which shows that the only structure-dependent form factors relevant for $K^{-}\to \ell^{-}\bar{\nu}_{\ell} \gamma$ decays are the vector ($F_V$) and the axial ($F_A$) form factors. In the following, we restrict ourselves to the kaon rest frame ($\bs{p}=\bs{0}$), and since for on-shell photons $E_{\gamma}=|\bs{k}|$, for the sake of simplifying the notation, we set $H^{\mu\nu}_{W}(\bs{k}) = H^{\mu\nu}_W(|\bs{k}|, \bs{k}, \bs{0})$.

As in our previous studies~\cite{Desiderio:2020oej}, we find it convenient to parametrize the form factors as a function of the following dimensionless variable
\be
x_{\gamma} = \frac{2p\cdot k}{m_K^2} = \frac{2E_{\gamma}}{m_K}, \qquad \qquad \qquad 0 \leq x_{\gamma} \leq 1 - \frac{m_{\ell}^2}{m_K^2} < 1,
\label{Eq:Def_xgamma}
\ee
where $m_{\ell}$ is the mass of the final-state charged lepton.

\section{Form Factors from Euclidean correlation functions}
\label{sec:FF_from_ECF}
The starting point of our calculation is the following Euclidean three-point correlation function, computed on a lattice with finite space-time volume $V=L^3 \times T$
\be
C_{3,W}^{\mu \nu}(t_{\gamma},  \bs{k}; t_W) =  \sum_{\bs{x}} \sum_{\bs{y}_{\gamma}}  \ e^{- i \bs{k}\cdot \bs{y}_{\gamma}} \bra{0}\hat{\mathrm{T}}[j^\nu_{W}(t_W, \bs{0})j^{\mu}_{\mathrm{em}}(t_{\gamma}+t_{W}, \bs{y}_{\gamma}) P^{\dagger}_K(0, \bs{x})] \ket{0}.
\label{Eq:Def3pt}
\ee 
In this equation, $P^{\dagger}_K(0, \bs{x})$ is a pseudoscalar interpolating operator with the quantum numbers of the kaon.  In our previous work~\cite{Desiderio:2020oej}, we showed that for real photon emission it is possible to determine the hadronic tensor $H^{\mu\nu}_W(\bs{k})$ in Eq.~(\ref{Eq:Def_hadronictensor}) by integrating the following amputated correlation function
\be
C_{W}^{\mu\nu}(t_\gamma,  \bs{k}; t_W) = e^{E_{\gamma}t_{\gamma}}~\mathcal{N}(t_{W}) ~C_{3, W}^{\mu\nu}(t_\gamma, \bs{k}; t_W )~,\qquad  \mathcal{N}(t_{W}) \equiv e^{m_K t_{W}} \frac{2m_K}{\bra{K^{-}(\bs{0})}P^{\dagger}_K(0)\ket{0}}
\label{Eq:Def_correlationfunction}
\ee
as
\be
H^{\mu\nu}_W(\bs{k}; t_W)  =  \int_{-\infty}^{\infty} d t_\gamma  \ C_{W}^{\mu\nu}(t_\gamma, \bs{k}; t_W ) = H^{\mu \nu}_W(\bs{k}) + \cdot \cdot \cdot \ .
\label{Eq:HmunufromCmunu}
\ee
The ellipses on the right-hand side of Eq.\,(\ref{Eq:HmunufromCmunu}) represent sub-leading terms that vanish 
exponentially in the limit $t_{W}\to\infty$, in which the initial-state kaon is fully 
isolated. We evaluate the correlation function in Eq.~(\ref{Eq:Def3pt}) as a function of $t_\gamma$ at a fixed value of $t_W$.  This approach differs from that adopted in our previous 
studies~\cite{Desiderio:2020oej,Frezzotti:2023ygt}, in which the correlation function was 
evaluated for all $t_{W}$ while performing the time-integration over $t_{\gamma}$ 
directly during the simulation. Both approaches have been studied in detail in 
Ref.~\cite{Giusti:2023pot}, where they are referred to as the 3d and 4d methods, respectively.

In this work, we employ the 3d method, which requires choosing sufficiently large values of 
$t_{W}$ so that the initial-state kaon is properly isolated (see 
Sec.~\ref{sec:tw-dependence} for a detailed discussion of finite-$t_{W}$ effects). On a 
lattice of temporal extent $T$, the time-integration over $t_{\gamma}$ in Eq.~(\ref{Eq:HmunufromCmunu}) is necessarily 
restricted to the finite interval $t_{\gamma} \in [-t_{W}, t_{\rm max}]$, where 
$t_{\rm max} \ll T - t_{W}$ in order to avoid around-the-world contributions from 
unwanted time orderings. We refer to the systematic error associated with truncating the 
integral as \textit{$t_{\gamma}$-integral truncation effects}. The 3d method allows us to improve the convergence of the integral by analytically continuing the 
correlator to times smaller (larger) than $-t_{W}$ $(t_{\rm max})$ under the assumption 
of ground-state dominance, as first proposed in Ref.~\cite{Tuo:2021ewr}, and pointed out in Ref.~\cite{Giusti:2023pot}. In 
Section~\ref{Sec:temporaltrunc}, we discuss this point in detail. Moreover, the 3d method is helpful to better control the exponential signal-to-noise problem that arises for $E_\gamma > m_{\pi}$ as observed in Ref.~\cite{Frezzotti:2023ygt}. The 3d method 
allows us to study cases in which the photon is off-shell, making it possible to access 
the rare kaon decay $K^{-} \to \ell' \bar{\ell'}\,\ell^{-}\,\bar{\nu}_{\ell}$%
~\cite{Gagliardi:2022szw,Tuo:2021ewr} without the need of generating additional correlation functions. 
It is also suitable for applying the spectral density reconstruction method developed in 
Ref.~\cite{Frezzotti:2023nun}, which is necessary in the region of large photon virtualities 
where issues related to analytic continuation to Euclidean spacetime arise. We will study 
the decay $K^{-} \to \ell' \bar{\ell'}\,\ell^{-}\,\bar{\nu}_{\ell}$ in a forthcoming paper.

The results in this work are obtained by setting the photon momentum to be along the $z$-axis, $\bs{k} = (0,  0,  k_z)$, using the strategy described in Sec.~\ref{sec:Lattice_setup} below.
The form factors $F_V$ and $F_A$ in Eq.~(\ref{Eq:RealHmunu}) for a given $x_{\gamma}$ can be then extracted from the large-$t_W$ limit of the following estimators: 
\begin{align}
\label{Eq:Estimators}
F_V(x_{\gamma}; t_W)=& \  \frac{iH_V^{21}(\bs{k}; t_W)}{k_z} = \frac{i}{k_z} \int_{-\infty}^{\infty} d t_\gamma  \ C^{21}_V(t_\gamma, \bs{k}; t_W)~, \\[10pt]
\label{Eq:EstimatorsFA}
F_A(x_{\gamma}; t_W) =& \  -\frac{1}{E_{\gamma}}\bigg[  H_A^{11}(\bs{k}; t_W)   - H_A^{11}(\bs{0}; t_W)  \bigg] = -\frac{1}{E_\gamma} \int_{-\infty}^{\infty} d t_\gamma \ \bigg[ C^{11}_A(t_\gamma, \bs{k}; t_W) - C_A^{11}(t_\gamma, \bs{0}; t_W) \bigg]   \nonumber \\[10pt]
\equiv& -\frac{1}{E_\gamma} \int_{-\infty}^{\infty} d t_\gamma \  \widetilde{C}^{11}_A(t_\gamma, \bs{k}; t_W)) \,.
\end{align}
The subtraction of the zero-momentum axial hadronic tensor,  $H_A^{11}( \bs{0}; t_W)$, allows us to isolate $F_A(x_{\gamma})$ from the point-like contribution proportional to $f_{K}$, without generating infrared-divergent cut-off effects, as discussed in Ref.~\cite{Desiderio:2020oej}.

Before closing this section, we would like to point out that from the zero-momentum correlator $C^{\mu\nu}_{A}(t_{\gamma}, \vec{0}; t_{W})$ it is possible to extract directly the axial form factor $F_{A}(x_{\gamma}=0)$. This has not been realized in previous calculations (using either the 3d or the 4d method). To show that this is the case, we start by rewriting $F_{A}(x_{\gamma}=0)$ as
\begin{align}
F_{A}(x_{\gamma} =0; t_{W}) = \lim_{E_{\gamma} \to 0} \left[ -\frac{1}{E_{\gamma}} \int_{-\infty}^{\infty} dt_{\gamma} \left[ ( e^{E_{\gamma}t_{\gamma}}-1)C_{A}^{11}(t_{\gamma}, \vec{0}; t_{W}) + \Delta C^{11}_{A}(t_{\gamma}, \vec{k}; t_{W}) \right]      \right]~,
\end{align}
where we have defined (see Eq.~(\ref{Eq:Def_correlationfunction}))
\begin{align}
\Delta C_{W}^{\mu\nu}(t_{\gamma}, \vec{k} ; t_{W}) =  e^{E_\gamma t_\gamma} \mathcal{N}(t_{W})\left[ C_{3, W}^{\mu\nu}(t_\gamma, \bs{k}; t_W) -  C_{3, W}^{\mu\nu}(t_\gamma, \bs{0}; t_W)\right]~.
\end{align}
The crucial point is that
\begin{align}
\label{eq:limit}
\lim_{E_{\gamma}\to 0} \Delta C_{A}^{11}(t_{\gamma}, \vec{k} ; t_{W}) / E_{\gamma} = 0~.
\end{align}
This can be easily seen by first noting that, by construction, $\Delta C_{A}^{11}(t_{\gamma}, \vec{0} ; t_{W})=0$. Moreover, since $\Delta C_{A}^{11}(t_{\gamma}, \vec{k} ; t_{W})$ is invariant under $\vec{k} \to -\vec{k}$, its Taylor expansion in the three-momentum $\vec{k}$, starts at order $O(E_{\gamma}^{2})$, which demonstrates Eq.~(\ref{eq:limit}).  It then follows immediately that
\begin{align}
\label{eq:FA_zero}
F_{A}(x_{\gamma}=0; t_{W}) =  \lim_{E_{\gamma} \to 0}  - \int_{-\infty}^{\infty} dt_{\gamma} \frac{ e^{E_{\gamma}t_{\gamma}}-1}{E_{\gamma}}~C_{A}^{11}(t_{\gamma}, \vec{0}; t_{W}) =  -\int_{-\infty}^{\infty} dt_{\gamma} \, t_{\gamma} \, C_{A}^{11}(t_{\gamma}, \vec{0}; t_{W})~,
\end{align}
namely that the axial form factor at $x_{\gamma}=0$ is nothing but the first moment of the zero-momentum correlator. We use the relation in Eq.~(\ref{eq:FA_zero}) to evaluate both the quark-connected and the quark-disconnected contributions to $F_A(x_{\gamma}=0)$. A similar relation cannot be derived for the vector form factor at $x_{\gamma}=0$, since $C^{21}_{V}(t_{\gamma}, \vec{0};t_{W})=0$, and 
\begin{align}
F_{V}(x_{\gamma}=0; t_{W})= \lim_{E_{\gamma}\to 0} i \int_{-\infty}^{\infty} dt_{\gamma} \frac{ \Delta C_{V}^{21}(t_{\gamma},\vec{k}; t_{W})}{k_{z}}~.
\end{align}
The vector form factor at zero momentum could however be computed using the method of Ref.~\cite{deDivitiis:2012vs}, which involves the insertion of the momentum-derivative of the Dirac operator (i.e., the point-split vector current) in the zero-momentum three-point function $C^{\mu\nu}_{3,V}(t_{\gamma},\vec{0};t_{W})$. However, this approach requires evaluating an additional (four-point) correlation function\,-\,a computation not undertaken in the present work.

\section{Details of the simulation}
\label{sec:Lattice_setup}
 To compute the form factors $F_{A}$ and $F_{V}$ on the lattice, we make use of the gauge configurations produced by the Extended Twisted Mass Collaboration (ETMC) with $N_{f}=2+1+1$ dynamical Wilson-clover twisted mass fermions~\cite{Frezzotti:2000nk,Frezzotti:2003xj} at the physical point\footnote{The simulated pion masses differ from the charged pion mass $M_{\pi}^{\pm} \simeq 139~{\rm MeV}$ by only $\simeq \pm 2~{\rm MeV}$. The induced effects are expected to be of the same order of the neglected isospin-breaking corrections, which are parametrically of order $O(\alpha_{\rm em}) \simeq O( \frac{m_d - m_u}{\Lambda_{QCD}}) \simeq O(1\%)$, thus smaller than the accuracy of the final results presented in Sec.~\ref{sec:numerical_results}.}.  This framework guarantees the automatic $\mathcal{O}(a)$ improvement of parity-even observables~\cite{Frezzotti:2003ni,Frezzotti:2004wz}. Essential information on the ensembles that we use for the present computation are collected in Table~\ref{tab:simudetails}, and we refer the reader to Ref.~\cite{ExtendedTwistedMassCollaborationETMC:2024xdf} for additional details.
 \begin{table}[t]
\begin{ruledtabular}
\begin{tabular}{lcccccc}
\textrm{ID} & $L/a$ & $M_{\pi}$ [\textrm{GeV}] & $a$ [\textrm{fm}] & $Z_{V}$ & $Z_{A}$ & $N_{g}$ \\
\colrule
\textrm{B48} & $48$ & $140.3(3)$ & $0.07948(11)$ &  $0.706354(54)$ & $0.74296(19)$  & $441$ \\
\textrm{B64} & $64$ & $140.2(3)$ & $0.07948(11)$ & $0.706354(54)$ & $0.74296(19)$  & $193$ \\
\textrm{B96} & $96$ & $140.1(3)$ & $0.07948(11)$ & $0.706406(52)$ & 
$0.74261(19)$  & $66$ \\
\textrm{C80} & $80$ & $136.7(3)$ & $0.06819(14)$ & $0.725440(33)$ &  $0.75814(13)$ & $172$  \\
\textrm{D96} & $96$ & $140.8(3)$ & $0.056850(90)$ & $0.744132(31)$ &  $0.77367(10)$  &  $132$
\end{tabular}
\end{ruledtabular}
\caption{$N_{f}=2+1+1$ ETMC gauge ensembles used for quark-connected contribution. We give the spatial extent in lattice units $L/a$, the pion mass $M_{\pi}$, the lattice spacing $a$, the vector ($Z_{V}$) and axial ($Z_{A}$) renormalization constants, and the number of gauge configurations $N_{g}$ analyzed. 
The quark-disconnected contribution is computed exclusively on the B96 ensemble employing 300 gauge configurations.
\label{tab:simudetails}}
\end{table}
We employ the mixed-action lattice setup introduced in~Ref.~\cite{Frezzotti:2004wz}, and described in the appendices of Ref.~\cite{ExtendedTwistedMassCollaborationETMC:2024xdf}. In this setup the action of the valence quarks is discretized in the so-called Osterwalder--Seiler (OS) regularization, namely
\begin{align}
\label{eq:tm_action}
S = \sum_{f=u,d,s,c}\sum_{x} \bar{q}_{f}(x) \left[ \gamma_{\mu}\bar{\nabla}_{\mu}[U] -ir_{f}\gamma^{5}(W^{\rm cl}[U] + m_{\rm cr}) + m_{f}  \right] q_{f}(x)~,
\end{align}
where $W^{\rm cl}[U]$ is the Wilson-clover term~\cite{Sheikholeslami:1985ij}, $m_{\rm cr}$ is the critical mass, $m_{f}$ the quark mass of the flavour $f$ (with $m_{u}=m_{d}=m_{l}$), and $r_{f}=\pm 1$ is the sign of the twisted-Wilson parameter  for the flavour $f$ ($r_{u,c}= -r_{d,s}=1$). 

At each lattice spacing, the mass $m_{s}$ of the strange quark has been set in order to reproduce $m_K= 494.6~{\rm MeV}$, following the prescription recommended for isospin-symmetric QCD in the latest FLAG review~\cite{FlavourLatticeAveragingGroupFLAG:2024oxs}. We performed simulations at three values of the lattice spacing $a$ in the range $0.08-0.056~{\rm fm}$. Moreover, at the coarsest lattice spacing ($a\simeq 0.08~{\rm fm}$) we have generated data on three different lattice volumes, with spatial extent $L\simeq 3.8~{\rm fm}$ (B48), $L\simeq 5.1~{\rm fm}$ (B64), and $L\simeq 7.7~{\rm fm}$ (B96), in order to perform a careful analysis of the finite-size effects (FSEs).  To interpolate the charged kaon we use the following pseudoscalar interpolator
\begin{align}
\label{eq:interpolators}
P_{K}^{\dag}(t,\vec{x}) = \sum_{\vec{y}}\bar{q}_{s}(t,\vec{x}) G^{N}_{t}(\vec{x},\vec{y}) \gamma_5 q_{u}(t,\vec{y}) \;, 
\end{align}
where $G_{t}
(\vec{x},\vec{y})$ is the Gaussian smearing operator
\begin{align}
G_{t}(\vec{x},\vec{y}) = \frac{1}{1+ 6\kappa}\left( \delta_{\vec{x},\vec{y}} + \kappa H_{t}(\vec{x},\vec{y})    \right)~,
\end{align}
where 
\begin{flalign}
H_{t}(\vec{x}, \vec{y}) = \sum_{\mu=1}^{3}\left( U^{\star}_{\mu}(t,\vec{x})\delta_{\vec{x}+\hat{\mu},\vec{y}} + U^{\star\dagger}_{\mu}(t,\vec{x}-\hat{\mu})\delta_{\vec{x}-\hat{\mu},\vec{y}}    \right)~,
\end{flalign}
and we have indicated by $U^{\star}_{\mu}(x)$ the APE-smeared links, defined as in Ref.~\cite{Becirevic:2012dc}.
We use $\kappa=0.4$, and fix on each ensemble the number of smearing-steps $N$ in Eq.~(\ref{eq:interpolators}) so as to obtain a smearing radius $r_{0}= a\sqrt{N}/\sqrt{\kappa^{-1}+6} \simeq 0.4~{\rm fm}$. We have found that this choice of the smearing radius provides the optimal overlap with the ground-state charged kaon~\cite{DiPalma:2024jsp}. Since we work in the rest frame of the kaon, the following two-point correlation function
\begin{align}
C_{\rm 2}(t) = \sum_{\vec{x}} T \langle 0 | P_{K}(t,\vec{x}) P_{K}^{\dag}(0) | 0 \rangle~,  
\end{align}
is used to determine the overlap factor $\langle K^{-}(\vec{0}) | P_{K}^{\dagger}(0) | 0 \rangle$, as well as the kaon mass $m_K$ appearing in Eq.~(\ref{Eq:Def_correlationfunction}).

\begin{figure}
    \centering
    \includegraphics[width=0.45
    \linewidth]{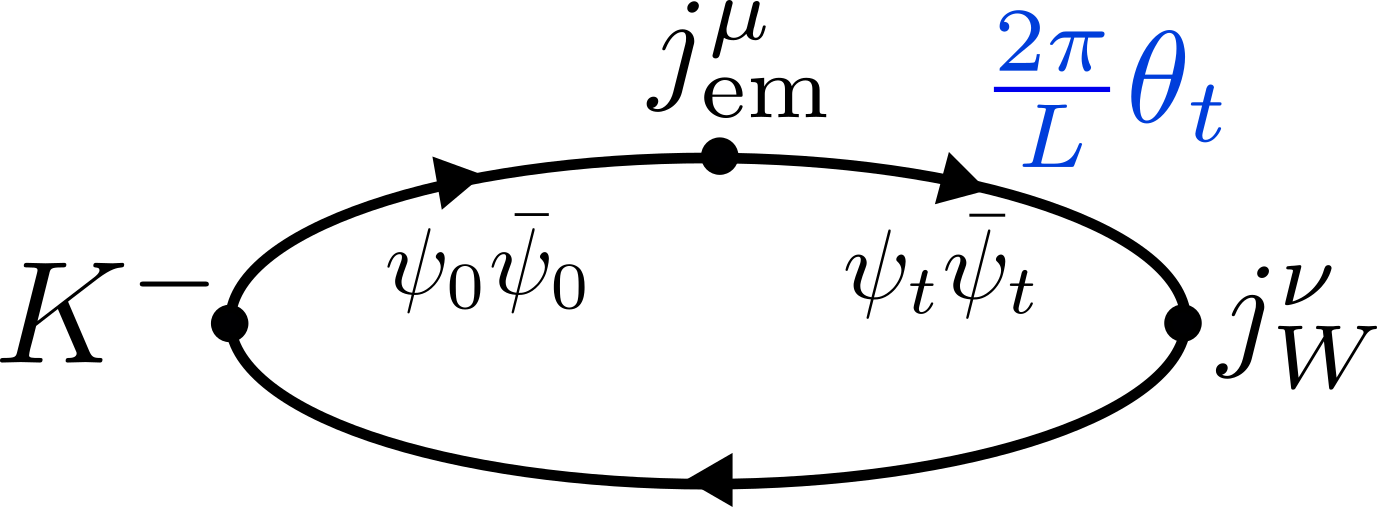}
    \caption{The diagram represents the quark-line connected contribution and illustrates our choice of the spatial boundary conditions, which allows us to set arbitrary values for the meson and photon spatial momenta. The diagram implicitly includes all orders in QCD.}
    \label{fig:twisted}
\end{figure}

We now discuss the strategy we have adopted to fix the momenta for the dominant quark-connected contribution. In this case, as explained in Ref.~\cite{Desiderio:2020oej}, we can use the so-called twisted boundary conditions~\cite{Sachrajda:2004mi,deDivitiis:2004kq} to choose arbitrary (non-quantised) values for the photon and kaon spatial momenta. This is done by building the two propagators, among which  the electromagnetic current is inserted, from the contraction of two different fields that we call $\psi_0(x)$ and $\psi_t(x)$. The two fields have the same mass and quantum numbers but satisfy different spatial boundary conditions, given by
\be
\psi_{0}(x + \bs{n}L) = \psi_{0}(x), \qquad \psi_{t}(x + \bs{n}L) = \exp [ i 2\pi \bs{n}\cdot \bs{\theta}_t/L ] \psi_{t}(x), \qquad \bs{\theta}_t=(0,0,\theta_t)~,
\label{Eq:Boundaryconditions}
\ee
where $\bs{n}$ is an integer valued three-vector, and $\theta_{t}$ is related to the lattice three-momentum of the photon by
\be
k_z = -\frac{2}{a} \sin \bigg[ \frac{a \pi}{L} \theta_t \bigg].
\label{Eq:Momenta}
\ee
In the diagram in Fig.~\ref{fig:twisted}, we explicitly show the propagator of the quark field $\psi_{t}$ which satisfies non-periodic spatial boundary conditions. As discussed in Refs.~\cite{Sachrajda:2004mi,deDivitiis:2004kq}, the use of twisted-boundary conditions induces small violations of unitarity, which are however exponentially vanishing with the lattice spatial extent $L$. 

For the quark-connected contribution, we compute $F_{V,A}(x_{\gamma})$ using five equally spaced values of $x_{\gamma}$, namely
\begin{align}
x_{\gamma}=0.1, \ 0.3, 
  \  0.5, \ 0.7, \ 0.9~,
\end{align}
as well as for $x_{\gamma}=0$\,\footnote{As shown in Sec.~\ref{sec:FF_from_ECF}, at $x_{\gamma}=0$ we are able to extract the axial form factor only.}. Since the vector and axial form factors depend smoothly on $x_{\gamma}$, performing the computations at only five non-zero values of $x_{\gamma}$ is justified. In the OS regularization of the valence fermionic action $S$ in Eq.~(\ref{eq:tm_action}), the weak hadronic current is given by
\begin{align}
j_{W}^{\nu}(x) = \bar{q}_{u}(x)\;\Gamma^{\nu}_{W}\;q_{s}(x) = \bar{q}_{u}(x)\;( Z_{A}\gamma^{\nu} - Z_{V}\gamma^{\nu}\gamma_{5})\; q_{s}(x)~, 
\end{align}
where $Z_{A/V}$ are the renormalization factors ensuring that the Ward identities are satisfied\,\footnote{Note that the renormalisation factors to be used in Twisted-Mass at maximal twist are chirally-rotated with respect to the ones of standard Wilson fermions. This is a consequence of the fact that the up-type and down-type quark fields in the action of Eq.~(\ref{eq:tm_action}) are discretised
with opposite values of the Wilson parameter.}, and which are reported in Tab.~\ref{tab:simudetails}. For the electromagnetic current, we use the exactly conserved point-split current given in Eq.~(B10) of Ref.~\cite{Desiderio:2020oej}.

The calculation of the quark-connected contribution has been performed on all the ETMC gauge ensembles of Tab.~\ref{tab:simudetails}, employing 72 (36) time-wall (spin-diluted) stochastic sources per gauge configuration to evaluate the Wick contraction corresponding to photon emission from the up-quark (strange-quark).
For the quark-disconnected contribution, which proved to be quite small (less than $2.5\%$ of the quark-connected contribution), we restricted the simulations to a single photon energy and a single gauge ensemble (the B96). For this contribution it is not possible to use twisted boundary conditions, and the photon momentum $\vec{k}$ is therefore limited to quantised values $\vec{k}= 2\pi\vec{n}/L$. We have evaluated the disconnected contribution for $\vec{k}=\vec{0}$ and for the first lattice momentum $\vec{k} = (0,0,2\pi/L)$, which for $L\simeq 7.7~$fm (the spatial linear extent of the B96 ensemble) corresponds to $x_{\gamma} \simeq 0.7$.
The quark-disconnected contribution to  $C^{\mu\nu}_{W}(t_{\gamma}, \vec{k}, t_W)$ in Eq.~(\ref{Eq:Def_correlationfunction}) is extracted from the quark-disconnected term 
\begin{align}
C^{\mu\nu}_{\mathrm{3,W;{\rm disc}}}(t_{\gamma}, \vec{k}; t_{W})
&= Z_{V}\,\sum_{\vec{x}}\sum_{\vec{y}_{\gamma}}\,
   e^{-i\,\vec{k}\cdot\vec{y}_{\gamma}}\,
  {\rm \hat{T}} \langle 0 | \!\Bigl[
   \wick[arrows={W->-,W-<-}, below]{%
     \c1 {\bar{q}_u}(x_W)\, \Gamma_{W}^{\nu}\,\c2 q_s(x_W)\;
     \c2 {\bar{q}_s}(x)\,\gamma^5\,\c1 q_u(x)
     {\sum_{q'=u,d,s,c}} \!\! e_{q'}\;
     \c1 {\bar{q}_{q'}}(y_\gamma)\,\gamma^\mu\,\c1 q_{q'}(y_\gamma)
   }
   \Bigr] | 0 \rangle \nonumber \\[12pt]
&= Z_{V}\,\sum_{\vec{x}}\sum_{\vec{y}_{\gamma}}
    e^{-i\vec{k}\cdot\vec{y}_{\gamma}}\,
    \biggl\langle
      \mathrm{Tr}\Bigl[
        \Gamma_{W}^{\nu}
        S_{s}(x,x_{W})\gamma^{5}S_{u}(x_{W},x)\Bigr]\times\!\!\!\!\sum_{q'=u,d,s,c}\!\!e_{q'}\;\mathrm{Tr}\Bigl[\gamma^{\mu}\,S_{q'}\bigl(y_{\gamma},y_{\gamma}\bigr)
      \Bigr]
    \biggr\rangle_{U}\!,
\label{eq:disco_corr}
\end{align}
of the three-point correlation function $C_{3,W}^{\mu\nu}(t_\gamma, \bs{k}; t_W)$, which we compute using a local interpolating operator for the kaon, $P_K^\dagger(x)=\bar{q}_s(x)\gamma_5q_u(x)$.
In Eq.~(\ref{eq:disco_corr}), the arrows specify the contraction of the quark fields, $S_{f}(x_1, x_2)$ is the quark propagator with flavour $f$ from $x_1$ to $x_2$, $x=(0,\vec{x})$, $x_{W}=(t_{W},\vec{0})$, $y_{\gamma}=(t_{\gamma}+t_{W},\vec{y}_{\gamma})$, and $\langle . \rangle_{U}$ indicates the average over the $\rm{SU}(3)$ gauge fields. The trace $\Tr[.]$ appearing in Eq.~(\ref{eq:disco_corr}) is intended over Dirac and colour indices. Note that, while the quark-connected contribution is calculated using the exactly conserved point-split electromagnetic current, the quark-disconnected contribution is computed using the local vector current (for which the one-end-trick discussed in Eq.~(\ref{eq:OET}) below can be readily implemented), which renormalizes with the renormalization constant $Z_{V}$.

The quark-disconnected term receives contribution from all active quark flavours. We have evaluated it by considering the $u$-, $d$- and $s$-quark contributions, but neglecting the contribution of the charm quark which is expected to be much suppressed due to the heavier mass. When the sum over $q'$ in Eq.~(\ref{eq:disco_corr}) is limited to the first three active flavours, the sum of the vector loops can be conveniently rewritten as
\begin{align}
\sum_{q'=u,d,s} e_{q'} \Tr \left[ \gamma^{\mu}  S_{q'} (y_{\gamma},y_{\gamma})\right] = \frac{1}{3} \Tr \left[ \gamma^{\mu} \left(  S_{u} (y_{\gamma},y_{\gamma}) - S_{s}(y_{\gamma},y_{\gamma}) \right)\right]~,
\end{align}
where we exploited the fact that $m_{u}=m_{d}=m_{l}$. As it is evident from the left-hand side of the previous equation, when neglecting the contribution from the charm quark, the disconnected term vanishes at the $\rm{SU}(3)$-symmetric point $m_{s}=m_{l}$. 

In contrast to the quark-connected contribution, the quark-disconnected contribution to the  hadronic tensor at zero momentum, $H_W^{\mu\nu}(\vec{0}; t_W)$, vanishes identically. This can be understood by viewing the Wick contraction in Eq.~(\ref{eq:disco_corr}) as a correlation function in a partially quenched theory, where the sea and valence quark fields are distinct but share the same mass (and critical mass), as well as the same electric charge. From the Ward--Takahashi identity of this partially quenched theory, it immediately follows that
\begin{equation}
\label{eq:zero_mom_disco}
\sum_{t_\gamma} 
C_{3,W;\mathrm{disc}}^{\mu\nu}(t_\gamma, \vec{0}; t_W) \;=\; 0.
\end{equation}
To see this explicitly, let $O(x_1, \dots, x_N)$ be a generic multi-local operator composed of valence quark fields, and define the sea-quark contribution to the electromagnetic current as
\begin{equation}
J_{q'}^\mu(x) \;=\; \bar{q}_{q'}(x)\,\gamma^\mu\,q_{q'}(x)\,.
\end{equation}
The homogeneous Ward--Takahashi identity satisfied by $J_{q'}^\mu(x)$ is
\begin{equation}
\partial_\mu \,C_{O}^\mu(x) \;=\; 0, \quad {\textrm{ where}} \quad C_{O}^\mu(x) \;=\; {\rm \hat{T}}\,\bigl\langle 0 \big|\,J_{q'}^\mu(x)\,O(x_1,\dots,x_N)\,\big|0\bigr\rangle.
\end{equation}
Then, multiplying both sides by $x^\rho$, integrating over all space-time, and performing an integration by parts, yields
\begin{equation}
\label{eq:WTI_disc}
0 \;=\; \int d^4x\,x^\rho\,\partial_\mu C_{O}^\mu(x)
\;=\; - \int d^4x\,C_{O}^\mu(x)\,\delta_\mu^\rho
\;=\; -\int d^4x\,C_{O}^\rho(x) \implies \int d^4x\,C_{O}^\rho(x) \;=\; 0~,
\end{equation}
for any operator $O$. The expression in Eq.~(\ref{eq:zero_mom_disco}) is simply a special case of this result with 
$O(x_1, x_2) = j_W^\nu(x_1)\,P_K^\dagger(x_2)$.
The argument above applies in the infinite-volume limit, where boundary terms vanish under integration by parts~.\footnote{In a finite volume, the Ward-Takahashi identity takes the form $\sum_\mu \nabla^+_\mu C^\mu_0(x) = 0$, where $\nabla^{+(-)}_\mu$ denotes the forward (backward) derivative. One can then show that Eq.~(\ref{eq:WTI_disc}) also holds on a lattice with periodic boundaries by replacing $x_{\rho}$ with the operator $\nabla^+_\rho / (\nabla^+_\rho \nabla^-_\rho + i \epsilon)$. Moreover, at finite lattice spacing, this result holds exactly only when using the point-split conserved current. In contrast, when employing the local current—as in our case—it remains valid only up to discretization effects.}

To evaluate the disconnected contribution, we have found it numerically convenient to use the so-called one-end-trick~\cite{ETM:2008zte,Dinter:2012tt}, generalized here to the up-strange quark doublet $\Psi=(q_{u},q_{s})$. The one-end-trick exploits the exact invariance of the action $S$ in Eq.~(\ref{eq:tm_action}) at $m_l=m_s=0$ under the axial transformation
\begin{align}
\Psi(x) \to  e^{i\alpha\gamma^{5}\tau_{1,2}} \Psi(x)~,\qquad \bar{\Psi}(x) \to \bar{\Psi}(x) e^{i\alpha\gamma^{5}\tau_{1,2}}~,
\end{align}
to derive the following algebraic identity valid for any fixed gauge configuration
\begin{align}
\label{eq:OET}
\Tr \left[ \gamma^{\mu} \left(  S_{u} (y_{\gamma},y_{\gamma}) - S_{s}(y_{\gamma},y_{\gamma}) \right)\right] =  \frac{(m_{l}+m_{s})}{2} \sum_{z} \Tr\left[    
 \gamma^{5} S_{u}(z,y_{\gamma})\gamma^{5}\gamma^{\mu} S_{s}(y_{\gamma},z) - \gamma^{5}S_{s}(z,y_{\gamma})\gamma^{5}\gamma^{\mu} S_{u}(y_{\gamma},z)     \right] ~.
\end{align}

As demonstrated in Ref.~\cite{Giusti:2019kff}, which first introduced a similar method for standard Wilson fermions,  more than an order-of-magnitude error reduction in the calculation of the difference of vector loops, with respect to the naive stochastic estimate of the left-hand side in Eq.~(\ref{eq:OET}), can be achieved by evaluating the right-hand side of the previous equation stochastically, performing the sum over $z$ using standard volume sources. We have employed $100$ volume sources and 300 gauge configurations to evaluate the right-hand side of Eq.~(\ref{eq:OET}). The first trace appearing in Eq.~(\ref{eq:disco_corr}), corresponding to the contraction of the kaon interpolating field with the weak current $j_{W}$, has been evaluated on the same set of gauge configurations employing $100$ wall-time (spin-diluted) stochastic sources placed at the time where the interpolator $P_{K}^{\dag}$ is inserted.
In contrast to the quark-connected contribution, which has been evaluated for all $t_{\gamma}$ at fixed $t_{W}$, in the case of quark-disconnected term we have access to all $t_{W}\in [0,T)$ and $t_W + t_{\gamma} \in [0,T)$, and we exploit this fact together with the following properties of the correlation function under time-reversal
\begin{align}
C^{\mu\nu}_{3,A}(t_{\gamma}, \vec{k}; t_W) &= C^{\mu\nu}_{3,A}(-t_{\gamma},\vec{k}; T-t_{W})~, \nonumber \\[10pt]
C^{\mu\nu}_{3,V}(t_{\gamma}, \vec{k}; t_W) &= -C^{\mu\nu}_{3,V}(-t_{\gamma},\vec{k}; T-t_{W})
\end{align}
to either symmetrize or anti-symmetrize $C_{3,W;{\rm disc}}^{\mu \nu}(t_{\gamma}, \bs{k}; t_W)$, which leads to a reduction of the errors by a factor of about $\sqrt{2}$. The fact that the correlation function is known for all $t_{W}$ allows us to monitor the isolation of initial-state kaon, without the need of performing additional simulations as in the case of the dominant quark-connected contribution, to be discussed extensively in the next section.  

\section{Numerical results}
\label{sec:numerical_results}

\begin{figure}[]
    \centering
\includegraphics[width=0.48\columnwidth]{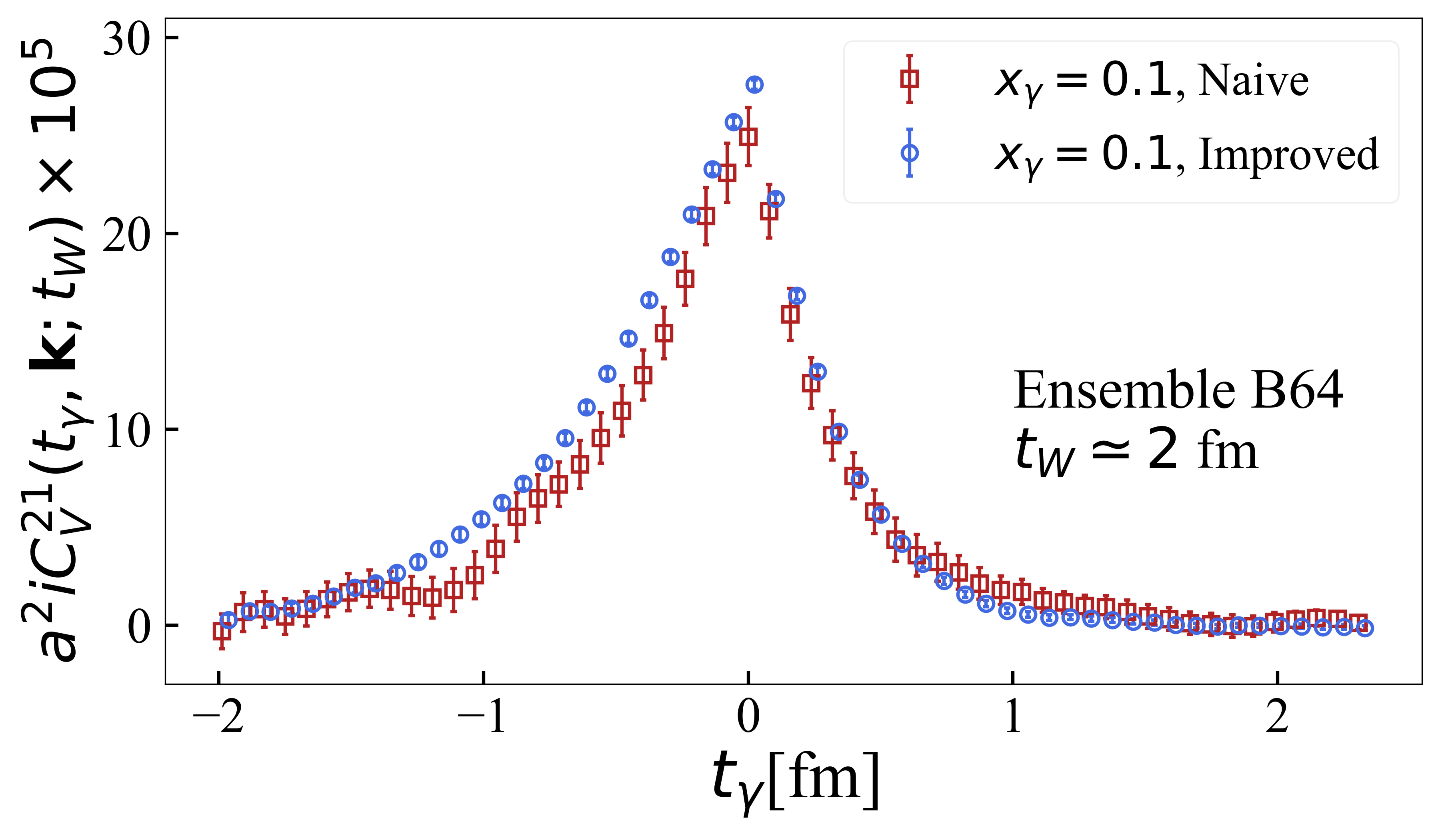}
\includegraphics[width=0.48\columnwidth]{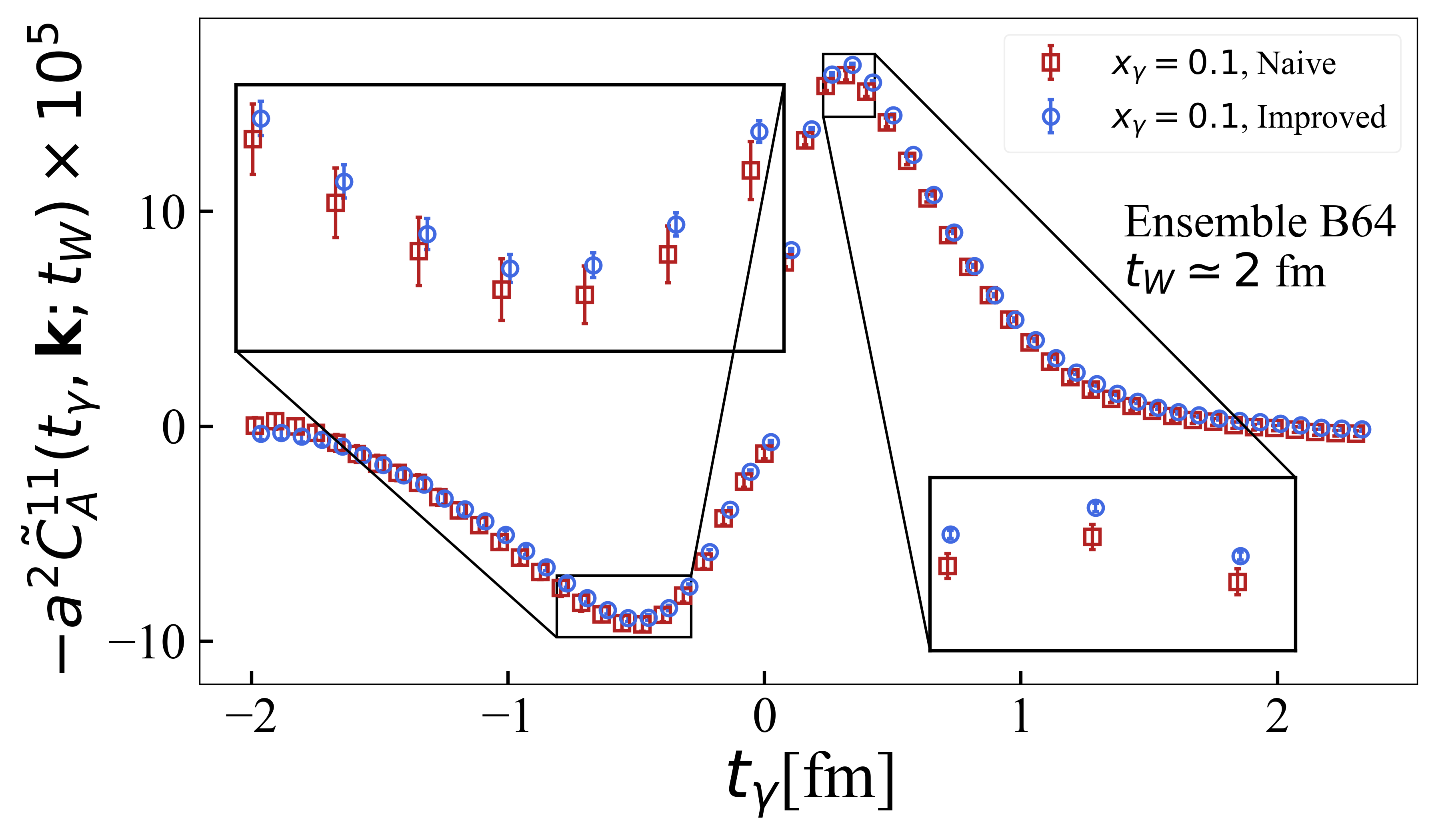}
\caption{Quark-connected vector (left panel) and axial (right panel) correlators, corresponding to photon emission by the up anti-quark, for $x_{\gamma} = 0.1$, as functions of $t_\gamma$. The data correspond to our determination on the B64 ensemble. The red squares and blue circles correspond to the \textit{naive} and \textit{improved} correlators of Eqs.~(\ref{Eq:CtildeAavg}) and~(\ref{Eq:CtildeV}), respectively (see text for details). In the figures the results obtained using the \textit{improved} correlators have been slightly shifted horizontally for better clarity.}
\label{Fig:Correlators}
\end{figure}

In Fig.~\ref{Fig:Correlators}, we show the quark-connected vector and axial correlators corresponding to photon emission by the up
anti-quark\footnote{The two contributions to the quark-connected part of the hadronic tensor
  in Eq.~(\ref{Eq:Def_hadronictensor}), due to photon emission by
  the up anti-quark and strange quarks, are obtained by considering the terms $j_{u}^{\mu}(x)$ and $j_{s}^{\mu}(x)$ in the electromagnetic
  current $j^{\mu}_{\rm em}(x)$ (see Eq.~(\ref{Eq:Def_em_current})).}, for $x_{\gamma} = 0.1$. 
The results, obtained on the B64 ensemble (see Tab.~\ref{tab:simudetails}), are shown as a function of $t_{\gamma}$ for a fixed value of $t_{W} \simeq 2~{\rm fm}$. The red squares in the figure correspond to the \textit{naive} correlators $C^{21}_{V}(t_{\gamma},\vec{k};t_{W})$ and $\widetilde{C}_A^{11}(t_\gamma, \bs{k}; t_W)$, whose integrals over $t_{\gamma}$ are proportional to $F_{V}$ and $F_{A}$, respectively, as in Eqs.~(\ref{Eq:Estimators}) and~(\ref{Eq:EstimatorsFA}). The blue circles in Fig.~\ref{Fig:Correlators} are obtained by averaging the axial correlator over opposite photon momenta $\vec{k}$ and $-\vec{k}$
\be
\widetilde{C}^{11}_A(t_\gamma, \bs{k}; t_W)  \to  \ \frac{1}{2} \bigg[ \widetilde{C}^{11}_A(t_\gamma, \bs{k}; t_W) + \widetilde{C}^{11}_A(t_\gamma, -\bs{k}; t_W)  \bigg],
\label{Eq:CtildeAavg}
\ee
and by considering the following zero-momentum-subtracted correlator in the vector channel:
\be
C_V^{21}(t_\gamma, \bs{k}; t_W) \to \  C_V^{21}(t_\gamma, \bs{k}; t_W) - C_V^{21}(t_\gamma, \bs{0}; t_W) e^{E_\gamma t_\gamma},
\label{Eq:CtildeV}
\ee
where we exploited the fact that $C_V^{21}(t_\gamma, \bs{0}; t_W)$ vanishes up to statistical errors.
At non-zero lattice spacing, the relation between the on-shell photon energy $E_{\gamma}$ and the three-momentum $\bs{k}$ is taken to be
\be
E_{\gamma} = \frac{2}{a} \sinh^{-1}\bigg( \frac{a\vert \bs{k} \vert}{2} \bigg).
\label{Eq:Egamma}
\ee
As the comparison between the red squares and the blue circles shows, using the two \textit{improved} correlators in Eqs.~(\ref{Eq:CtildeAavg}) and~(\ref{Eq:CtildeV}) allows for
a significant reduction of the statistical noise.

\begin{figure}[]
    \centering
\includegraphics[width=0.44\columnwidth]{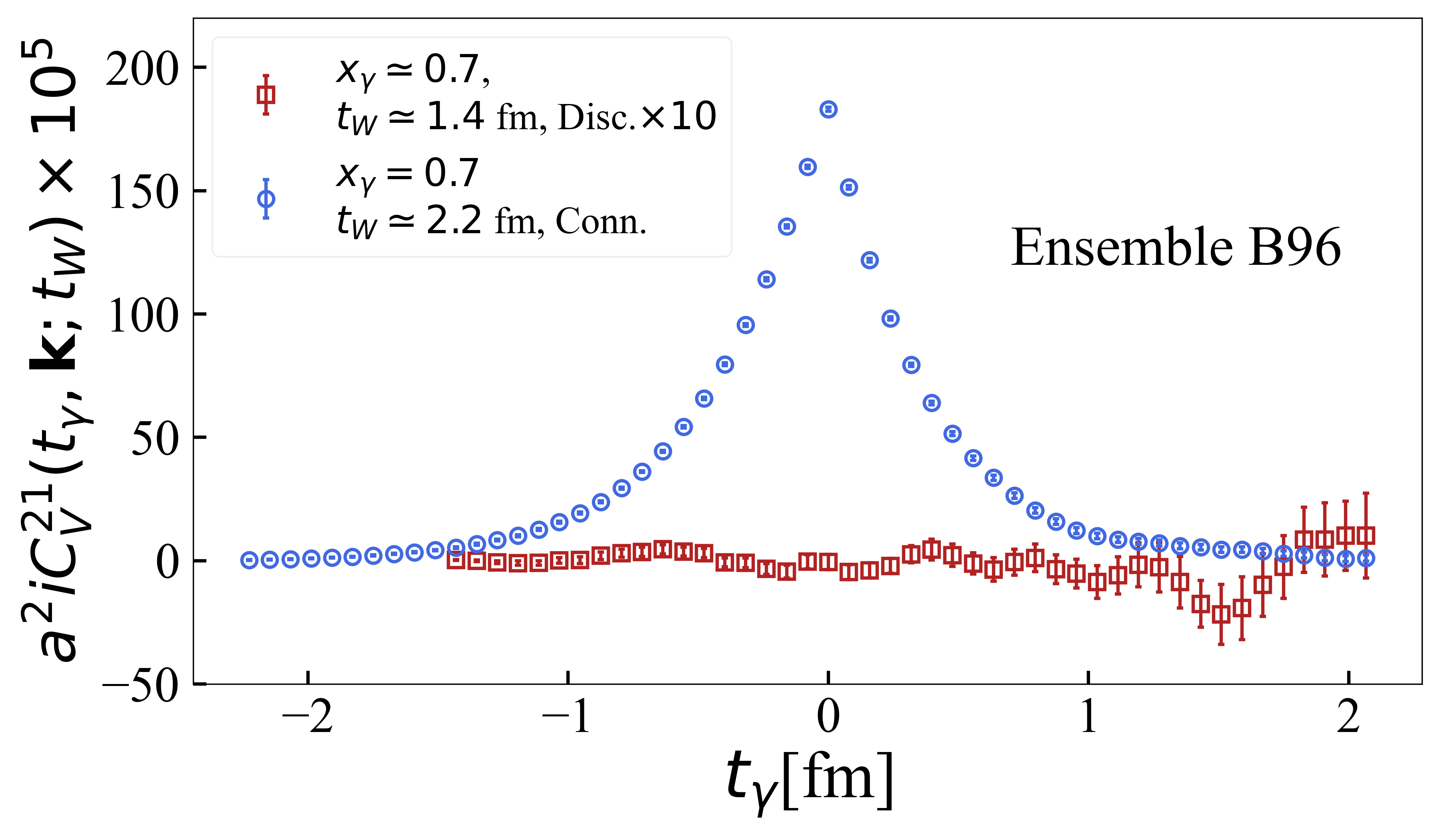}
\includegraphics[width=0.44\columnwidth]{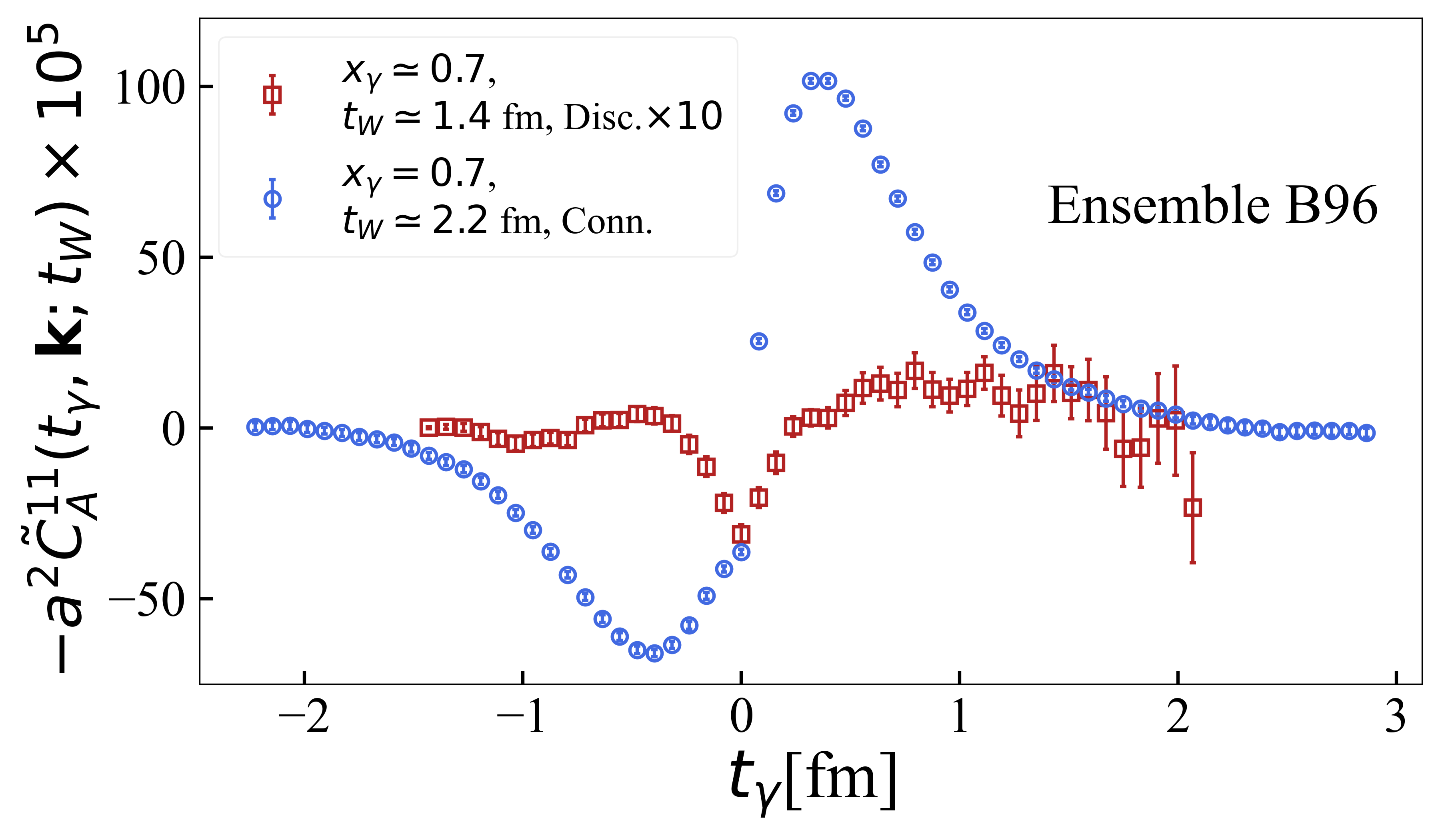}
\caption{Vector (left panel) and axial correlators (right panel) as a function of $t_\gamma$. In each of the two panels, we show both the quark-connected  (blue circles) and the quark-disconnected (red squares) contributions. In the figure, the quark-disconnected contributions have been multiplied by a factor of 10 for visualization purposes. The quark-connected contributions shown in the figure correspond to photon emission by the up anti-quark. The results correspond to our determination on the B96 ensemble,  with $x_\gamma \simeq 0.7$.    }
\label{Fig:DiscCorrelators}
\end{figure}
For $C_V^{21}(t_\gamma, \bs{k}; t_W)$ (left panel of Fig.~\ref{Fig:Correlators}) at $x_{\gamma} = 0.1$, the zero-momentum subtraction reduces the statistical errors by approximately a factor of 6 in the region contributing the most to the integral, $t_\gamma \simeq 0$.
This reduction occurs because, although $C_{V}^{21}(t_{\gamma}, \vec{0}; t_{W})$ vanishes in the limit of infinite statistics, this is not true when only a finite number of gauge configurations  is used.
As a result, $C_{V}^{21}(t_{\gamma}, \vec{k}; t_{W})$ carries a momentum-independent error which, for small values of $\vec{k}$, can exceed the signal itself (the signal is proportional to $k_z$).
The correlated subtraction in Eq.~(\ref{Eq:CtildeV}) removes this unwanted large component of the error by construction, resulting in the improvement illustrated in Fig.~\ref{Fig:Correlators}.
As expected, we found that the reduction in the error becomes progressively smaller as the photon's momentum is increased. We have also investigated averaging the results over opposite momenta, $\vec{k}$ and $-\vec{k}$, in the vector channel, but found no notable reduction in the error compared when compared to the zero-momentum subtraction. \\

For $C_A^{11}(t_\gamma, \bs{k}; t_W)$ (right panel of Fig.~\ref{Fig:Correlators}), at $x_{\gamma}=0.1$ and in the $t_{\gamma}$-region contributing most to the integral, we observe an error reduction of about a factor of $3$ when performing the average over opposite momenta $\vec{k}$ and $-\vec{k}$. The reason behind the improvement, is likely related to the cancellation of error components which are odd in $\bs{k}$ and presumably particularly large in twisted-mass QCD due to due fact that parity-symmetry is broken by cutoff effects. Similar to the vector form factor, the improvement observed for the axial form factor becomes progressively less significant as the photon momentum increases. In the results presented in the following, the \textit{improved} estimators will always be employed~\footnote{Averaging over opposite photon momenta in the case of photon emission by the strange quark does not significantly improve  the precision of the axial correlator, as we explicitly tested for $x_{\gamma}=0.1$. Therefore, the improved estimator in Eq.~(\ref{Eq:CtildeAavg}) is only adopoted in the case of photon emission by the up antiquark.}.

\begin{figure}[]
    \centering
\includegraphics[width=0.48\columnwidth]{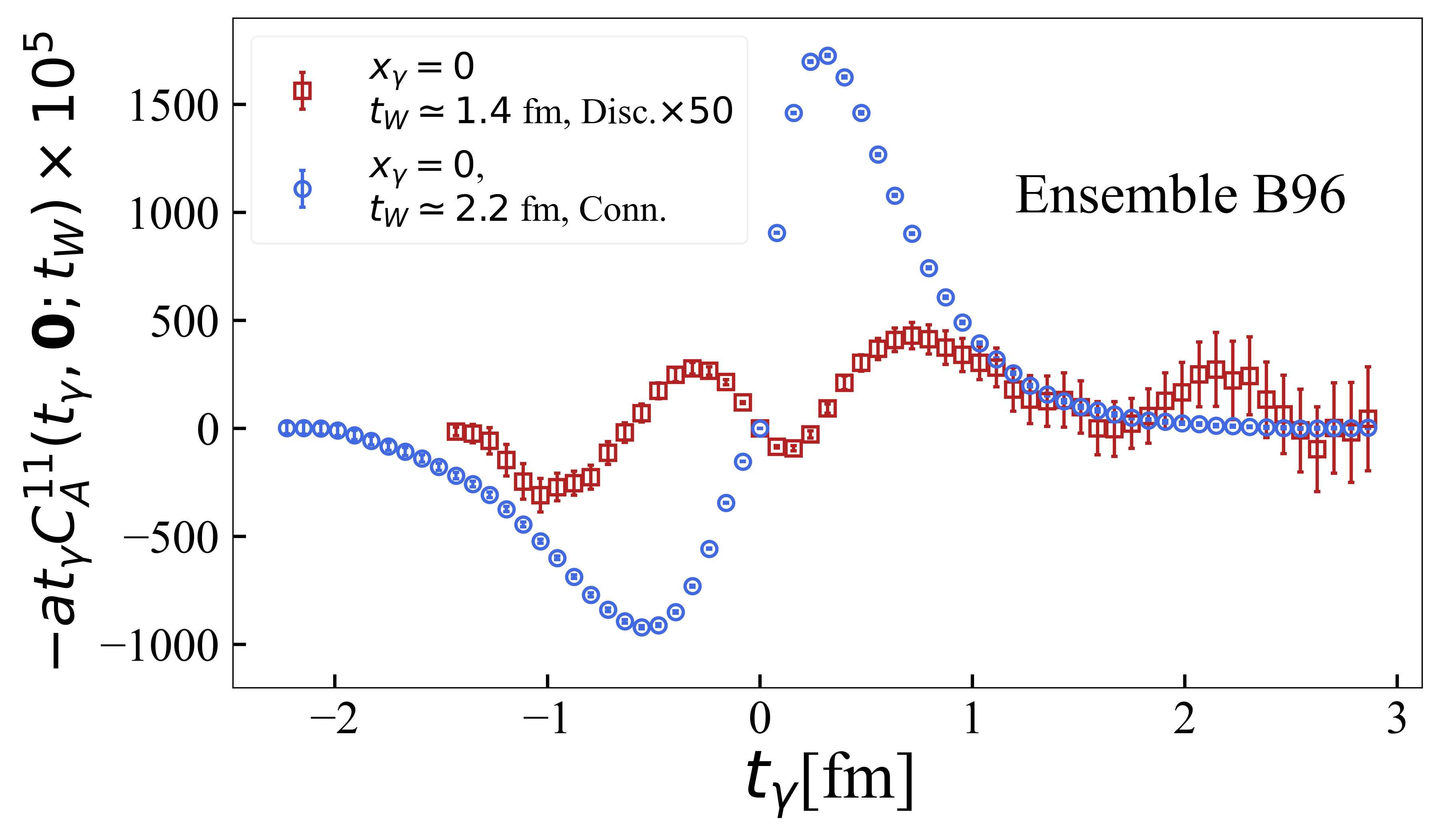}
\caption{Same as in Fig.~\ref{Fig:DiscCorrelators} for $-t_{\gamma} \times C_{A}^{11}(t_{\gamma}, \vec{0}; t_{W})$. As shown in Eq.~(\ref{eq:FA_zero}), the axial form factor at $x_{\gamma}=0$ is obtained by integrating $-t_{\gamma} \times C_{A}^{11}(t_{\gamma}, \vec{0}; t_{W})$ over $t_{\gamma}$. The quark-disconnected contribution has been multiplied by a factor of 50 for visualization purposes.}
\label{Fig:DiscCorrelatorsxg0}
\end{figure}

In Fig.~\ref{Fig:DiscCorrelators}, we compare the quark-connected and 
quark-disconnected contributions to the vector and 
axial correlators, 
as obtained on the B96 ensemble for 
$\displaystyle x_{\gamma} \simeq 0.7$. 
As shown in the figure, the disconnected contributions are suppressed 
by more than an order of magnitude relative to the quark-connected ones 
and display substantially larger relative uncertainties. The quark-disconnected contribution to 
$\displaystyle \widetilde{C}_A^{11}(t_\gamma,\bs{k}; t_W)$ 
(right panel of Fig.~\ref{Fig:DiscCorrelators}) is clearly non-zero, although it has opposite signs in the regions  $t_{\gamma} > 0$ and $t_{\gamma} < 0$, which leads to 
significant cancellations after integrating over $t_{\gamma}$
(a similar behaviour is observed for the quark-connected contribution). 
The disconnected contribution to the vector correlator 
(left panel of Fig.~\ref{Fig:DiscCorrelators}) is even more suppressed 
than the axial one, and remains compatible with zero  for all $\displaystyle t_{\gamma}$. In Fig.~\ref{Fig:DiscCorrelatorsxg0}, we present the quark-connected (blue circles) and quark-disconnected (red squares) contributions to $t_{\gamma} \times C_A^{11}(t_\gamma, \bs{0};t_W)$. As shown in Eq.~(\ref{eq:FA_zero}), the  axial form factor at $ x_\gamma = 0$ is obtained from the integral of $t_{\gamma} \times C_A^{11}(t_\gamma, \bs{0};t_W)$ over $t_\gamma$. Although the disconnected contribution is suppressed relative to the connected one, a clear signal is visible.

We now describe in detail our estimates of the three most relevant sources of systematic errors: those arising from i)~the truncation of the integral over $t_{\gamma}$, ii)~finite-$t_{W}$ effects and iii)~finite-size effects. We then discuss the extrapolation of our results to the continuum limit.

\subsection{Effects from the truncation of the integration over $t_{\gamma}$}
\label{Sec:temporaltrunc}

To discuss our strategy for handling the effects from truncating the integration over $t_{\gamma}$, we start by separating the contributions from the two different time-orderings to $H_{W}^{\mu\nu}(\vec{k}; t_{W})$ as 
\be
H^{\mu\nu}_W(\vec{k}; t_{W}) = H^{\mu\nu}_{W,1}(\vec{k}; t_{W}) + H^{\mu\nu}_{W,2}(\vec{k}; t_{W}),
\label{Eq:TOseparation}
\ee
where 
\be
    H^{\mu\nu}_{W,1}(\vec{k}; t_{W}) = \int_{-\infty}^0 d t_\gamma ~ C^{\mu\nu}_W(t_\gamma, \bs{k}; t_{W}), \qquad \qquad  H^{\mu\nu}_{W,2}(\bs{k}; t_{W}) = \int_{0}^{\infty} d t_\gamma ~ C^{\mu\nu}_W(t_\gamma,\bs{k}; t_{W}).
\label{Eq:TOdefinitions}
\ee
In what follows, we refer to the time-ordering $t_\gamma < 0$ (i.e., when the electromagnetic current acts before the weak current) as the \emph{first time-ordering}, and the time-ordering $t_\gamma > 0$ (i.e., when the electromagnetic current acts after the weak current) as the \emph{second time-ordering}. 

As already pointed out in Sec.~\ref{sec:FF_from_ECF}, the lattice correlator $C^{\mu\nu}_{W}(t_{\gamma}, \bs{k}; t_{W})$ can be used only for times $t_{\gamma} \in [-t_{W}, t_{\gamma}^{\rm max}]$ where $t_{\gamma}^{\rm max}\ll T$ in order to avoid significant around-the-world effects (a safe choice is $t_{\gamma}^{\rm max} = T/2 - t_{W}$). Thus, one must ensure that the contributions from the missing tails of the correlator to $H_{W,1}^{\mu\nu}(\bs{k}; t_{W})$ and $H_{W,2}^{\mu\nu}(\bs{k};t_{W})$ are small compared to their statistical errors.

It was first pointed out in Ref.~\cite{Tuo:2021ewr} that one can accelerate the convergence of the lattice integral towards its asymptotic value by analytically continuing
$C_{W}^{\mu\nu} (t_{\gamma}, \vec{k}; t_{W})$
into the region inaccessible directly using lattice data, under the assumption of ground-state dominance.
We begin by illustrating this strategy for the first time-ordering contribution.

The central idea is to introduce an intermediate time $t_{\gamma}^{*} \in [0, t_{W}]$ and define
\begin{equation}
H^{\mu\nu}_{W,1}(t_{\gamma}^{*}, \vec{k}; t_W) 
\,=\,
\int_{-\infty}^{- t_{\gamma}^*} d t_\gamma  \; C^{\mu\nu}_{W; \mathrm{asy}}(t_\gamma, \vec{k};  t_W)\,
\;+\; \int_{- t_{\gamma}^*}^{0} d t_\gamma  \; C^{\mu\nu}_W(t_\gamma,\vec{k}; t_{W})\,.
\label{Eq:1TO-rewrite}
\end{equation}
The second integral on the right-hand side of Eq.~\eqref{Eq:1TO-rewrite} is computed directly from the lattice data.\footnote{Throughout this work, we use continuum notation. However, on the lattice, the integral over $t_\gamma$ is understood as a finite sum over $t_\gamma = n a$, with $n \in \mathbb{Z}$. By convention, we include the point $t_{\gamma}=0$ in the definition of the first time-ordering.}
For the first integral, we assume that for $t_{\gamma} \leq -t_{\gamma}^{*}$, the correlator is dominated by the lightest intermediate state, given by the following asymptotic (asy) approximation:
\begin{equation}
C^{\mu\nu}_{W; \mathrm{asy}}(t_\gamma, \vec{k}; t_{W}) 
\,\equiv\, 
\exp \Big[\big(\mathcal{E}_1 + E_{\gamma} - m_K\big)\,\big(t_\gamma + t_\gamma^*\big) \Big]\,
C^{\mu\nu}_W\big(-t_\gamma^*, \vec{k}; t_W \big),
\quad 
t_\gamma \leq -t_\gamma^*~,
\label{Eq:LargeTimeCorr} 
\end{equation}
where $\mathcal{E}_1$ is the energy of the lightest hadronic intermediate state (with momentum $\vec{k}_{1} = -\vec{k}$) propagating between the electromagnetic and weak currents in the first time-ordering.
Substituting Eq.~\eqref{Eq:LargeTimeCorr} into the first integral of Eq.~\eqref{Eq:1TO-rewrite} yields
\begin{equation}
H^{\mu\nu}_{W,1}(t_{\gamma}^{*}, \vec{k}; t_W) 
\,=\,
 \frac{1}{E_\gamma + \mathcal{E}_1 - m_K}\,
C^{\mu\nu}_W\big(-t_\gamma^*, \vec{k}; t_W\big)\,
\;+\; \int_{- t_\gamma^*}^{0} dt_{\gamma} \; C^{\mu\nu}_W(t_\gamma, \vec{k}; t_W)\,.
\label{Eq:1TODiscretization}
\end{equation}
The first term estimates the contribution of the tail of the correlator assuming ground-state dominance, thereby improving convergence. We can then compute the estimators of the form factors using Eqs.~(\ref{Eq:Estimators}) and~(\ref{Eq:EstimatorsFA}) for every  $t_\gamma^*$ as 
\begin{align}
F_{A,1}(t_{\gamma}^{*},x_{\gamma}; t_{W} ) =  -\frac{1}{E_{\gamma}}\bigg[ H_{A,1}^{11}(t_{\gamma}^{*}, \bs{k}; t_W)   - H_{A,1}^{11}(t_{\gamma}^{*}, \bs{0}; t_W)  \bigg]~, \qquad F_{V,1}(t_{\gamma}^{*},x_{\gamma}; t_{W} ) =  \frac{i}{k_{z}}H_{V,1}^{21}(t_{\gamma}^{*}, \bs{k}; t_W)\,.
\end{align}

The next step is to examine the stability of 
$F_{V/A,1}(t_{\gamma}^{*}, x_{\gamma}; t_{W})$
as a function of the switching time $t_{\gamma}^{*}$.
Once a plateau is observed (i.e., a sufficiently flat behavior in $t_{\gamma}^{*}$),
a constant fit is performed to extract the  value of the contribution from the first time-ordering to the axial and vector form factors. 

The procedure for the second time-ordering is analogous.
One introduces $t_{\gamma}^{*} \in [0, t_{\gamma}^{\rm max}]$ and defines
\begin{align}
\label{Eq:2TO-rewrite}
H^{\mu\nu}_{W,2}(t_{\gamma}^{*}, \vec{k}; t_W) 
&=\,
\int_{0}^{t_{\gamma}^{*}} d t_\gamma  \; C^{\mu\nu}_W(t_\gamma,\vec{k}; t_{W}) 
\;+\; 
\int_{t_{\gamma}^{*}}^{\infty} d t_\gamma  \; C^{\mu\nu}_{W; \mathrm{asy}}(t_\gamma, \vec{k};  t_W), \\[6pt]
C^{\mu\nu}_{W; \mathrm{asy}}(t_\gamma, \vec{k}; t_{W}) 
&\equiv\,
\exp \Big[-\big(\mathcal{E}_2 - E_{\gamma}\big)\,\big(t_\gamma - t_\gamma^*\big) \Big]\,
C_W^{\mu\nu}\big(t_\gamma^*, \vec{k} ; t_W\big),
\quad 
t_\gamma \geq t_\gamma^*~,
\end{align}
where $\mathcal{E}_2$ is the energy of the lightest hadronic state (with momentum $\vec{k}_{2} = \vec{k}$) propagating between the weak and electromagnetic currents in the second time-ordering. 
The states with energies $\mathcal{E}_1$ and $\mathcal{E}_2$ will be identified below.
Performing the second integral in the right-hand side of Eq.~\eqref{Eq:2TO-rewrite} gives
\begin{equation}
\label{Eq:2TODiscretization}
H^{\mu\nu}_{W,2}(t_{\gamma}^{*}, \vec{k}; t_W) 
\,=\,
\int_{0}^{t_{\gamma}^{*}} dt_\gamma \; C^{\mu\nu}_W(t_\gamma, \vec{k}; t_W) 
\;+\;
\frac{1}{\mathcal{E}_2 - E_{\gamma}} \; 
C^{\mu\nu}_W\big(t_\gamma^*, \boldsymbol{k}; t_W\big)\,.
\end{equation}
As for the contribution from the first time ordering discussed above, we then define
\begin{equation}
F_{A,2}(t_{\gamma}^{*},x_{\gamma}; t_{W} ) =  -\frac{1}{E_{\gamma}}\bigg[ H_{A,2}^{11}(t_{\gamma}^{*}, \bs{k}; t_W)   - H_{A,2}^{11}(t_{\gamma}^{*}, \bs{0}; t_W)  \bigg]~, \qquad F_{V,2}(t_{\gamma}^{*},x_{\gamma}; t_{W} ) =  \frac{i}{k_{z}}H_{V,2}^{21}(t_{\gamma}^{*}, \bs{k}; t_W)~,
\end{equation}
and then monitor the stability of $F_{V/A,2}(t_{\gamma}^{*}, x_{\gamma}; t_{W})$ as a function of $t_{\gamma}^{*}$.

\begin{figure}[]
    \centering
\includegraphics[width=1.\columnwidth]{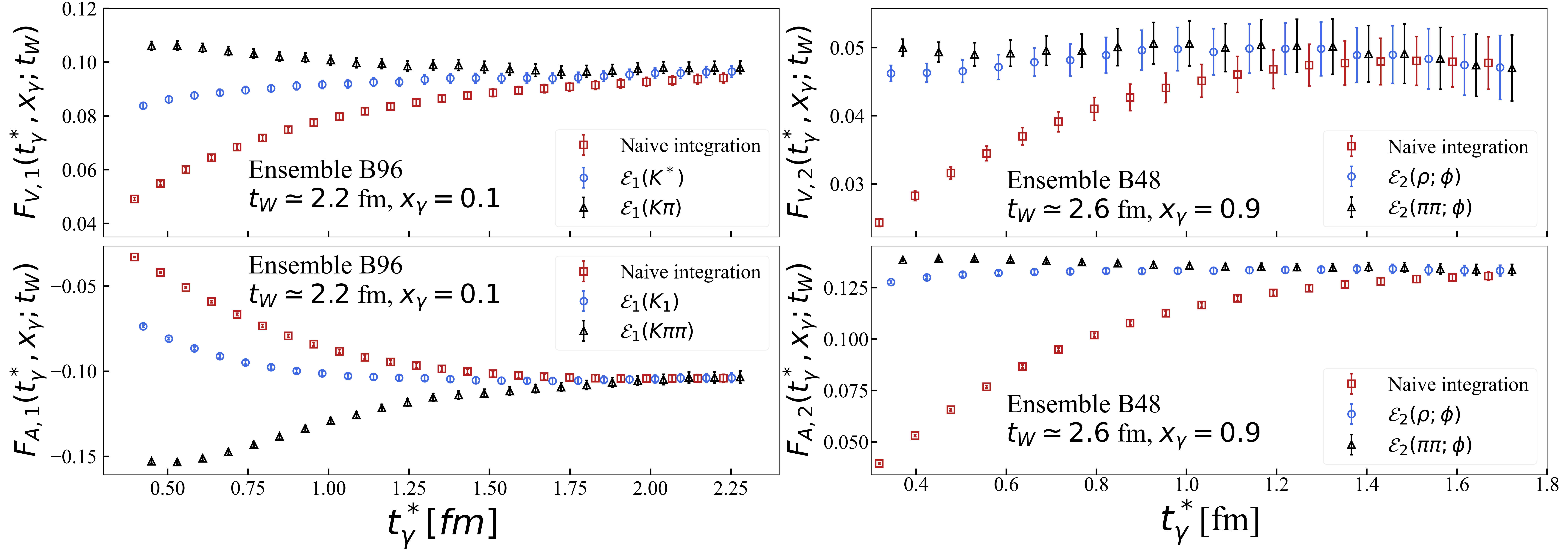}
\caption{
Quark-connected contribution to the vector (top panels) and axial (bottom panels)
form factors, shown as functions of the switching time $t_\gamma^*$. In the left (right) panels we show the contribution from the first (second) time-ordering. The blue circles and black triangles correspond 
to results obtained using the analytic-continuation method, where the ground-state 
energy is set to either the resonance energy ($K^{*}, K_{1}, \rho$) treated as 
a stable state (blue circles) or the energy of its decay product (black triangles), as described in the text. 
The red squares correspond to the results obtained without using the analytic continuation method. 
For clarity, the data points are slightly shifted horizontally.}
\label{Fig:FF_continuation}
\end{figure}
The formulae in Eqs.~(\ref{Eq:1TODiscretization}) and (\ref{Eq:2TODiscretization}) can be straightfowardly modified in the case of the axial form factor at $x_{\gamma}=0$, where Eq.~(\ref{eq:FA_zero}) must be used.

As for the choice of the ground-state energies $\mathcal{E}_{1}$ and $\mathcal{E}_{2}$, we proceed as follows.
In the first time-ordering, the propagating intermediate states have $\bar{u}s$ flavor quantum numbers.
Moreover, in the vector and axial form factors, these states carry quantum numbers $J^{P} = 1^{-}$ and $J^{P} = 1^{+}$, respectively.
Consequently, for the first time-ordering, the lowest-lying intermediate state in the vector (axial) channel is the $K^{*}$ ($K_{1}$) resonance.
Both $K^{*}$ and $K_{1}$ are unstable under QCD interactions. The $K^{*}$ decays into a $K\pi$ final state, whereas the $K_{1}$ primarily decays into  $K^*\pi$ and $\rho K$ states that subsequently decay into  $K\pi\pi$ states.

We consider two different choices for $\mathcal{E}_{1}$.
First, we treat the $K^{*}$ and $K_{1}$ resonances as stable and set
\begin{equation}
\label{eq:E1}
\mathcal{E}_{1} 
= 
\sqrt{m_{K^*}^{2} + \lvert \vec{k} \rvert^{2}} 
\quad (\text{vector channel}), 
\qquad
\mathcal{E}_{1} 
= 
\sqrt{m_{K_{1}}^{2} + \lvert \vec{k} \rvert^{2}}
\quad (\text{axial channel}).
\end{equation}
Alternatively, in the vector channel, we consider a second value for $\mathcal{E}_{1}$ obtained by replacing $m_{K^*}$ in Eq.~(\ref{eq:E1}) with the energy of a non-interacting $K\pi$ system, where the kaon and pion each carry the lowest back-to-back quantized momentum ($2\pi/L$).
Similarly, in the axial channel, we considered a second option for $\mathcal{E}_{1}$ obtained by replacing $m_{K_1}$ with the energy of a non-interacting $K\pi\pi$ system, where the kaon and one of the two pions carry back-to-back momentum $\lvert \vec{p}_{\pi} \rvert = \lvert \vec{p}_{K} \rvert = 2\pi/L$, while the other
pion is at rest.
Among the various ways to distribute momenta among the kaon and the two pions, this configuration yields the lowest energy.

In the second time-ordering, the propagating states are flavor-neutral, as they
share the same quantum numbers as the electromagnetic current $j_{\mathrm{em}}^{\mu}$. When only the quark-connected contribution is considered the lowest-lying intermediate state is either the $\rho$- or
the $\phi$-resonance, depending on whether the photon is emitted by the up anti-quark
or by the strange quark.
For the quark-disconnected contribution, the lowest-lying intermediate state
is the $\rho$-resonance. 

Both the $\rho$- and the $\phi$-resonances are not QCD-stable and
decay into $\pi\pi$ and (mainly into) $KK$ states, respectively.
\begin{figure}[t]
    \centering
\includegraphics[width=0.9\columnwidth]{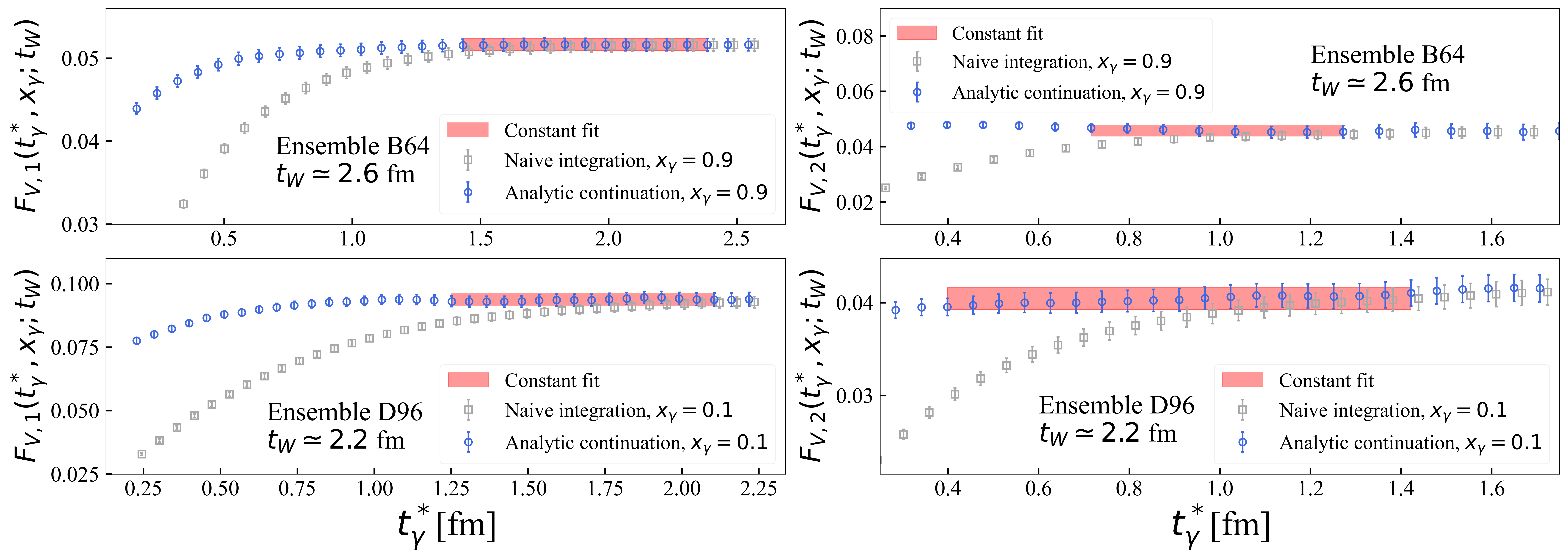}
\caption{
Quark-connected contribution to $F_{V,1}(t_\gamma^*, x_\gamma; t_W)$ (left panels) and $F_{V,2}(t_\gamma^*, x_\gamma; t_W)$ (right panels)  as functions of the switching time $t_\gamma^*$ (see text for details). The results correspond to our determination on the B64 ensemble for $x_{\gamma}=0.9$ (top panels) and on the D96 ensemble for $x_{\gamma}=0.1$ (bottom panels).
The results obtained using the analytic continuation method are shown as blue circles. The red bands indicate the results of a constant fit in the region where $F_{V, 1/2}(t_\gamma^*, x_\gamma; t_W)$ show a plateaux. The results from naive integration (grey squares) are also shown for reference. To improve readability, data points have been slightly shifted horizontally. }
\label{fig:plateaus_FV_conn}
\end{figure}
\begin{figure}[t]
    \centering
\includegraphics[width=0.9\columnwidth]{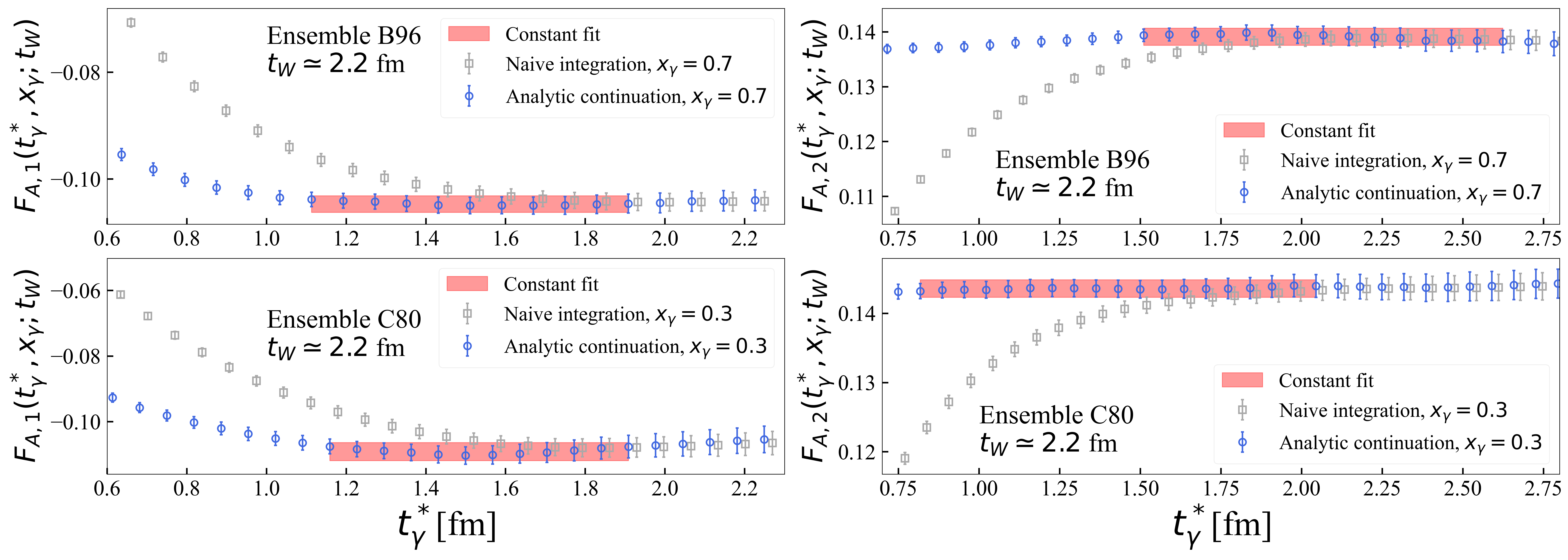}
\caption{Quark-connected contribution to $F_{A,1}(t_\gamma^*, x_\gamma; t_W)$ (left panels) and $F_{A,2}(t_\gamma^*, x_\gamma; t_W)$ (right panels)  as functions of the switching time $t_\gamma^*$ (see text for details). The results correspond to our determination on the B96 ensemble for $x_{\gamma}=0.7$ (top panels) and on the C80 ensemble for $x_{\gamma}=0.3$ (bottom panels).
The results obtained using the analytic continuation method are shown as blue circles. The red bands indicate the results of a constant fit in the region where $F_{A, 1/2}(t_\gamma^*, x_\gamma; t_W)$ show a plateaux. The results from naive integration (grey squares) are also shown for reference. To improve readability, data points have been slightly shifted horizontally. }
\label{fig:plateaus_FA_conn}
\end{figure}
However, in the case of the $\phi$-resonance, the spatial lattice extent $L$ is not
large enough to allow the $\phi \rightarrow KK$ decay, which is kinematically forbidden. Therefore, when focusing on the quark-connected contribution in which the photon
is emitted by the strange quark, we set
\begin{equation}
  \mathcal{E}_{2} \;=\; \sqrt{\,m_{\phi}^{2} \;+\; \lvert \vec{k} \rvert^{2}}.
\end{equation}
In all other cases, we either set
\begin{align}
  \mathcal{E}_{2} \;=\; \sqrt{\,m_{\rho}^{2} \;+\; \lvert \vec{k} \rvert^{2}}
\end{align}
or replace $m_{\rho}$ with the energy of a non-interacting $\pi\pi$ system,
where the two pions carry back-to-back momenta $2\pi/L$.

In the panels of Fig.~\ref{Fig:FF_continuation}, we show, for illustration, the connected contribution to
$F_{V/A,1}(t_{\gamma}^{*}, x_{\gamma} = 0.1; t_W)$ (on the B96 ensemble) 
and to $F_{V/A,2}(t_{\gamma}^{*}, x_{\gamma} = 0.9; t_W)$ (on the B48 ensemble) 
as functions of $t_{\gamma}^{*}$. 
\begin{figure}[]
    \centering
\includegraphics[width=0.9\columnwidth]{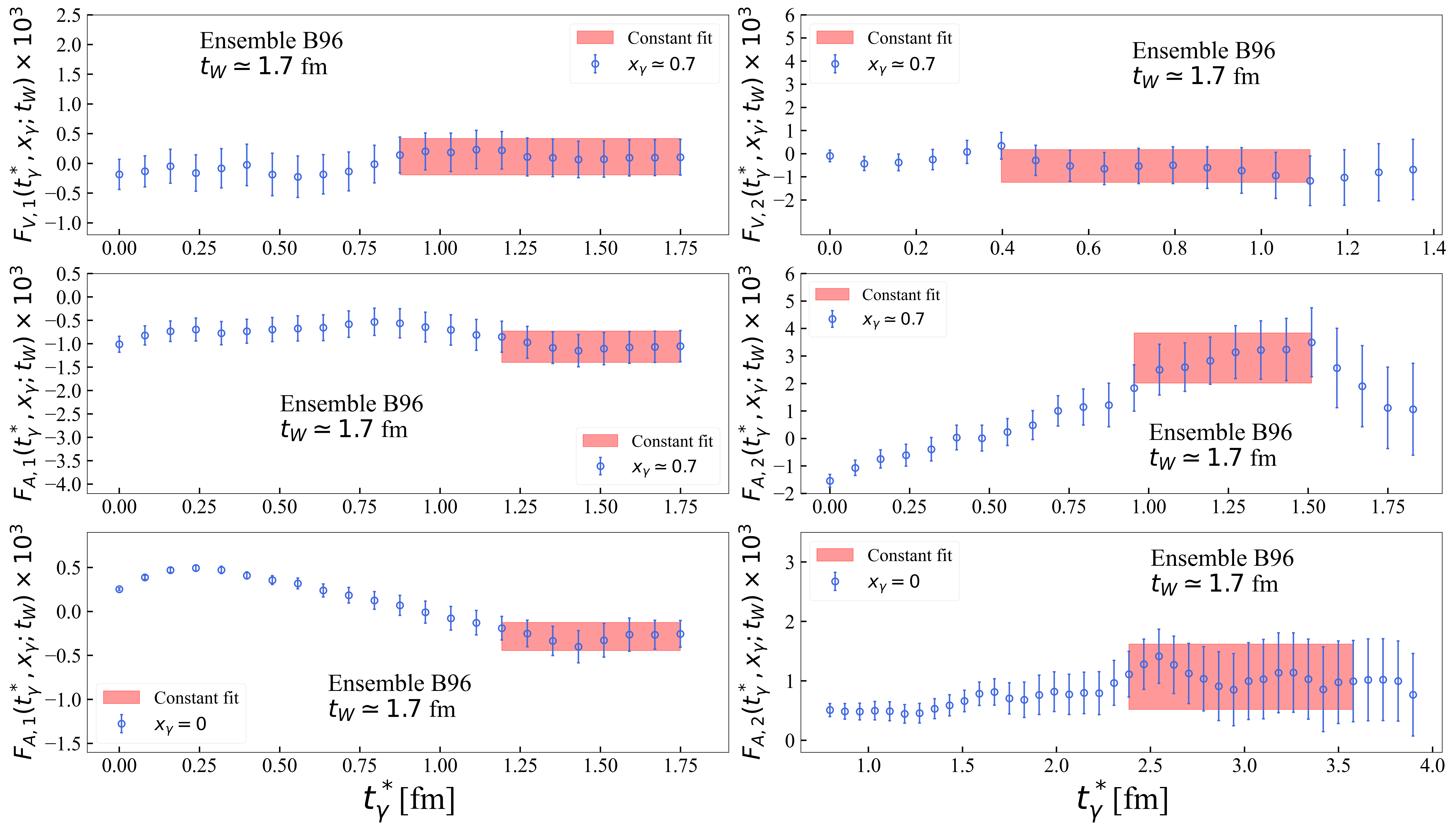}
\caption{Quark-disconnected contribution to the vector (top panels) and axial (bottom panels) form factors, shown as functions of the switching time $t_\gamma^*$ (see text for details) and for $t_W \simeq 1.7~\mathrm{fm}$. In the left (right) panels, we show the contribution from the first (second) time-ordering. The form factors are computed on the B96 ensemble for $x_\gamma \simeq 0.7$ and, for the axial form factor, also at $x_\gamma = 0$. The red bands correspond to the results of a constant fit to the data in a region where they display a plateau. For improved readability, data points have been slightly shifted horizontally.}
\label{fig:plateaus_F_disc}
\end{figure}
The results obtained on all other ensembles and $x_{\gamma}$ values are very similar to the ones shown in the figures. The blue circles and black triangles correspond 
to results obtained using the analytic-continuation method, where the ground-state 
energy is set to either the resonance energy ($K^{*}, K_{1}, \rho$) treated as 
a stable state (blue circles) or the energy of its decay product (black triangles), as described above. The squares represent the results obtained without analytic continuation, i.e., by setting
\begin{align}
  C^{\mu\nu}_{W;\mathrm{asy}}(t_{\gamma}, \vec{k}; t_W) \;=\; 0
\end{align}
in Eqs.~(\ref{Eq:1TODiscretization}) and~(\ref{Eq:2TODiscretization}). As seen in the figure, the analytic-continuation method converges very rapidly, 
exhibiting stable plateaux for $t_{\gamma}^{*} \simeq 1\text{--}1.5\,\mathrm{fm}$. Moreover, the blue circles and black triangles are in 
excellent agreement within the plateau-region, which is reassuring. By contrast, the 
naive integration (red squares) converges only at much larger $t_{\gamma}^{*}$, 
and in some cases convergence is not fully reached within the displayed range. We 
also find slightly better convergence when using the resonance energies for 
$\mathcal{E}_{1}$ and $\mathcal{E}_{2}$, and hence adopt this choice in what 
follows.

In Figs.~\ref{fig:plateaus_FV_conn} and \ref{fig:plateaus_FA_conn}, we illustrate how the quark-connected contributions to the form factors $F_{V}$ (Fig.~\ref{fig:plateaus_FV_conn}) and $F_{A}$ (Fig.~\ref{fig:plateaus_FA_conn}) are extracted from the plateaux of the estimators $F_{V/A,1/2}\bigl(t_{\gamma}^{*}, x_{\gamma}; t_{W}\bigr)$.
The results are shown at fixed values of $t_{W}$ for several choices of $x_{\gamma}$ (see the legends in the figures). Blue circles (gray squares) represent results obtained with (without) the analytic continuation. Where the blue circles show a clear plateaux in $t_{\gamma}^{*}$, we perform a constant fit, the outcome of which is depicted by the coloured bands. In all cases, the asymptotic regime begins at times of about $1$--$1.5~\mathrm{fm}$, allowing us to determine the form factors with high confidence.

In Fig.~\ref{fig:plateaus_F_disc}, we present the corresponding plots for the quark-disconnected contribution, obtained on the B96 ensemble. The figure shows our results for the vector and axial form factors at $x_{\gamma}\simeq 0.7$, as well as the axial form factor $F_{A}(x_{\gamma}=0)$, obtained from the zero-momentum disconnected correlator via Eq.~\eqref{eq:FA_zero}. Results are shown for $t_{W} \simeq 1.7~\mathrm{fm}$. A similar behaviour has been observed also for other choices of $t_W$. As can be seen, the contributions from both the first and second time-orderings to the vector form factor are consistent with zero within their statistical uncertainties. These uncertainties are, however, significantly smaller than those found in the dominant connected contribution.  For the axial form factor on the other hand, the contributions from both time-orderings are non-zero for the two values of $x_\gamma$ considered in this paper. Notably, the relative uncertainties at $x_{\gamma}=0$ are much smaller than at $x_{\gamma}\simeq 0.7$.

\subsection{Finite-$t_{W}$ effects}
\label{sec:tw-dependence}
Having presented our strategy for determining the axial and vector form factors 
$F_{A}(x_{\gamma};t_{W})$ and $F_{V}(x_{\gamma}; t_{W})$ 
at fixed $t_{W}$, we now turn to the systematic errors associated 
with the $t_{W} \to \infty$ extrapolation required to isolate the 
initial-state kaon. We begin by focusing on the dominant quark-connected contribution. 
As discussed earlier, the quark-connected component of the correlation function
$C_{3;W}^{\mu\nu}(t_{\gamma}, \vec{k}; t_{W})$ 
is evaluated for all $t_{\gamma}$ at fixed values of 
$t_{W}$.

\begin{figure}[]
    \centering
\includegraphics[width=1.\columnwidth]{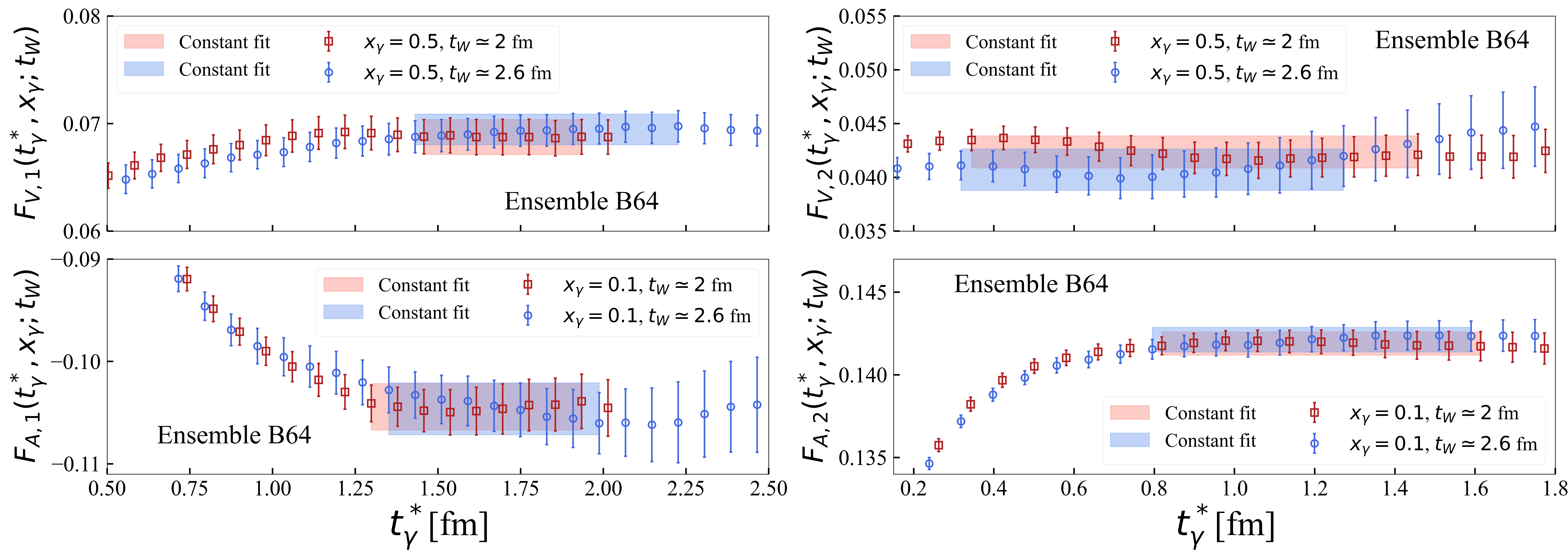}
\caption{Quark-connected contribution to the vector (top panels) and axial (bottom panels) form factors, shown as functions of the switching time $t_\gamma^*$ (see text for details). In the left (right) panels, we show the contribution from the first (second) time-ordering.
Results are presented for the B64 ensemble with $x_\gamma = 0.5$ for $F_{V,1/2}(t_\gamma^*, x_\gamma; t_W)$ and $x_\gamma = 0.1$ for $F_{A,1/2}(t_\gamma^*, x_\gamma; t_W)$. Red squares and blue circles correspond to $t_W \simeq 2~\mathrm{fm}$ and $t_W \simeq 2.6~\mathrm{fm}$, respectively. The coloured bands represent constant fits performed on the data points of the corresponding colors.}
\label{fig:tw_dependence}
\end{figure}

At an initial stage of our simulations, we computed the quark-connected contribution to 
$C_{3;W}^{\mu\nu}(t_{\gamma}, \vec{k}; t_{W})$
for multiple values of $t_{W}$, but with limited statistics 
($O(50)$ gauge configurations), on the B64 ensemble. 
The goal was to estimate how large $t_{W}$ must be so that any 
systematic error arising from the (necessarily) imperfect isolation of the initial-state kaon remains 
much smaller than the statistical uncertainty. We observed 
practically no $t_{W}$-dependence for 
$t_{W} \geq 2\,\mathrm{fm}$. Consequently, we chose to perform 
the full-statistics calculations on the B64 ensemble for all $x_{\gamma}$ 
at two specific values of $t_{W}$, namely 
$t_{W}\simeq 2\,\mathrm{fm}$ and 
$t_{W}\simeq 2.6\,\mathrm{fm}$. The resulting form factors 
$F_{V,1/2}(t_{\gamma}^{*}, x_{\gamma}; t_{W})$ and 
$F_{A,1/2}(t_{\gamma}^{*}, x_{\gamma}; t_{W})$ for these two values 
of $t_{W}$ are shown in Fig.~\ref{fig:tw_dependence} for selected 
$x_{\gamma}$. As the figure illustrates, the two sets of results 
agree very well.

\begin{figure}[]
    \centering
\includegraphics[width=0.91\columnwidth]{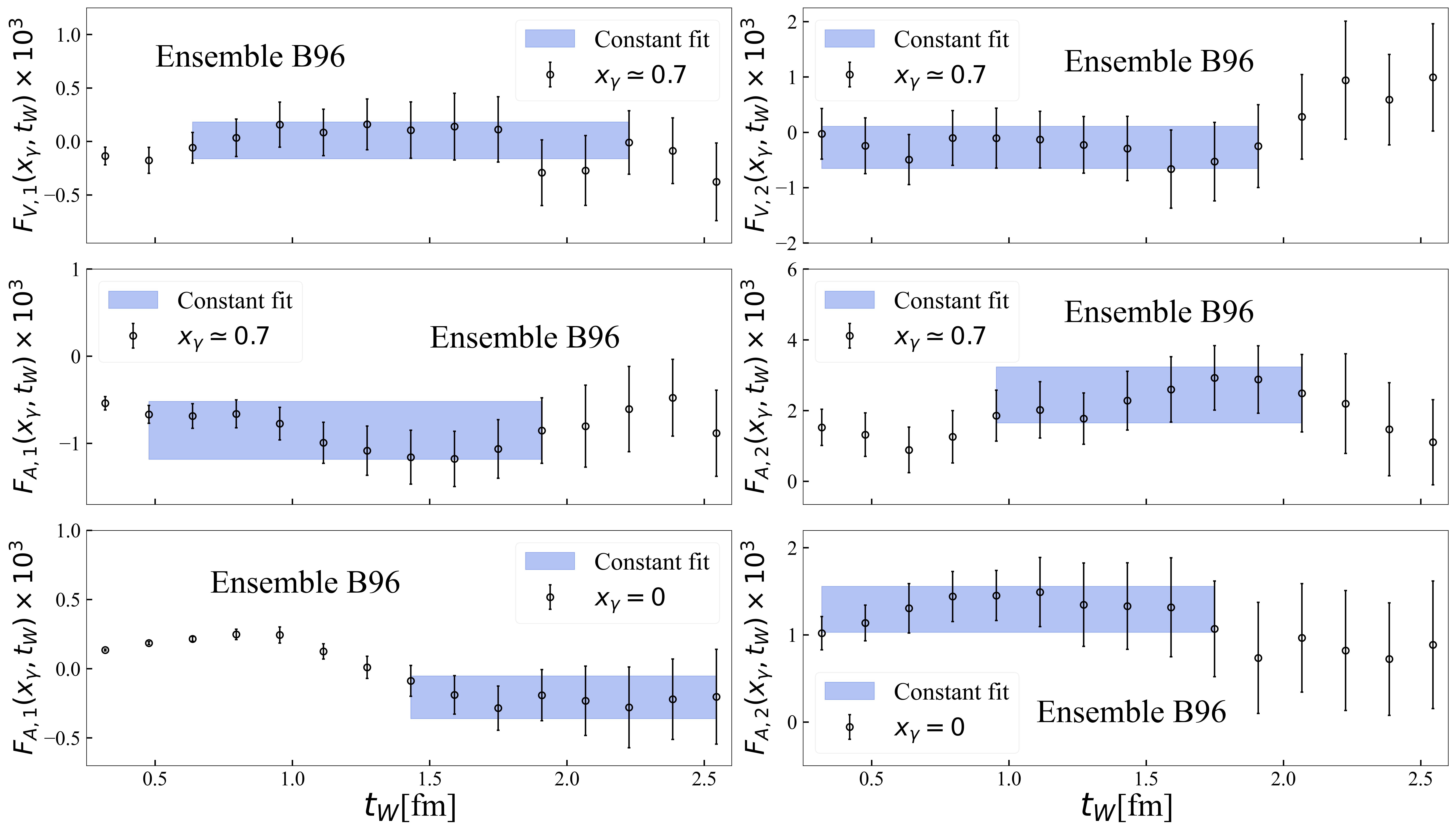}
\caption{Quark-disconnected contribution to the vector (top panels) and axial (bottom panels) form factors, as a function of $t_W$. In the left (right) panels, we show the contribution from the first (second) time-ordering. The results are obtained on the B96 ensemble at $x_\gamma = 0.7$ for $F_{V,1/2}(x_\gamma, t_W)$ and at $x_\gamma = 0, 0.7$ for $F_{A, 1/2}( x_\gamma, t_W)$. The blue bands indicate the result of a constant fit to the data where they display a plateau.}
\label{fig:Disctw_dependence}
\end{figure}
To quantify the systematic error due to finite-$t_{W}$ effects, we define the following quantity:
\begin{equation}
\label{eq:syst_tw}
\Sigma_{V/A}(x_{\gamma}) 
= \Delta_{V/A}(x_{\gamma}) \,\mathrm{erf}\!\Bigg(\frac{\Delta_{V/A}(x_{\gamma})}{\sqrt{2}\,\sigma_{\Delta_{V/A}}(x_{\gamma})}\Bigg)~
\end{equation}
where $\mathrm{erf}(x)$ is the error function and
\begin{equation}
\Delta_{V/A}(x_{\gamma}) 
= \bigl|\,
    F_{V/A}(x_{\gamma}; t_{W} \simeq 2.6~\mathrm{fm}) 
  - F_{V/A}(x_{\gamma}; t_{W} \simeq 2.0~\mathrm{fm}) \,
\bigr|~.
\end{equation}
In Eq.~(\ref{eq:syst_tw}) $\sigma_{\Delta_{V/A}}(x_{\gamma})$ is the statistical 
uncertainty on $\Delta_{V/A}(x_{\gamma})$. The quantity 
$\Sigma_{V/A}(x_{\gamma})$ corresponds to the  spread 
between the results at $t_{W}\simeq 2\,\mathrm{fm}$ and 
$t_{W}\simeq 2.6\,\mathrm{fm}$, weighted by the probability 
that the difference is not a mere statistical fluctuation.

On the B64 ensemble, our final central value for each form factor is taken to be 
the average of the results at $t_{W}\simeq 2\,\mathrm{fm}$ and 
$t_{W}\simeq 2.6\,\mathrm{fm}$. The total error is obtained by 
adding in quadrature the statistical error and 
$\Sigma_{V/A}(x_{\gamma})$. For all other ensembles, we have limited our simulations to a single $t_{W}\simeq 2.2-2.3\,\mathrm{fm}$. To account 
for possible finite-$t_{W}$ effects there, we add in quadrature 
to the statistical error the 
systematic error $\Sigma_{V/A}(x_{\gamma})$ determined on the B64 ensemble.

We have compared $\Sigma_{V/A}(x_{\gamma})$ to the statistical uncertainties of the form factors, obtained on all the 5 ensembles, and for every value of $x_\gamma$ considered, amounting to a total of 55 comparisons. In nearly all cases, $\Sigma_{V/A}(x_{\gamma})$ is small: in about
90\% of cases, it is below the statistical uncertainty itself; in 98\% of cases, it is below twice the statistical uncertainty; and it is never larger than 2.5  times the statistical uncertainty.

In the case of the quark-disconnected contribution, the finite-$t_W$ effects 
can be examined rather straightforwardly. As discussed in 
Sec.~\ref{sec:Lattice_setup}, we evaluate the correlation function $C^{\mu\nu}_{\rm 3,W ; {\rm disc}}(t_{\gamma}, \vec{k}; t_{W})$ in Eq.~(\ref{eq:disco_corr}) for all $t_{\gamma}$ and 
$t_{W}$. Figure~\ref{fig:Disctw_dependence} illustrates the $t_{W}$-dependence of the axial $F_{A, 1/2}(x_\gamma, t_W)$
and vector $F_{V, 1/2}(x_\gamma, t_W)$ form factors at $x_{\gamma} \simeq 0.7$, as well as the axial form 
factor at $x_{\gamma} = 0$, on the B96 ensemble. Results are shown for both the 
first and second time-ordering contributions. 
From the figure, we observe intervals containing approximately 8–10 data points (corresponding to a length of about $1~\mathrm{fm}$) where no appreciable dependence on $t_W$ is detected. Consequently, the form factors can be reliably extracted by performing constant fits within these intervals.

\subsection{Finite-size effects}
\label{sec:finite_vol}
\begin{figure}[]
    \centering
\includegraphics[width=1.\columnwidth]{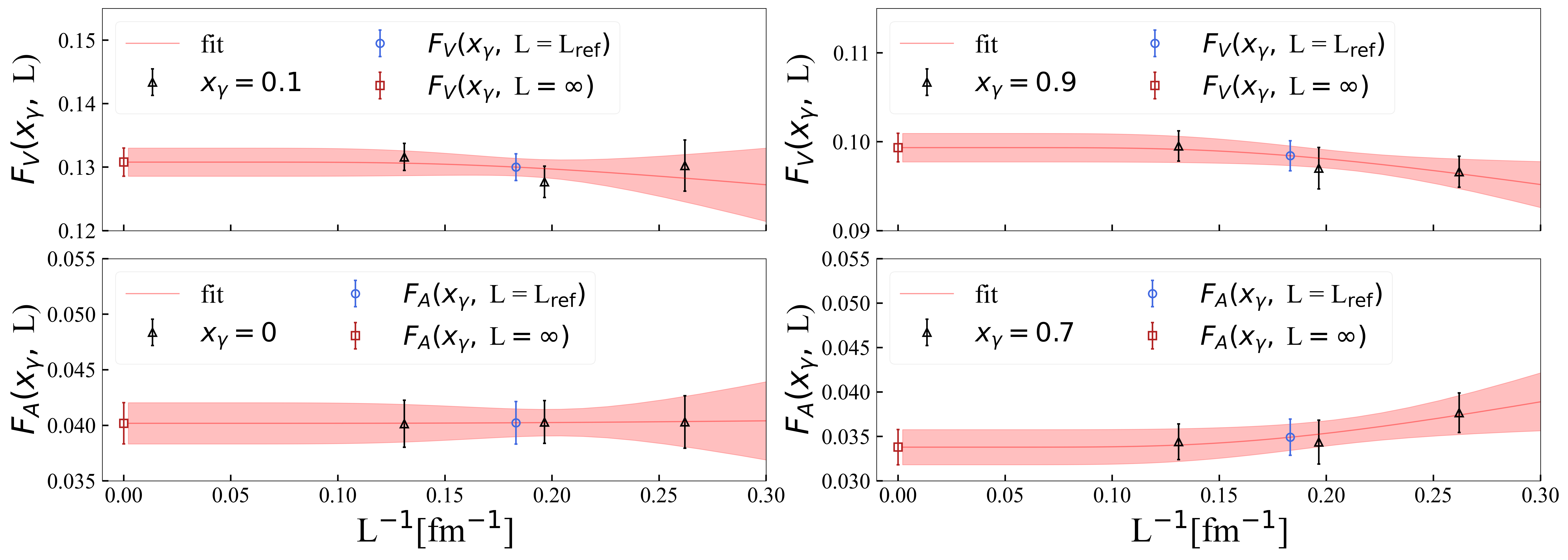}
\caption{Infinite volume extrapolation of the quark-connected contribution to the vector (top panels) and axial (bottom panels) form factors. Results are shown for $F_V(x_\gamma, L)$ at $x_\gamma = 0.1, 0.9$, and for $F_A(x_\gamma, L)$ at $x_\gamma = 0, 0.7$. From larger to smaller values of the inverse of the spatial lattice extent $L^{-1}$, the black triangles represent form factors computed on the B48, B64, and B96 ensembles. The blue circles correspond to the result of the interpolation at the reference volume $V_{\rm ref}= L_{\rm ref}^{3}$, with $L_{\rm ref}=5.46~{\rm fm}$. The coloured bands indicate the fit to the black triangles using the ansatz in Eq.~(\ref{eq:FSE_ansatz}). The red squares represent the form factors extrapolated to the infinite volume limit ($L \to \infty$).}
\label{fig:inf_vol_extr}
\end{figure}
We now discuss our strategy for the infinite-volume extrapolation. We begin with 
the dominant quark-connected contribution, which has been computed on all ETMC 
ensembles listed in Tab.~\ref{tab:simudetails}. As shown in the table, the C80 
and D96 ensembles both have (within uncertainties) a spatial lattice extent of 
$L \simeq 5.46\,\mathrm{fm}$. At the coarsest lattice spacing 
($a \simeq 0.0796\,\mathrm{fm}$), we have three ensembles with spatial lattice 
extents of $L \simeq 3.8\,\mathrm{fm}$ (B48), $L \simeq 5.1\,\mathrm{fm}$ 
(B64), and $L \simeq 7.7\,\mathrm{fm}$ (B96).

Our approach is as follows. First, we use the form factors obtained on the B48, 
B64, and B96 ensembles to interpolate 
the results at $a\simeq 0.0796~{\rm fm}$ to a reference spatial lattice extent $L_{\mathrm{ref}} = 5.46\,\mathrm{fm}$. 
Next, we perform the continuum-limit extrapolation (discussed in the next 
section) at this fixed volume. After performing the continuum-limit extrapolation, we add the finite-volume corrections required to reach the infinite-volume limit.

We denote by $F_{A}(x_{\gamma}, L)$ and $F_{V}(x_{\gamma}, L)$ the axial and 
vector form factors evaluated on a lattice of spatial extent $L$. To both 
interpolate the form factors at $a \simeq 0.0796\,\mathrm{fm}$ to 
$L = L_{\mathrm{ref}}$ and to the infinite volume limit, 
we fit, separately for each $x_{\gamma}$, the B48, B64, and B96 data using the following Ansatz:
\begin{equation}
\label{eq:FSE_ansatz}
F_{A}(x_{\gamma}, L) \;=\; C_{A}(x_{\gamma}) \;+\; D_{A}(x_{\gamma})\,\exp(-M_{\pi}L), 
\qquad
F_{V}(x_{\gamma}, L) \;=\; C_{V}(x_{\gamma}) \;+\; D_{V}(x_{\gamma})\,\exp(-M_{\pi}L).
\end{equation}
Here, $C_{A}(x_{\gamma})$, $D_{A}(x_{\gamma})$, $C_{V}(x_{\gamma})$, and $D_{V}(x_{\gamma})$ are free parameters (there is a different set of fit parameters for each $x_{\gamma})$. The results obtained using  
this Ansatz are shown in Fig.~\ref{fig:inf_vol_extr} for selected values 
of $x_{\gamma}$. The data used for the fits are already inclusive of the systematic error associated with the finite-$t_{W}$ effects, estimated as discussed in the previous section.  As can be seen, the dependence on $L$ is mild, and the B64 
and B96 data agree within uncertainties. The blue circles in the figure indicate 
the results of the interpolation to $L = L_{\mathrm{ref}}$. As a conservative choice, we inflated the errors on the interpolated points so that their uncertainties do not become smaller than the most precise among the B48, B64 and B96 data-points. We have also tested variations of 
the Ansatz in Eq.~(\ref{eq:FSE_ansatz}), for example by including a prefactor $\bigl(1/L\bigr)^n$ in front 
of the exponential terms (with $n = 1$ or $3/2$), or by replacing $M_{\pi}L$ with 
$\sqrt{2}\,M_{\pi}L$ or $2\,M_{\pi}L$. None of these modifications 
significantly changed the final results.

After the continuum-limit extrapolation (see the next section) is performed at 
$L = L_{\mathrm{ref}}$, in order to extrapolate to the infinite-volume limit, we add the following finite-volume corrections
\begin{equation}
\Delta_{L} F_{A}(x_{\gamma}) \equiv F_{A}(x_{\gamma}, \infty) 
   \;-\; F_{A}(x_{\gamma}, L_{\mathrm{ref}}),
\qquad
\Delta_{L} F_{V}(x_{\gamma}) \equiv F_{V}(x_{\gamma}, \infty) 
   \;-\; F_{V}(x_{\gamma}, L_{\mathrm{ref}}),
\end{equation}
which are related to the fit parameters in Eq.~(\ref{eq:FSE_ansatz}) through
\begin{align}
\Delta_{L} F_{A}(x_{\gamma}) = -D_{A}(x_\gamma)\exp{-M_{\pi}L_{\rm ref}}~, \qquad  \Delta_{L} F_{V}(x_{\gamma}) = -D_{V}(x_\gamma)\exp{ -M_{\pi}L_{\rm ref}}~.
\end{align}

For the quark-disconnected contribution, we have data only on the B96 ensemble. 
Since the lattice extent of the B96 is relatively large 
($L \simeq 7.7\,\mathrm{fm}$), and given that the disconnected contribution has 
significantly larger statistical uncertainties than the connected part, we treat 
the B96 results as effectively in the infinite-volume limit.

\subsection{Continuum-limit extrapolation}
The final step in our analysis is the continuum-limit extrapolation. We start by discussing the quark-connected contribution. In this case, we use for the extrapolations the results from three lattice spacings, $a \simeq 0.057\,\mathrm{fm}$ (D96), $a \simeq 0.068\,\mathrm{fm}$ (C80), and $a \simeq 0.0796\,\mathrm{fm}$ (B-ensembles), all corresponding to a spatial volume of linear extent $L_{\mathrm{ref}} = 5.46\,\mathrm{fm}$. We denote by $F_{A}^{L_{\mathrm{ref}}}(x_{\gamma}, a)$ and $F_{V}^{L_{\mathrm{ref}}}(x_{\gamma}, a)$ the axial and vector form factors, respectively, evaluated at a lattice spacing $a$ and at the spatial extent $L_{\mathrm{ref}}$.

We follow two different approaches. In the first, we perform the continuum-limit extrapolation for both axial and vector form factors, and separately for each simulated value of $x_{\gamma}$,  using the following linear Ansatz in $a^2$
\begin{equation}
F_{A}^{L_{\mathrm{ref}}}(x_{\gamma}, a) = R_{A}(x_{\gamma}) + B_{A}(x_{\gamma})\,a^{2}, 
\qquad 
F_{V}^{L_{\mathrm{ref}}}(x_{\gamma}, a) = R_{V}(x_{\gamma}) + B_{V}(x_{\gamma})\,a^{2},
\label{Eq:continuum_ansatz}
\end{equation}
where distinct fit parameters, $R_{A}(x_{\gamma})$, $B_{A}(x_{\gamma})$, $R_{V}(x_{\gamma})$, and $B_{V}(x_{\gamma})$ are used for each $x_{\gamma}$ value.
Two fits are performed. In one, the slope parameters $B_{A}(x_{\gamma})$ and $B_{V}(x_{\gamma})$ are set to zero, thus reducing the fit to a constant form, and we restrict the fit to the two finest lattice spacings (D96 and C80). The errors of the continuum-extrapolated values from this constant fit are inflated so that they do not become smaller than the most precise of the two data points (either D96 or C80). In the other fit, the full linear Ansatz is retained by including the slope parameters, and the fit is performed including the data from the coarsest lattice spacing as well. These two fits are then combined via the Bayesian Akaike Information Criterion (BAIC)~\cite{Neil:2022joj}. Let $x_{1}$ and $x_{2}$ be the outcomes of the two fits. The final central value is given by
\begin{equation}
\bar{x} = w_{1}\,x_{1} + w_{2}\,x_{2},
\end{equation}
and the total error is
\begin{equation}
\sigma^{2} = \sigma_{\mathrm{stat}}^{2} + w_{1}(x_{1}-\bar{x})^{2} + w_{2}(x_{2}-\bar{x})^{2},
\end{equation}
where $\sigma_{\mathrm{stat}}$ is the statistical error of $\bar{x}$, and $w_{1}$ and $w_{2}$ are weights normalized to one and given by
\begin{equation}
w_{i} \propto \exp\Bigl[-\bigl(\chi_{i}^{2} + 2\,N_{i}^{\mathrm{pars}} - 2\,N_{i}^{\mathrm{data}}\bigr)\Bigr]~,\qquad i=1,2~.
\end{equation}
Here, $\chi_{i}^{2}$ denotes the chi-squared of the $i$-th fit, while $N_{i}^{\mathrm{pars}}$ and $N_{i}^{\mathrm{data}}$ are its number of free parameters and data points, respectively. 
\begin{figure}[]
    \centering
\includegraphics[width=1.\columnwidth]{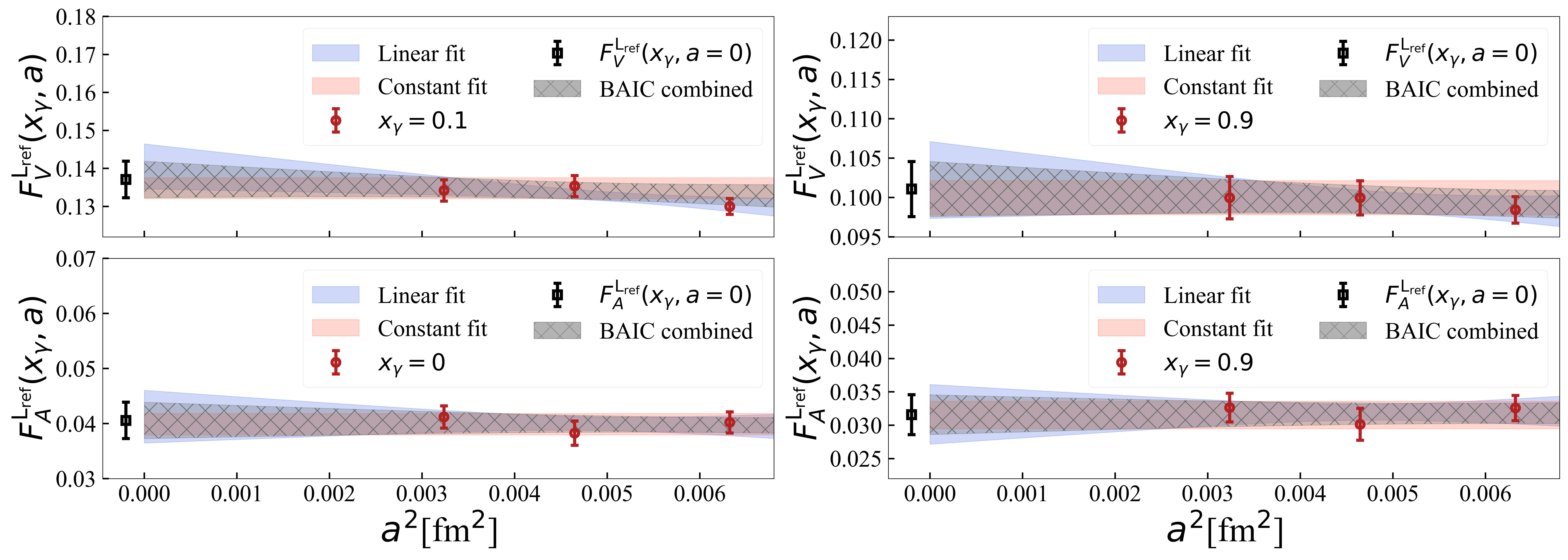}
\caption{Continuum-limit extrapolation of the quark-connected contribution to the vector (top) and axial (bottom) form factors. The extrapolation is performed at fixed reference volume $V_{\rm ref}= L_{\rm ref}^{3}$, with $L_{\mathrm{ref}} = 5.46 \ \mathrm{fm}$. The results are shown for $F_V^{L_{\mathrm{ref}}}(x_\gamma, a)$ at $x_\gamma = 0.1, 0.9$ and for $F_A^{L_{\mathrm{ref}}}(x_\gamma, a)$ at $x_\gamma = 0, 0.9$. The red circles correspond, in order of increasing lattice spacing $a$, to the results obtained on the D96 and C80 ensembles, and to the values obtained at $a\simeq 0.0796~{\rm fm}$  after the volume interpolation. The red bands represent the result of a constant fit to the two finest lattice spacings, while the blue bands correspond to a linear fit including all the red circles. The two fits are combined using the BAIC, which yields the meshed bands. Finally, the black squares represent the final results
of the continuum-limit extrapolation at spatial extent $L=L_{\rm ref}$.}
\label{fig:cont_extr}
\end{figure}
\begin{figure}[]
    \centering
\includegraphics[width=0.48\columnwidth]{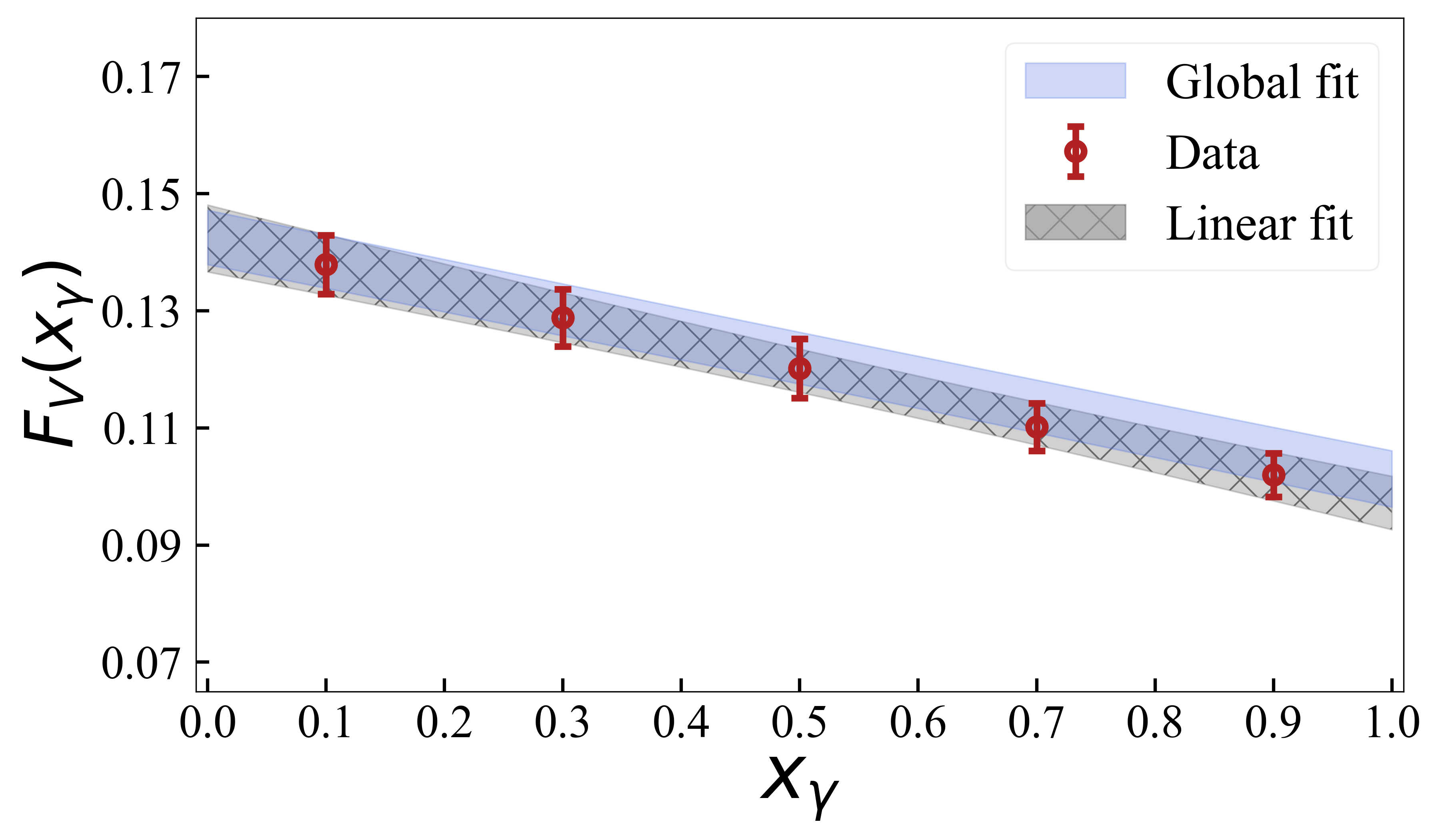}
\includegraphics[width=0.48\columnwidth]{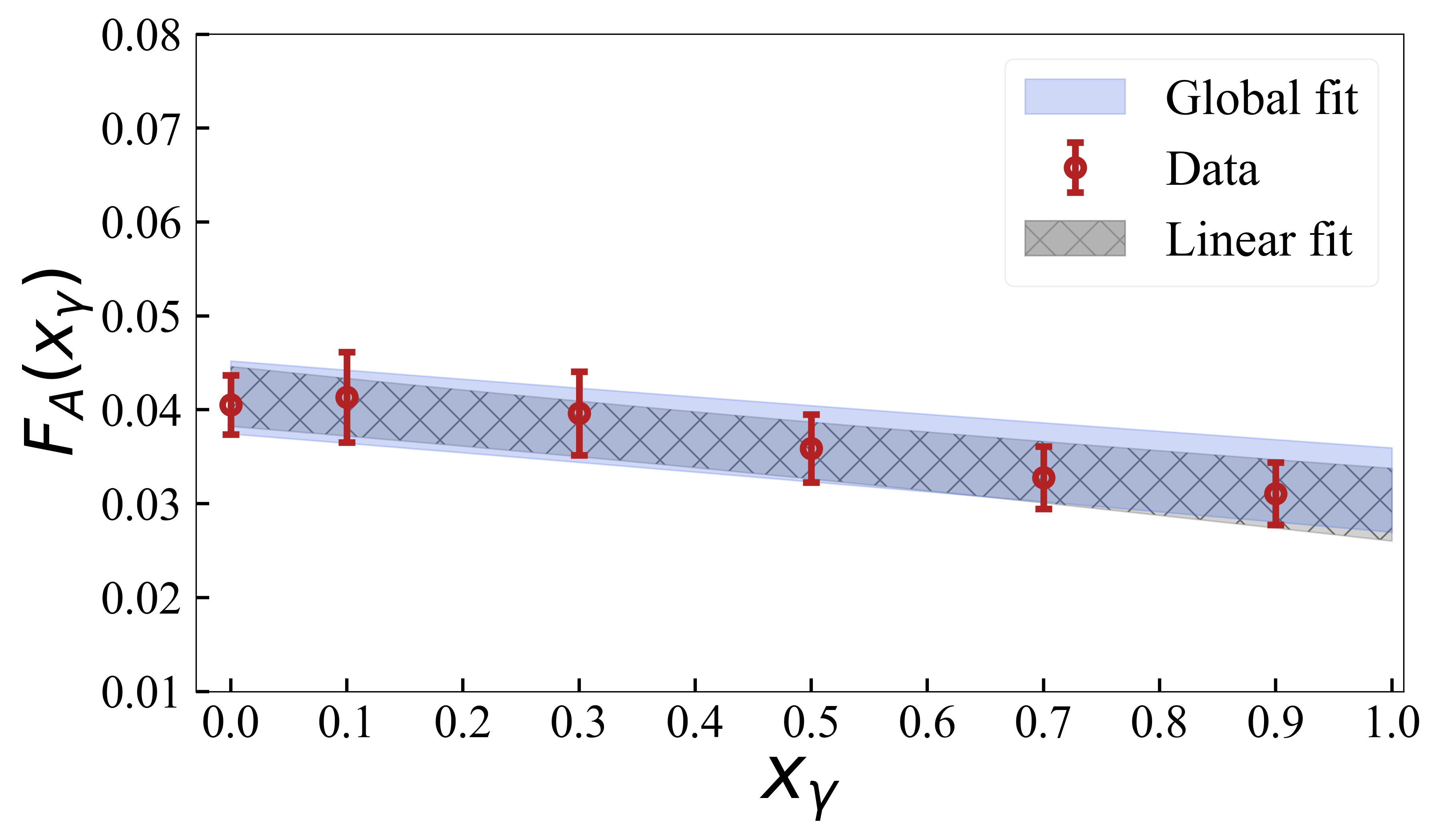}
\caption{Quark-connected contribution to the vector (left panel) and axial (right panel) form factors as functions of $x_\gamma$. The red circles represent the results obtained after performing the continuum and infinite-volume limits separately for each $x_\gamma$ (see Eq.~(\ref{Eq:continuum_ansatz})). In both panels, two different fitting strategies are compared. The meshed band corresponds to a linear fit of the red circles. The blue band represents a global fit, where data from all photon momenta and ensembles are simultaneously fitted using a linear ansatz in both $x_\gamma$ and the lattice spacing $a^2$ (see Eq.~(\ref{eq:global_fit})).}
\label{fig:xg_dep_FF}
\end{figure}
Results obtained for selected values of $x_{\gamma}$ are shown in Fig.~\ref{fig:cont_extr}. In each panel, the red band represents the constant fit to the two finest lattice spacings (with inflation of its uncertainty as described above), the blue band represents the linear fit to all three lattice spacings, and the meshed band indicates the BAIC-weighted average. As the figure shows, no significant cutoff effects are observed within the quoted uncertainties, which justifies the inclusion of a constant fit in the BAIC. The reduced $\chi^2$ values are always close to unity.

The second approach to the extrapolation to the continuum limit is a global fit, that simultaneously describes the $x_{\gamma}$ and $a$ dependence of the form factors, using
\begin{equation}
\label{eq:global_fit}
F_{A}^{L_{\mathrm{ref}}}(x_{\gamma}, a) = R_{A,1} + R_{A,2}\,x_{\gamma} + B_{A}\,a^{2},
\quad\qquad 
F_{V}^{L_{\mathrm{ref}}}(x_{\gamma}, a) = R_{V,1} + R_{V,2}\,x_{\gamma} + B_{V}\,a^{2},
\end{equation}
where $R_{W,1
}, R_{W,2}, B_{W}$, with $W=\left\{ V,A \right\}$, are free-fit parameters. The simple parameterization in Eq.~(\ref{eq:global_fit}) accurately describes the data, and including an additional term proportional to $a^2 x_{\gamma}$ does not change the results (the corresponding fit parameter turns out to be zero within uncertainties). 

Fig.~\ref{fig:xg_dep_FF} compares the two strategies. The red circles correspond to the results obtained performing separate continuum-limit extrapolations for each value of $x_{\gamma}$, while the blue band is the outcome of the global fit. In both cases we have added the finite-volume correction $\Delta_{L}F_{A}(x_{\gamma})$ and $\Delta_{L}F_{V}(x_{\gamma})$, discussed in Sec.~\ref{sec:finite_vol}. The meshed band is a linear fit in $x_{\gamma}$ to the red circles. Both methods  agree well within uncertainties, indicating the robustness of the continuum-limit extrapolation.

\begin{figure}[]
    \centering
\includegraphics[width=1.\columnwidth]{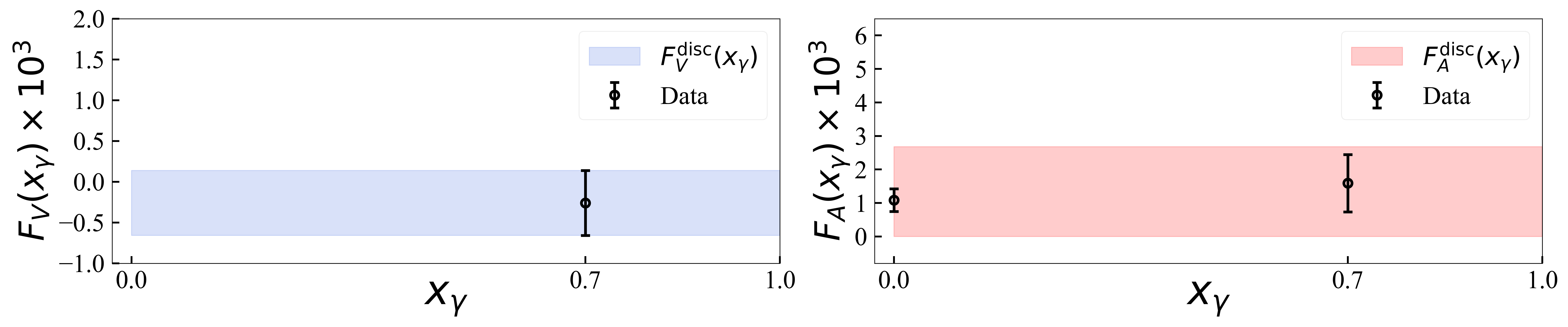}
\caption{Final results for the quark-disconnected contribution to the vector (left panel) and axial (right panel) form factors. Black circles are obtained on the B96 Ensemble. The blue and red bands represent the final results for $F_V^{\mathrm{disc}}(x_\gamma)$ and $F_A^{\mathrm{disc}}(x_\gamma)$, respectively. Since the form factors were computed at a single lattice spacing, a 100\% uncertainty was assigned to the axial form factor. For the vector 
form factor, we used its own uncertainty, as its relative uncertainty already exceeds 100\%.}
\label{fig:disco}
\end{figure}

For the disconnected contribution, we have results only for the B96 ensemble. Fig.~\ref{fig:disco} shows $F_{V}$ and $F_{A}$ at $x_{\gamma} = 0.7$, as well as $F_{A}$ at $x_{\gamma} = 0$. At $x_{\gamma} = 0.7$, the value of $F_{V}$ is compatible with zero within uncertainties, while $F_{A}$ has an uncertainty of approximately $50\%$. At $x_{\gamma} = 0$, the error on $F_{A}$ is about $35\%$. Given these relatively large statistical errors, cutoff effects are expected to be subleading. We take the blue and red bands in Fig.~\ref{fig:disco} as our final estimates for the disconnected contribution to $F_{V}$ (blue band) and $F_{A}$ (red band), namely 
\be
F_V^{\mathrm{disc}}(x_{\gamma}) = -0.3(4) \times 10^{-3}, \qquad F_A^{\mathrm{disc}}(x_{\gamma}) = 1.3(1.3) \times 10^{-3}.
\ee
Note that the disconnected contribution is very small, and negligible within the current uncertainties of the quark-connected contribution. 

In Tab.~\ref{tab:final_res}, we present our final results for $F_{A}(x_{\gamma})$ and $F_{V}(x_{\gamma})$, which include the contributions from the quark-disconnected term and the finite-volume corrections. To facilitate their use in phenomenological analyses, we also provide a linear parameterization of the axial and vector form factors:
\begin{equation}
F_{A}(x_{\gamma}) = F_{A}^{0} + F_{A}^{1} \, x_{\gamma},
\quad
F_{V}(x_{\gamma}) = F_{V}^{0} + F_{V}^{1} \, x_{\gamma}.
\label{Eq:F_vs_xg_linear}
\end{equation}
\begin{table}[] 
\centering 
\resizebox{\linewidth}{!}{\begin{tabular}{lcccccc} \hline \hline &  &  &  & & &   \\ [-2. ex] 
 & $x_\gamma=0$ & $x_\gamma=0.1$ & $x_\gamma=0.3$ & $x_\gamma=0.5$ & $x_\gamma=0.7$ & $x_\gamma=0.9$  \\ [1. ex] \hline &  &  &  & & &  \\ [-2. ex] 
$F_V(x_\gamma) \times 10$ & -- & $1.376 [50] (48)(14)(4)$ & $1.285 [49] (44)(21)(4)$ & $1.199 [51] (49)(13)(4)$ & $1.099 [41] (38)(14)(4)$ & $1.017 [37] (35)(13)(4)$ \\ [1. ex] &  &  &  & & &  \\ [-2. ex]$F_A(x_\gamma) \times 10^2$ & $4.19 [34] (33)(0)(13)$ & $4.27 [50] (43)(21)(13)$ & $4.10 [46] (39)(22)(13)$ & $3.72 [39] (34)(14)(13)$ & $3.41 [36] (31)(11)(13)$ & $3.24 [36] (30)(14)(13)$ \\ [1. ex] \hline \hline
\end{tabular}} 
\caption{Final results for the vector and axial form factors for all simulated  values of $x_\gamma$. For each $x_\gamma$, the total uncertainty is given in square brackets. A detailed breakdown of the uncertainties is provided in parentheses: the first term represents the statistical error, the second accounts for finite-size effects, and the third corresponds to the quark-disconnected contribution.}
\label{tab:final_res} 
\end{table} 
where
\begin{align}
\label{Eq:slopes}
F_{A}^{0} &= 0.0428(34), 
\quad
F_{A}^{1} = -0.0116(40), 
\quad
\mathrm{corr}(F_{A}^{0}, F_{A}^{1}) = -0.37, \\[8pt]
F_{V}^{0} &= 0.1421(57),
\quad
F_{V}^{1} =  -0.0452(72),
\quad
\mathrm{corr}(F_{V}^{0}, F_{V}^{1}) = -0.77,
\end{align}

and $\mathrm{corr}(F_{A}^{0}, F_{A}^{1})$ and $\mathrm{corr}(F_{V}^{0}, F_{V}^{1})$ denote the correlation coefficients between the parameters $F_{A}^{0}, F_{A}^{1}$ and $F_{V}^{0}, F_{V}^{1}$, respectively.

\begin{figure}[]
    \centering
\includegraphics[width=0.48\columnwidth]{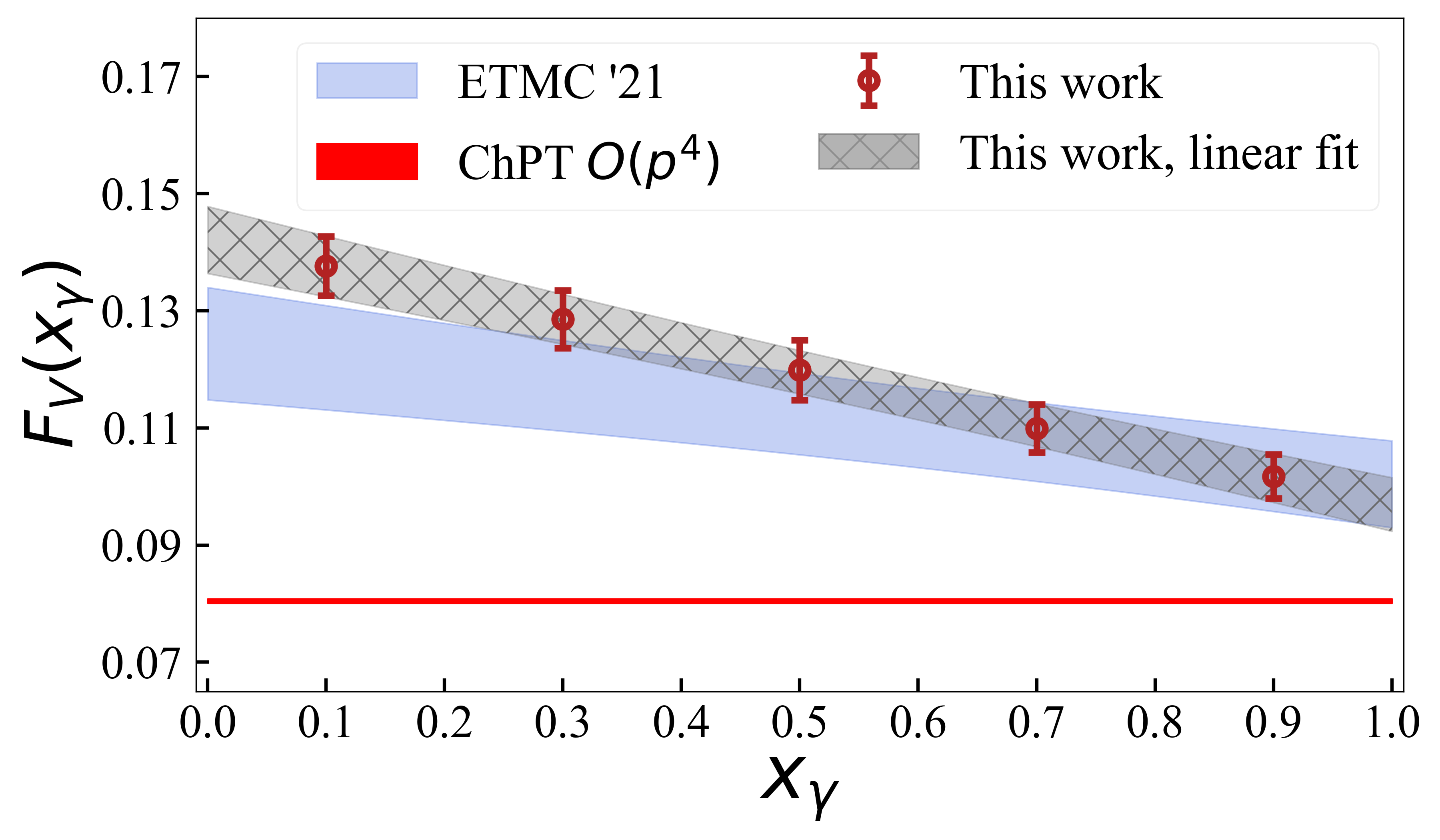}
\includegraphics[width=0.48\columnwidth]{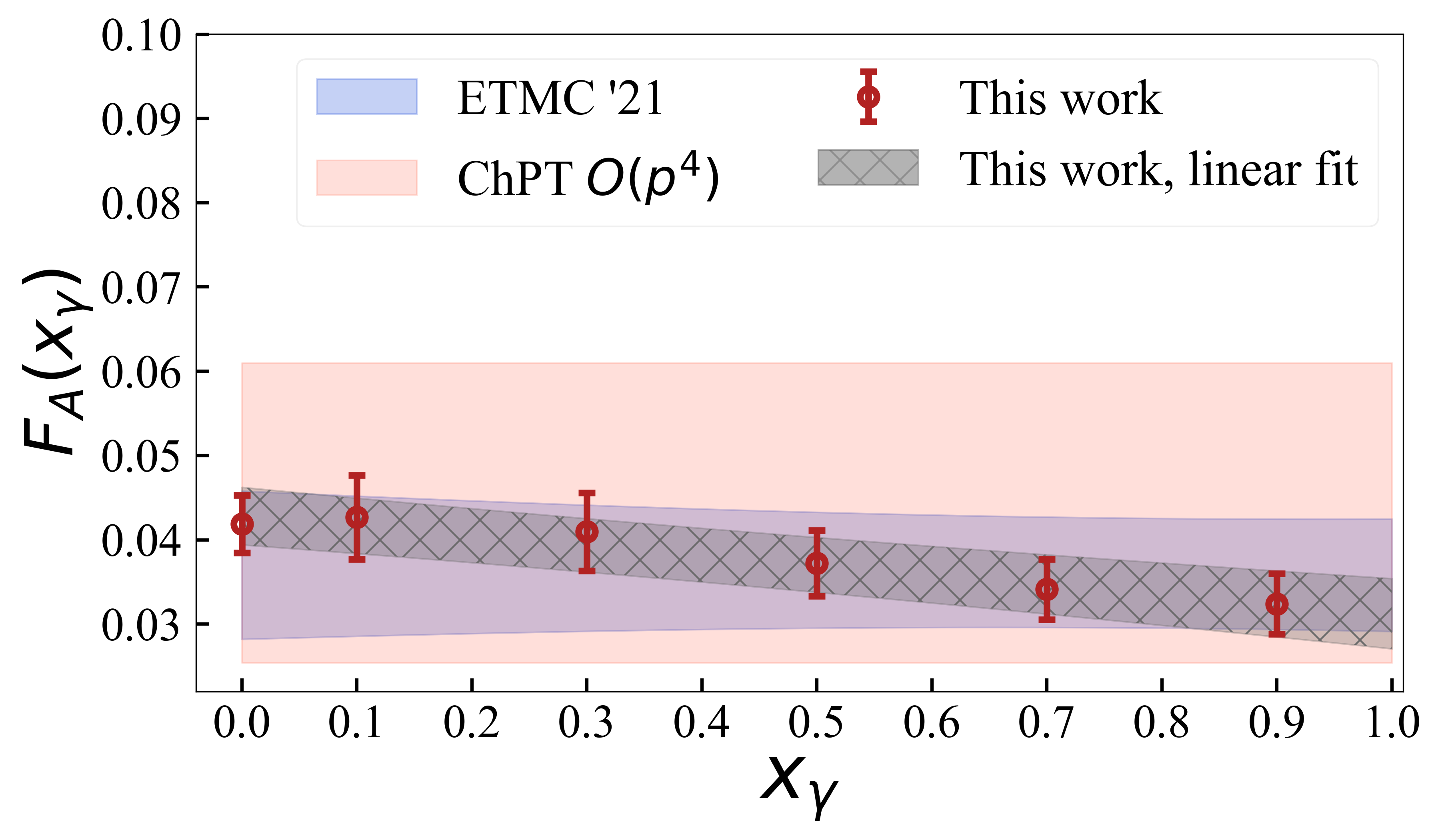}
\caption{Comparison between $F_{V}(x_{\gamma})$ (left panel) and $F_{A}(x_{\gamma})$ (right panel), as obtained in this work (meshed grey band), in our ETMC '21 paper~\cite{Desiderio:2020oej} (blue band), and in ChPT at $O(p^4)$ (red band).}
\label{fig:xg_final}
\end{figure}

In Fig.~\ref{fig:xg_final}, we compare our final results for $F_{V}$ and $F_{A}$ with those obtained in our previous work~\cite{Desiderio:2020oej}, as well as with the ChPT predictions at $O(p^4)$, given by
\begin{align}
\label{eq:ChPT_FF}
F_{V}^{\rm ChPT}(x_{\gamma}) = \frac{m_{K}}{4\pi^{2}f_{K}}~, \qquad F_{A}^{\rm ChPT}(x_{\gamma}) = \frac{8m_{K}}{f_{K}}\left( L_{9}^{r} + L_{10}^{r} \right)~.
\end{align}
Adopting the values  $L_{9}^{r}+L_{10}^{r} = 0.0017(7)$~\cite{Bijnens:2014lea}, $m_K = 494.6 \  \mathrm{MeV}$ and $f_K = 155.7(7) \  \mathrm{MeV}$\,\cite{FlavourLatticeAveragingGroupFLAG:2024oxs}, the form factors in Eq.~(\ref{eq:ChPT_FF}) from ChPT are
\be
F_{V}^{\rm ChPT}(x_{\gamma}) = 0.08046(36), \qquad  F_{A}^{\rm ChPT}(x_{\gamma}) = 0.043(18)~.
\label{Eq:ChPT_FFres}
\ee
Our results for $ F_A(x_\gamma) $ show a good agreement with both our previous determination and with the $O(p^4) $ ChPT predictions. However, the accuracy is greatly improved and the presence of a non-zero slope is now more evident, as it can be seen by comparing the result for $F_{A}^{1}$ in Eq.~(\ref{Eq:slopes}) with our previous value of $ F_A^1 = -0.0012(74)$. 
For the vector form factor $ F_V(x_\gamma) $, as illustrated in the figure, our results are compatible with the previous analysis within at most $ 1.5 \sigma $ in the small $ x_\gamma $ region. However, we observe a steeper slope in magnitude compared to our earlier work, where we obtained $ F_V^1 = -0.024(10) $, which differs from our result in Eq.~(\ref{Eq:slopes}) by approximately $ 1.7 \sigma $. Additionally, both the new and previous results lie above the corresponding ChPT predictions. Note that the errors have been reduced by approximately a factor of two with respect to our previous work.

\section{Comparison with experimental data}
\label{sec:exp_comparison}
In this section, we use our results for the vector and axial form factors $F_{V}$ and $F_{A}$ to evaluate the branching fractions for $K^{-} \to \ell^{-}\bar{\nu}_{\ell}\gamma$\footnote{Throughout this section, all relations regarding the decay process $K^{-} \to \ell^{-} \bar{\nu}_{\ell} \gamma$ equally apply to its charge conjugate counterpart, $K^{+} \to \ell^{+} \nu_{\ell} \gamma$.}, and compare them with available experimental data. The starting point is the double-differential decay rate that, following Refs.~\cite{Desiderio:2020oej,Frezzotti:2020bfa}, can be written as
\begin{equation}
    \frac{d^2\Gamma(K^- \to \ell^- \bar{\nu}_\ell \gamma)}{dx_\gamma dx_\ell} \equiv  \frac{d^2\Gamma_1}{dx_\gamma dx_\ell} = 
        \frac{\alpha_{\mathrm{em}}}{4\pi} \Gamma^{(0)} ~ \left[ \frac{d^2R_1^{\mathrm{pt}}}{dx_\gamma dx_\ell} +
        \frac{d^2R_1^{\mathrm{SD}}}{dx_\gamma dx_\ell} + \frac{d^2R_1^{\mathrm{INT}}}{dx_\gamma dx_\ell} \right] ~ , ~ 
    \label{eq:dGamma}
\end{equation}
where the subscript $1$ denotes the number of photons in the final state, while $x_\gamma$ and $x_\ell$ are the photon and lepton kinematical variables, the latter being defined as
\be
    x_\ell \equiv \frac{2p \cdot p_\ell}{m_K^2} - r_\ell^2\,,
    \label{eq:xgammaxell}
\ee 
where $p_{\ell}$ is the four-momentum of the final-state lepton of mass $m_{\ell}$, and $r_{\ell} = m_{\ell}/m_{K}$. In the kaon's rest frame $x_{\gamma} = 2E_{\gamma}/m_{K}$, while $x_\ell = 2 E_\ell / m_K - r_\ell^2$, where $E_\ell$ is the energy of the final-state lepton. In Eq.\,(\ref{eq:dGamma}) the quantity $\Gamma^{(0)}$ is the leptonic decay rate at tree level, given by
\be
    \Gamma^{(0)} = \frac{G_F^2 \vert V_{us} \vert^2 f_K^2}{8\pi} m_K^3 r_\ell^2 \left(1 - r_\ell^2 \right)^2 ~ 
    \label{eq:Gamma_QCD}
\ee
where $G_F$ is the Fermi constant, and $V_{us}$ the relevant CKM matrix element. The other entries on the right-hand side of Eq.~(\ref{eq:dGamma}) are
\begin{align}
    \label{eq:d2Gamma_pt}
    \frac{d^2R_1^{\mathrm{pt}}}{dx_\gamma dx_\ell} & = \frac{2}{(1 - r_\ell^2)^2} f_{\mathrm{pt}}(x_\gamma, x_\ell)\,, 
    \\[4mm]
    \frac{d^2R_1^{\mathrm{SD}}}{dx_\gamma dx_\ell} & \equiv 
        \frac{m_K^2}{2f_K^2 r_\ell^2(1 - r_\ell^2)^2} f_{\mathrm{SD}}^+(x_\gamma, x_\ell) \left[ F^+(x_\gamma) \right]^2  
        +  \frac{m_K^2}{2f_K^2 r_\ell^2(1 - r_\ell^2)^2} f_{\mathrm{SD}}^-(x_\gamma, x_\ell) \left[ F^-(x_\gamma) \right]^2 ,\label{eq:d2Gamma_SD}  \\[5mm]
    \frac{d^2R_1^{\mathrm{INT}}}{dx_\gamma dx_\ell} & \equiv  
        - \frac{2m_K}{f_K\, (1-r_\ell^2)^2} f_{\mathrm{INT}}^+(x_\gamma, x_\ell) F^+(x_\gamma)
         -  \frac{2m_K}{f_K\, (1-r_\ell^2)^2} f_{\mathrm{INT}}^-(x_\gamma, x_\ell) F^-(x_\gamma), 
        \label{eq:d2Gamma_INT}
\end{align}
where $f_{\rm pt}, f_{\rm SD}^{\pm}$ and $f_{\rm INT}^{\pm}$ are analytic kinematical functions whose explicit expressions are given in Eqs.~(7)-(11) of Ref.~\cite{Frezzotti:2020bfa}. The term in Eq.~(\ref{eq:d2Gamma_pt}) corresponds to the 
point-like contribution to the decay rate (often referred to as the \textit{inner-bremsstrahlung} term), while the term in Eq.~(\ref{eq:d2Gamma_SD}) to the structure-dependent contribution, which depends on the following combinations of the axial and vector form factors:
\begin{align}
\label{eq:Fpm}
F^{+}(x_{\gamma}) = F_{V}(x_{\gamma}) + F_{A}(x_{\gamma})~,\qquad F^{-}(x_{\gamma}) = F_{V}(x_{\gamma}) - F_{A}(x_{\gamma})~.
\end{align}
Finally, the term in Eq.~(\ref{eq:d2Gamma_INT}) is the interference between the point-like and structure-dependent contribution.

In absence of any kinematical cut, one has
\begin{align}
\label{eq:xellmin}
x_{\ell}^{\rm min} \equiv 1 - x_\gamma + x_\gamma \frac{r_\ell^2} {1 - x_\gamma } \leq x_{\ell} \leq 1~,
\end{align}
and defining  
\begin{align}
 \frac{d R_1^{r}}{dx_\gamma} \equiv \int_{x_{\ell}^{\rm min}}^{1}  dx_{\ell}  \, \frac{d^{2} R_1^{r}}{dx_\gamma dx_{\ell}}~,\qquad\ 
 R_{1}^{r}(\Delta E_{\gamma}) \equiv \int_{2\frac{\Delta E_{\gamma}}{m_{K}}}^{1-r_{\ell}^{2}} dx_{\gamma}\,  \frac{d R_1^{r}}{dx_\gamma} ~,\qquad   r=\left\{ {\rm pt}, {\rm SD}, {\rm INT} \right\}~,
\end{align}
one has, in the limit 
$x_{\gamma} \to 0$, that $dR_1^{\mathrm{SD}} / dx_\gamma \propto x_\gamma^3$ and $dR_1^{\mathrm{INT}} / dx_\gamma \propto x_\gamma$, while $dR_1^{\mathrm{pt}} / dx_\gamma \propto 1 / x_\gamma$~\cite{Frezzotti:2020bfa}. The point-like term
gives rise, in the soft photon limit $\Delta E_{\gamma} \to 0$, to a logarithmically divergent contribution proportional to $\log{\left(\Delta E_{\gamma}\right)}$ and is therefore the dominant contribution for sufficiently small values of $\Delta E_\gamma$\!
\footnote{In the limit $\Delta E_{\gamma} \to 0$, the infrared divergence in the leptonic decay with a real photon in the final state is cancelled by the $O(\alpha_{\mathrm{em}})$ virtual photon contribution to the purely leptonic decay amplitude, through the Bloch-Nordsieck
mechanism\,\cite{PhysRev.52.54}. The inclusive leptonic decay rate $K^{-}\to \ell^{-} \bar{\nu}_{\ell} (\gamma)$ is infrared finite.}.
However, it is also chirally suppressed with respect to the SD contribution by the factor $r_{\ell}^{2} = (m_{\ell}/m_{K})^{2}$. Unlike the point-like contribution, the SD contribution is small at small photon energies, then grows reaching a maximum at some value of the photon energy which depends on the specific channel considered, and then decreases to zero at the edge of phase space, i.e. for $x_{\gamma} = 1-r_{\ell}^{2}$. Therefore, for a sufficiently large photon energy cut-off $\Delta E_{\gamma}$ and a small value of $r_{\ell}$, the SD contribution is the dominant one. The modulus of the final-state lepton’s three-momentum, $|\vec{p}_{\ell}|$, and the angle $\theta_{\ell\gamma}$ between the photon and lepton (both measured in the kaon rest frame) are related to $x_{\gamma}$ and $x_{\ell}$ by

\begin{align}
|\vec{p}_{\ell}| = \frac{m_{K}}{2}\,\sqrt{\bigl( x_{\ell} + r_{\ell}^{2}\bigr)^2 - 4 r_l^2}~,\qquad 
\cos \theta_{\ell\gamma} (x_\gamma, x_\ell) 
= \frac{x_{\ell} + r_{\ell}^{2} - 2 \;-\; 2\,(x_{\ell}-1)/x_{\gamma}}
{\sqrt{\bigl(r_{\ell}^{2} + x_{\ell}\bigr)^{2} - 4\,r_{\ell}^{2}}}.
\label{eq:moml_theta}
\end{align}

In the following, we use Eqs.~\eqref{eq:dGamma}--\eqref{eq:moml_theta} to compare our results with experimental data, imposing the same kinematical cuts on $x_{\gamma}, x_{\ell}, |\vec{p}_{\ell}|,$ and $\theta_{\ell\gamma}$ that each experiment employs. For a more detailed description of the explicit formulae used to evaluate the branching fractions in the presence of different types of kinematical cuts, we refer to Ref.~\cite{Frezzotti:2020bfa}.

\subsection{$K^- \to e^- \bar{\nu}_e \gamma$: comparison with the experimental results from KLOE and E36}
\begin{figure}
    \centering  \includegraphics[width=0.75\linewidth]{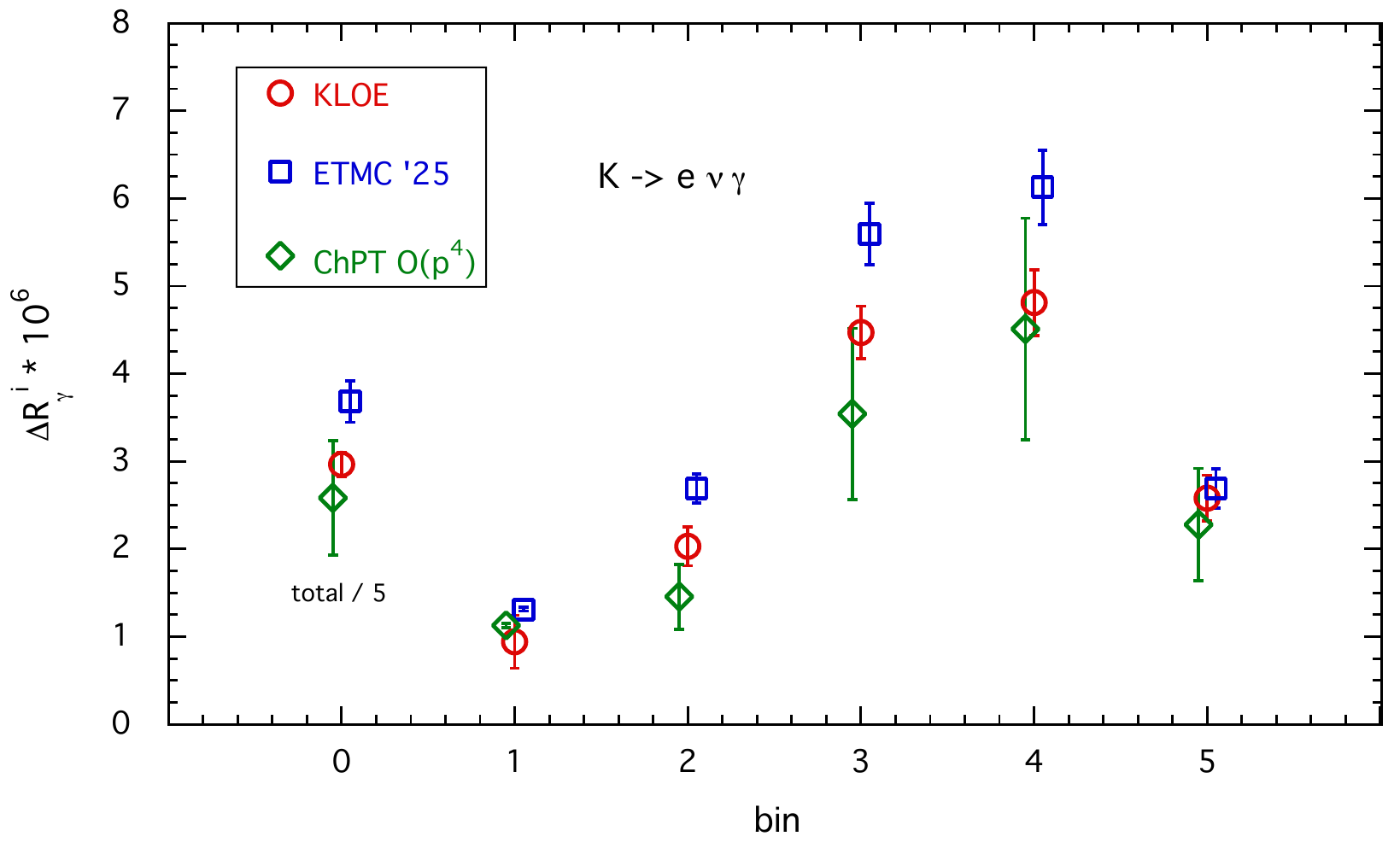}
    \caption{Comparison between our  lattice results (blue squares) for $\Delta R_{\gamma}^{i}$, the experimental data from KLOE (red circles), and the prediction from ChPT at $O(p^4)$ (green diamonds). The bin labelled ``0" represents the total rate given in the last column of Tab.\,\ref{tab:KLOE} divided by the number of bins.}
    \label{fig:KLOEbins}
\end{figure}
In order to compare our results with the KLOE~\cite{KLOE:2009urs} experimental data, we consider the following differential branching fraction for the decay $K^- \to e^- \bar{\nu}_e \gamma$ as a function of the photon energy $E_\gamma$:
\begin{equation}
\label{Eq:dR}
    \frac{d R_\gamma}{d E_\gamma} = \frac{1}{\Gamma(K^- \to \mu^- \bar{\nu}_\mu (\gamma))} \frac{d \Gamma (K^- \to e^- \bar{\nu}_e \gamma)}{dE_\gamma},
\end{equation}
with
\begin{equation}
    \frac{d \Gamma (K^- \to e^- \bar{\nu}_e \gamma)}{dE_\gamma} = \frac{2}{m_K} \int_{x_e^{\mathrm{cut}}}^{1} d x_e \ \frac{d^2\Gamma_1}{dx_\gamma dx_e},
    \label{Eq:dRgamma}
\end{equation}
where the double-differential decay rate on the right-hand side is given in Eq.~(\ref{eq:dGamma}). The lower integration limit, $x_e^{\mathrm{cut}}$ corresponds to an experimental cut on the electron momentum, namely $| \bs{p}_e | > | \bs{p}_e^{\mathrm{cut}} |$, and it is given by 
\begin{equation}
    \qquad x_e^{\mathrm{cut}} = \frac{2}{m_K} \sqrt{ m_e^2 + | \bs{p}_e^{\mathrm{cut}} |^2 } - r_e^2.
    \label{xe_min}
\end{equation}
The KLOE Collaboration has measured~\cite{KLOE:2009urs} the differential branching fraction in Eq.~(\ref{Eq:dRgamma}) under the experimental condition
\be
\vert \bs{p}_e \vert > \vert \bs{p}_e^{\mathrm{cut}} \vert = 200~\mathrm{MeV} \Rightarrow x_e^{\mathrm{cut}} \simeq 0.8, 
\label{Eq:KOLEcut}
\ee
and integrated it over five photon energy bins:
\begin{equation}
    \Delta R_\gamma^i = \int_{E_\gamma^i}^{E_\gamma^{i+1}} d E_\gamma \ \frac{d R_\gamma}{d E_\gamma},
    \qquad E_\gamma^i \in \{10, 50, 100, 150, 200, 250\}~\mathrm{MeV}.
    \label{Eq:DRgammai}
\end{equation}
The integrated branching fraction, so-called $R_{\gamma}$, is then given by
\begin{equation}
\label{eq:Rgamma_def}
    R_\gamma = \int_{10~\mathrm{MeV}}^{250~\mathrm{MeV}} d E_\gamma \ \frac{d R_\gamma}{d E_\gamma} = \sum_{i} \Delta R^i_\gamma.
\end{equation}
Since we work at first-order in $\alpha_{\mathrm{em}}$, we can replace $\Gamma(K^- \to \mu^- \bar{\nu}_\mu (\gamma))$  in the denominator of Eq.~(\ref{Eq:dR}) with its tree-level expression in Eq.~(\ref{eq:Gamma_QCD}). The neglected radiative corrections to $\Gamma(K^- \to \mu^- \bar{\nu}_\mu (\gamma))$ are estimated to be at the level of few permille.

\begin{table}[] 
\centering 
\resizebox{15cm}{!}{
\begin{tabular}{ccccccc} \hline \hline &  &  &  & &   \\ [-2. ex] 
bin & $1$ & $2$ & $3$ & $4$ & $5$ & tot \\ [1. ex] \hline &  &  &  & &  \\ [-2. ex] 
$E_\gamma [\mathrm{MeV}]$ &  $10-50$& $50-100$ & $100-150$ & $150-200$ & $200-250$ & $10-250$ \\ [1. ex] &  &  &     \\ [-2. ex] 
$\Delta R_\gamma^{\mathrm{exp}, i} \times 10^6$ & $0.94(30)(3)$ & $2.03(22)(2)$ & $4.47(30)(3)$ & $4.81(37)(4)$ & $2.58(26)(3)$ & $14.83(66)(13)$ \\ [1. ex]&  &  &  & &      \\ [-2. ex] 
$\Delta R_\gamma^{\mathrm{th}, i} \times 10^6$  & $1.31 (2)$ & $2.69 (16)$ & $5.60 (36)$  & $6.13 (46)$ & $2.69 (24)$ & $18.4 (1.2) $
\\ [1. ex] \hline \hline
\end{tabular}}
\caption{Comparison between KLOE experimental data ($\Delta R_\gamma^{\mathrm{exp}, i}$) and our predictions ($\Delta R_\gamma^{\mathrm{th}, i}$) for each of the
energy bins considered by the KLOE experiment.}
\label{tab:KLOE} 
\end{table} 
\begin{figure}
    \centering
    \includegraphics[width=0.7\linewidth]{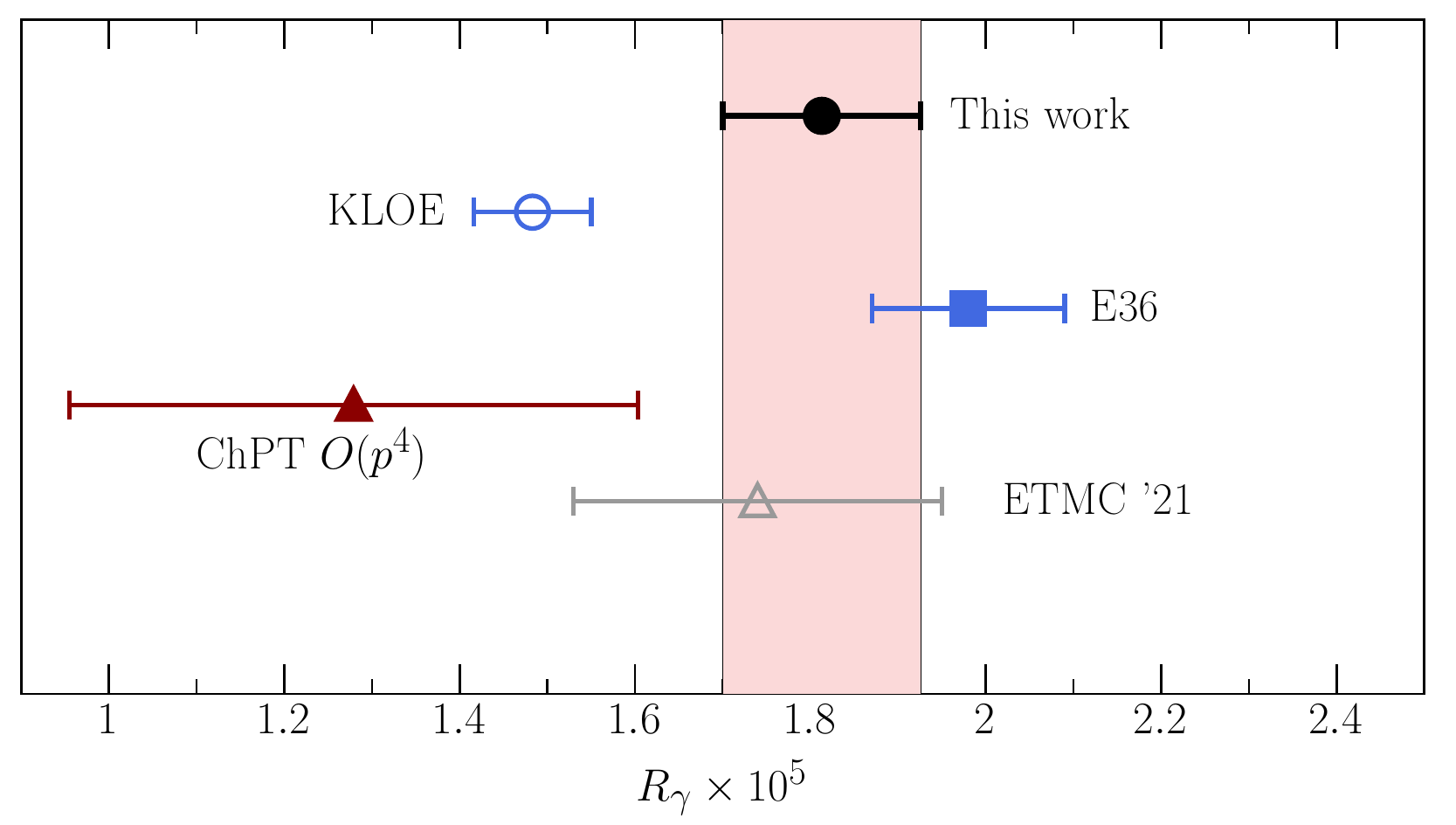}
    \caption{Comparison between the values of $R_{\gamma}$ obtained in the present work (black circle), and the determination by the KLOE (empty blue circle) and the E36 (filled blue square) Collaborations. In the figure, we also show the ChPT predictions at $O(p^4)$ (filled red triangle), as well as the results from our previous work~\cite{Desiderio:2020oej} denoted in the figure as ETMC '21 (empty gray triangle).}
    \label{fig:Rgamma_comp}
\end{figure}

In Fig.~\ref{fig:KLOEbins}, we compare our estimates (blue squares) with the ChPT predictions at $O(p^4)$ (green diamonds) and the KLOE experimental results (red circles). The data presented in the figure are also provided in Tab.~\ref{tab:KLOE}.  \\ 
Our results are consistently larger than the KLOE data in all bins except the last one. The largest difference, approximately $2.4\sigma$, occurs in the second, third and fourth bins, which correspond to photon energies $E_\gamma \in [50, 200]~\mathrm{MeV}$. For all the bins except the first (i.e. for $i > 1$), the dominant contribution to $\Delta R_\gamma^i$ is the structure-dependent term related to the $F^+(x_\gamma)$ combination of the form factors in Eq.~(\ref{eq:Fpm}). 
Note that our previous work already showed our predictions to be slightly larger than the KLOE results: with the uncertainties now reduced by a factor of two and central values practically unchanged, the significance of the difference is more pronounced. We provide a quantitative measure of the overall level of agreement between the KLOE experimental data and our results by considering the following reduced $\chi^2$-variable 
\be
\hat{\chi}^2_{\mathrm{KLOE}} = \frac{1}{N_{\mathrm{bins}}}  \sum_{i,j=1}^{N_{\mathrm{bins}}} (\Delta R^{\mathrm{exp}, i}_{\gamma} - \Delta R^{\mathrm{th}, i}_{\gamma}) C_{ij}^{-1} (\Delta R^{\mathrm{exp}, j}_{\gamma} - \Delta R^{\mathrm{th}, j}_{\gamma}),
\label{Eq:KLOE_redchisq}
\ee
where $N_{\mathrm{bins}}=5$ is the total number of bins, while $C_{ij}$ is the covariance matrix of our theoretical predictions (the correlations between the experimental data are not available). We get
\be
\hat{\chi}^2_{\mathrm{KLOE}} = 2.3,
\label{eq:KLOEchisqres}
\ee
to be compared with the value obtained in our previous work $\hat{\chi}^2_{\mathrm{KLOE}} = 0.7$.

A new measurement of the $R_{\gamma}$ ratio in Eq.~(\ref{eq:Rgamma_def}) was recently published by the E36 Collaboration at J-PARC~\cite{J-PARCE36:2021yvz,J-PARCE36:2022wfk}. In Fig.~\ref{fig:Rgamma_comp}, we compare our prediction for $R_{\gamma}$ with the experimental measurements by the KLOE and E36 Collaborations.

For completeness, we also show the value predicted by ChPT at $O(p^4)$ (filled red triangle) and the result from our previous work~\cite{Frezzotti:2020bfa}, indicated by an empty gray triangle. As the figure illustrates, the new E36 measurement aligns more closely with our prediction, whereas the KLOE determination is $2.6\,\sigma$ lower. These findings highlight the need for further experimental improvements to clarify the source of the observed differences.

\subsection{$K^- \to \mu^- \bar{\nu}_\mu \gamma:$ comparison with the experimental results from E787}
\begin{figure}[]
    \centering
    \includegraphics[width=0.75\linewidth]{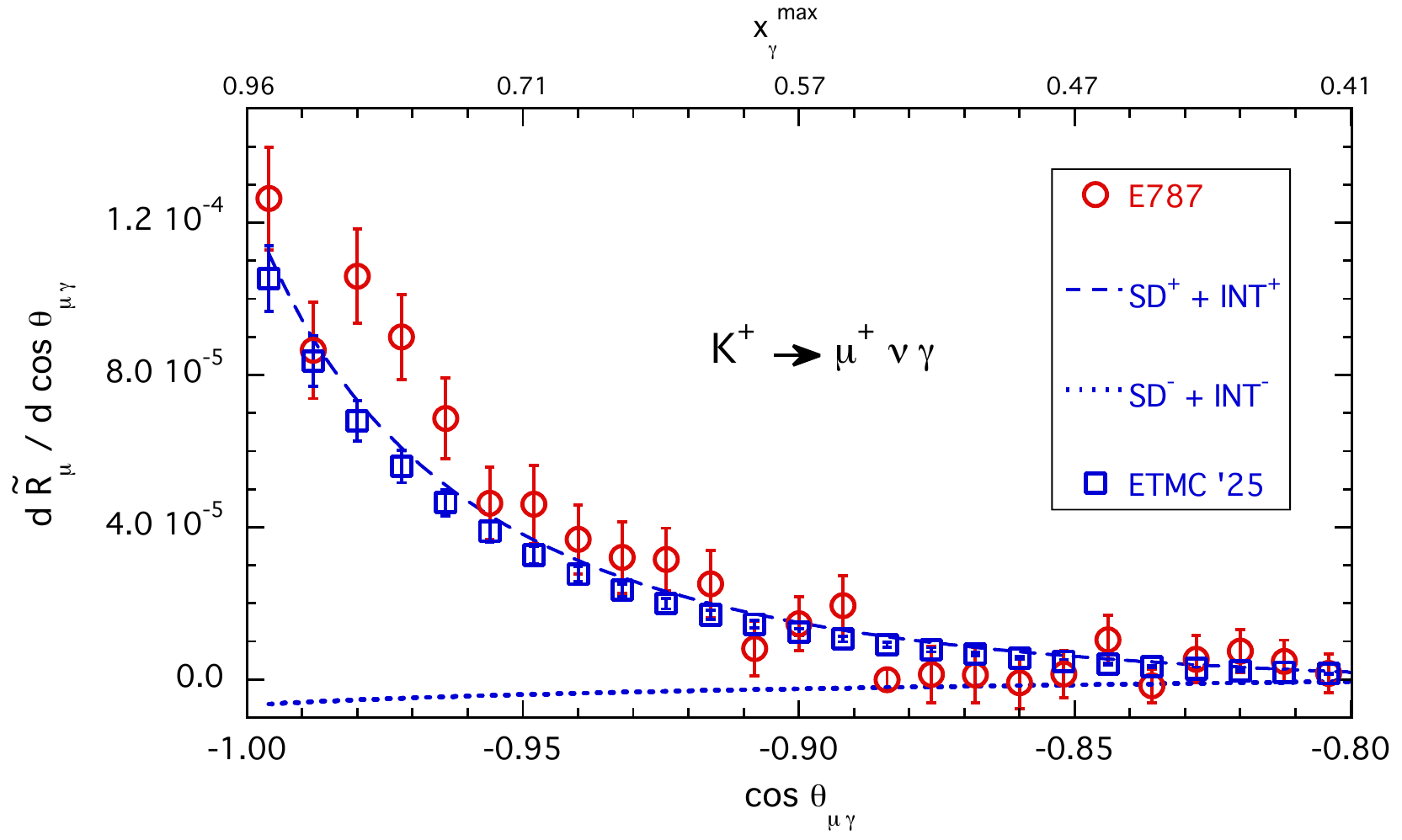}
    \caption{Comparison between E787 experimental data (red circles) and our theoretical 
predictions (blue squares) for the differential decay rate 
$\displaystyle d\widetilde{R}_\mu/d \cos \theta_{\mu \gamma}$. 
For completeness, the figure also displays the individual contributions from the 
form factors $F^\pm$, labeled as $\mathrm{SD}^\pm + \mathrm{INT}^\pm$. In the figure, $x_{\gamma}^{\mathrm{max}}$ stands for the maximum value that can be reached by $x_\gamma$ given the corresponding value of $\cos \theta_{\mu \gamma}$.}
    \label{fig:E787}
\end{figure}
The E787 experiment~\cite{E787:2000ehx} has measured the  $K^-\to \mu^- \bar{\nu}_\mu \gamma$ decay at fixed values of the angle $\theta_{\mu\gamma}$ between the muon and the photon (as measured in the kaon's rest frame). Specifically, they have provided results for the following ratio
\begin{equation}
\frac{dR_\mu}{d \cos \theta_{\mu \gamma}} = \frac{1}{\Gamma(K^- \to \mu^- \bar{\nu}_\mu (\gamma))} \frac{d \Gamma(K^- \to \mu^- \bar{\nu}_\mu \gamma)}{d \cos \theta_{\mu\gamma}},
\label{Eq:E787DR}
\end{equation}
where
\be
\frac{d \Gamma(K^- \to \mu^- \bar{\nu}_\mu \gamma)}{d \cos \theta_{\mu\gamma}} =   \int_{x_\gamma^{\mathrm{cut}}}^{1 - r_\mu^2} d x_\gamma ~ 
        \int_{ \max( x_{\mu}^{\rm min}, x_{\mu}^{\rm cut}) }^1 d x_\mu \left[ \frac{d^2\Gamma_1}{dx_\gamma dx_\mu} \right]           \times  \delta\left[ \cos \theta_{\mu \gamma} - \cos \theta_{\mu \gamma} (x_\gamma, x_\mu)\right] ,
\label{Eq:E787dgamma}
\ee
and where $\cos \theta_{\mu \gamma} (x_\gamma, x_\mu)$ and $x_{\mu}^{\rm min}$ are given respectively in Eq.~(\ref{eq:moml_theta}) and Eq.~(\ref{eq:xellmin}). In the previous equation $x_{\gamma}^{\rm cut}$ and $x_{\mu}^{\rm cut}$ correspond to the kinematical cut imposed by the E787 experiment respectively on the photon and muon energy, given by
\be
\begin{aligned}
E_\gamma > 90~\mathrm{MeV} \longrightarrow    \ x_\gamma^{\mathrm{cut}} \simeq 0.36,\qquad \quad
E_\mu > 243~\mathrm{MeV} \longrightarrow  \ x_\mu^{\mathrm{cut}} \simeq 0.94.
\end{aligned}
\label{Eq:E787cuts}
\ee
The ratio $dR_\mu/d \cos \theta_{\mu \gamma}$ in Eq.~(\ref{Eq:E787DR}) includes both the structure-dependent and the point-like contributions. Since the latter is a purely kinematic factor, it can be subtracted from the experimental data without introducing any additional uncertainty. The subtraction of the point-like contribution leads to
\begin{equation}
\frac{d\widetilde{R}_\mu}{d \cos \theta_{\mu \gamma}} = \frac{dR_\mu}{d \cos \theta_{\mu \gamma}} - \frac{dR^{\mathrm{pt}}_\mu}{d \cos \theta_{\mu \gamma}}~,
\label{Eq:E787sub}
\end{equation}
where $d R^{\mathrm{pt}}_\mu/d \cos \theta_{\mu \gamma}$ is obtained by retaining only the point-like term in the double-differential decay-rate in Eq.~(\ref{Eq:E787dgamma}).

The comparison between the E787 data for $d\widetilde{R}_\mu/ d \cos \theta_{\mu \gamma}$ and our theoretical predictions is shown in Fig.~\ref{fig:E787} and presented in Tab.~\ref{tab:E787},  for all the values of $\cos \theta_{\mu \gamma}$ considered in the E787 experiment. 
\begin{table}[t]	 
\begin{minipage}[t]{0.475\linewidth}
{\small
\begin{tabular}{ccc}   
 \hline \hline  &  &   \\ [-2. ex] 
$\cos \theta_{\mu \gamma}$ & ~$ d \widetilde{R}^{\mathrm{exp}}_\mu / d \cos \theta_{ \mu \gamma} \times 10^4$~ & ~$d \widetilde{R}^{\mathrm{th}}_\mu / d \cos \theta_{\mu \gamma} \times 10^4$ \\
[1. ex] \hline &  &    \\ [-2. ex] 
$-0.996$ & $1.26~(14)$ & $1.053 ~(87)$ \\ [1. ex]  &  & \\ [-2. ex] 
$-0.988$ & $0.87~(13)$ & $0.836 ~(66)$ \\ [1. ex]  &  & \\ [-2. ex]  
$-0.980$ & $1.06~(12)$ & $0.680 ~(52)$  \\ [1. ex]  &  & \\ [-2. ex] 
$-0.972$ & $0.90~(11)$ & $0.559 ~(42)$ \\ [1. ex]  &  & \\ [-2. ex]
$-0.964$ & $0.69~(11)$ & $0.465 ~(35)$ \\ [1. ex]  &  & \\ [-2. ex] 
$-0.956$ & $0.463~~(94)$ & $0.389 ~(28)$ \\ [1. ex]  &  &    \\ [-2. ex] 
$-0.948$ & $0.46~(10)$ & $0.327 ~(24)$ \\ [1. ex]  &  &    \\ [-2. ex] 
$-0.940$ & $0.368~~(91)$ & $0.276 ~(20)$ \\ [1. ex]  &  &    \\ [-2. ex] 
$-0.932$ & $0.320~~(94)$ & $0.234 ~(17)$ \\ [1. ex]  &  &    \\ [-2. ex] 
$-0.924$ & $0.315~~(82)$ & $0.199 ~(14)$ \\ [1. ex]  &  &    \\ [-2. ex] 
$-0.916$ & $0.251~~(88)$ & $0.170 ~(12)$ \\ [1. ex]  &  &    \\ [-2. ex] 
$-0.908$ & $0.081~~(71)$ & $0.145 ~(10)$ \\ [1. ex]  &  &    \\ [-2. ex] 
 &  & \\ 
 [1. ex] \hline \hline
\end{tabular}
}
\end{minipage}
\begin{minipage}[t]{0.475\linewidth}
{\small
\begin{tabular}{ccc}

 \hline \hline  &  &   \\ [-2. ex] 
 $\cos \theta_{\mu \gamma}$ & ~$ d \widetilde{R}^{\mathrm{exp}}_\mu / d \cos \theta_{ \mu \gamma} \times 10^4$~ & ~$d \widetilde{R}^{\mathrm{th}}_\mu / d \cos \theta_{\mu \gamma} \times 10^4$ \\
[1. ex] \hline &  &    \\ [-2. ex] 
$-0.900$ & $~~0.146~(71)$ & $0.1241 ~(88)$ \\ [1. ex]  &  &    \\ [-2. ex] 
$-0.892$ & $~~0.194~(79)$ & $0.1062 ~(75)$ \\ [1. ex]  &  &    \\ [-2. ex] 
$-0.884$ & $-0.001~(28)$ & $0.0909 ~(64)$ \\ [1. ex]  &  &    \\ [-2. ex] 
$-0.876$ & $~~0.013~(74)$ & $0.0777 ~(55)$ \\ [1. ex]  &  &    \\ [-2. ex] 
$-0.868$ & $~~0.011~(74)$ & $0.0663 ~(47)$ \\ [1. ex]  &  &    \\ [-2. ex] 
$-0.860$ & $-0.009~(68)$ & $0.0565 ~(40)$ \\ [1. ex]  &  &    \\ [-2. ex] 
$-0.852$ & $~~0.014~(62)$ & $0.0479 ~(34)$ \\ [1. ex]  &  &    \\ [-2. ex] 
$-0.844$ & $~~0.104~(65)$ & $0.0405 ~(29)$ \\ [1. ex]  &  &    \\ [-2. ex] 
$-0.836$ & $-0.017~(44)$ & $0.0339 ~(25)$ \\ [1. ex]  &  &    \\ [-2. ex] 
$-0.828$ & $~~0.053~(62)$ & $0.0282 ~(21)$ \\ [1. ex]  &  &    \\ [-2. ex] 
$-0.820$ & $~~0.074~(56)$ & $0.0232 ~(17)$ \\ [1. ex]  &  &    \\ [-2. ex] 
$-0.812$ & $~~0.047~(56)$ & $0.0187 ~(14)$ \\ [1. ex]  &  &    \\ [-2. ex] 
$-0.804$ & $~~0.016~(50)$ & $0.0148 ~(11)$  \\ 
 [1. ex] \hline \hline
\end{tabular}
}
\end{minipage}
\caption{Comparison between the E787 experimental distribution,
$\displaystyle d \widetilde{R}^{\mathrm{exp}}_\mu / d\cos \theta_{\mu \gamma}$,
with our theoretical predictions,
$\displaystyle d \widetilde{R}^{\mathrm{th}}_\mu / d\cos \theta_{\mu \gamma}$,
over the entire range of $\cos \theta_{\mu \gamma}$ values considered
in the E787 experiment.}
\label{tab:E787}
\end{table} 
As in the previous section, our predictions are obtained by using the tree-level expression for $\Gamma(K^- \to \mu^- \bar{\nu}_\mu (\gamma))$ (given in Eq.~(\ref{eq:Gamma_QCD})) in the denominator of Eq.~(\ref{Eq:E787DR}). Overall, our results exhibit good agreement with the experimental data, with the exception of the region of large angles, where we observe a difference of about $2 \sigma$.  
In this kinematical region, our estimate of $ d\widetilde{R}_\mu / d \cos \theta_{\mu \gamma} $ is primarily determined by the contribution of the form factor $F^+(x_\gamma)$, labeled in the figure as $\mathrm{SD}^+ + \mathrm{INT}^+$. While the relative contribution from $F^-(x_\gamma)$ (denoted as $\mathrm{SD}^- + \mathrm{INT}^-$ in the figure) is generally small, it becomes more significant at small angles, reaching  20--30\%  for $\cos\theta_{\mu\gamma} \simeq -0.8$.
As in the previous subsection, to quantify the overall level of agreement between our predictions and the E787 measurements, we introduce the reduced $\chi^{2}$-variable
\be
\hat{\chi}^2_{\mathrm{E787}} = \frac{1}{N_{\theta}}  \sum_{ij=1}^{N_\theta} \bigg ( \frac{d\widetilde{R}^{\mathrm{exp},i}_\mu}{d \cos \theta_{\mu \gamma}} - \frac{d\widetilde{R}^{\mathrm{th},i}_\mu}{d \cos \theta_{\mu \gamma}} \bigg ) C_{ij}^{-1} \bigg(\frac{d\widetilde{R}^{\mathrm{exp},j}_\mu}{d \cos \theta_{\mu \gamma}} - \frac{d\widetilde{R}^{\mathrm{th},j}_\mu}{d \cos \theta_{\mu \gamma}} \bigg ).
\label{Eq:E787chisq}
\ee
In the previous equation, $C$ is the correlation matrix (which again includes only correlations among the theoretical predictions), and the indices $i$ and $j$ run over all values of $\theta_{\mu \gamma}$ given in Tab.~\ref{tab:E787}, which total number is $N_{\theta}=25$. We obtain
\be
\hat{\chi}^2_{\mathrm{E787}} = 1.7,
\label{Eq:E787chisqres}
\ee
which is very close to the value obtained in our previous work, namely $\hat{\chi}^2_{\mathrm{E787}} = 1.6$.

\subsection{$K^- \to \mu^- \bar{\nu}_\mu \gamma:$ comparison with the experimental results from ISTRA+ and OKA}
\begin{table}[!hbt]	 
\begin{minipage}[t]{0.475\linewidth}
{\footnotesize
\setlength{\tabcolsep}{7pt}
\begin{tabular}{cccc}
 \hline \hline  &  & &  \\ [-2. ex] 
strip & $x_\gamma^i < x_\gamma < x_\gamma^{i+1}$ & $x_\mu^{\mathrm{cut}, i} < x_\mu < X_\mu^{\mathrm{cut}, i}$ & $ \cos \theta^{\mathrm{cut}, i}_{\mu \gamma}$
\\
[1. ex] \hline &  & &   \\ [-2. ex] 
01 & $0.05 < x_\gamma < 0.10$ & $0.85 < x_\mu  < 1.05$ & $- 0.8$ \\ [1. ex]  &  &   & \\ [-2. ex] 
02 & $0.10 < x_\gamma < 0.15$ & $0.85 < x_\mu  < 1.05$ & $- 0.8$ \\ [1. ex]  &  & &   \\ [-2. ex] 
03 & $0.15 < x_\gamma < 0.20$ & $0.80 < x_\mu  < 0.95$ & $- 0.8$ \\ [1. ex]  &  &  &  \\ [-2. ex] 
04 & $0.20 < x_\gamma < 0.25$ & $0.75 < x_\mu  < 0.90$ & $- 0.2$ \\ [1. ex]  &  &   & \\ [-2. ex] 
05 & $0.25 < x_\gamma < 0.30$ & $0.70 < x_\mu  < 0.85$ & $- 0.3$ \\ [1. ex]  &  &  &  \\ [-2. ex] 
06 & $0.30 < x_\gamma < 0.35$ & $0.67 < x_\mu  < 0.82$ & $- 0.4$ \\ [1. ex]  &  &  &  \\ [-2. ex] 
07 & $0.35 < x_\gamma < 0.40$ & $0.60 < x_\mu  < 0.80$ & $- 0.3$ \\ [1. ex]  &  &  &  \\ [-2. ex] 
08 & $0.40 < x_\gamma < 0.45$ & $0.57 < x_\mu  < 0.80$ & $- 0.5$ \\ [1. ex]  &  &  &  \\ [-2. ex] 
09 & $0.45 < x_\gamma < 0.50$ & $0.52 < x_\mu  < 0.75$ & $- 0.7$ \\ [1. ex]  &  &  &  \\ [-2. ex] 
10 & $0.50 < x_\gamma < 0.55$ & $0.47 < x_\mu  < 0.70$ & $- 1.0$ \\ [1. ex]  &  &  &  \\ [-2. ex] 
11 & $0.55 < x_\gamma < 0.60$ & $0.43 < x_\mu  < 0.65$ & $- 1.0$ \\ [1. ex] \hline \hline
\end{tabular}
}
\end{minipage}
\hfill
\begin{minipage}[t]{0.475\linewidth}
{\footnotesize
\setlength{\tabcolsep}{7pt}
\begin{tabular}{cccc}
 \hline \hline  &  &   \\ [-2. ex] 
strip & $x_\gamma^i < x_\gamma < x_\gamma^{i+1}$ & $x_\mu^{\mathrm{cut}, i} < x_\mu < X_\mu^{\mathrm{cut}, i}$ & $ \cos \theta^{\mathrm{cut}, i}_{\mu \gamma}$
\\
[1. ex] \hline &  & &   \\ [-2. ex] 
01 & $0.10 < x_\gamma < 0.15$ & $0.84 < x_\mu  < 0.96$ & $- 0.8$ \\ [1. ex]  &  &   & \\ [-2. ex]  
02 & $0.15 < x_\gamma < 0.20$ & $0.80 < x_\mu  < 0.96$ & $- 0.2$ \\ [1. ex]  &  &   & \\ [-2. ex] 
03 & $0.20 < x_\gamma < 0.25$ & $0.75 < x_\mu  < 0.95$ & $- 0.2$ \\ [1. ex]  &  &   & \\ [-2. ex]  
04 & $0.25 < x_\gamma < 0.30$ & $0.70 < x_\mu  < 0.92$ & $- 0.4$ \\ [1. ex]  &  &   & \\ [-2. ex] 
05 & $0.30 < x_\gamma < 0.35$ & $0.65 < x_\mu  < 0.88$ & $- 0.4$ \\ [1. ex]  &  &   & \\ [-2. ex] 
06 & $0.35 < x_\gamma < 0.40$ & $0.61 < x_\mu  < 0.85$ & $- 0.5$ \\ [1. ex]  &  &   & \\ [-2. ex] 
07 & $0.40 < x_\gamma < 0.45$ & $0.57 < x_\mu  < 0.83$ & $- 0.5$ \\ [1. ex]  &  &   & \\ [-2. ex] 
08 & $0.45 < x_\gamma < 0.50$ & $0.53 < x_\mu  < 0.81$ & $- 0.6$ \\ [1. ex]  &  &   & \\ [-2. ex] 
09 & $0.50 < x_\gamma < 0.55$ & $0.49 < x_\mu  < 0.78$ & $- 0.6$ \\ [1. ex]  &  &   & \\ [-2. ex] 
10 & $0.55 < x_\gamma < 0.60$ & $0.45 < x_\mu  < 0.75$ & $- 0.6$ \\ [1. ex] &  &   & \\ [-2. ex] 
 & & & \\ [1. ex] 
\hline \hline

\end{tabular}
}
\end{minipage}
\caption{Summary of the kinematic cuts applied by the ISTRA+ experiment (left table) and the OKA experiment (right table).}
\label{tab:ISTRA+OKA}
\end{table} 
The ISTRA+ and OKA experiments have measured the decay rate for $K^- \to \mu^- \bar{\nu}_\mu \gamma$, in specific regions of the phase-space (strips) defined by kinematic constraints on $x_\gamma$, $x_\mu$ and $\cos \theta_{\mu \gamma}$, given by   
\be
i\mathrm{-th \ strip}: \qquad x_\gamma^i < x_\gamma < x_\gamma^{i+1}, \qquad x_{\mu}^{\mathrm{cut}, i} < x_\mu < X_{\mu}^{\mathrm{cut}, i}, \qquad \cos \theta_{\mu \gamma} > \cos \theta_{\mu \gamma}^{\mathrm{cut}, i} \ .
\label{Eq:ISTRAOKAstrips}
\ee
The values of $x_\gamma^i$, $x_\mu^{\mathrm{cut}, i}$, $X_{\mu}^{\mathrm{cut}, i}$ and $\cos \theta_{\mu \gamma}^{\mathrm{cut}, i}$ adopted by the ISTRA+ and OKA Collaborations are reported in Tab.~\ref{tab:ISTRA+OKA}. \\ 
The double-differential decay rate in Eq.~(\ref{eq:dGamma}) integrated over the $i$-th strip reads
\be
 \Gamma_\mu^i =  \int_{x_\gamma^i}^{x_\gamma^{i+1}} d x_\gamma
        \int_{x_\mu^{*, i}}^{X_\mu^{*,i}} d x_\mu \left[ \frac{d^2\Gamma_1}{dx_\gamma dx_\mu} \right] 
         \times  \Theta\left[  \cos \theta_{\mu \gamma} (x_\gamma, x_\mu) - \cos \theta^{\mathrm{cut}, i}_{\mu \gamma} \right], 
\label{Eq:ISTRA+OKAdr}
\ee
where $\Theta(x)$ is the Heaviside function. The integration limits are given by  
\be
x_\mu^{*,i} = \max( x_\mu^{\mathrm{min}}, x_\mu^{\mathrm{cut},i} ), \qquad X_\mu^{*,i} = \min(1, X_\mu^{\mathrm{cut},i}),
\label{Eq:ISTRA+OKAcuts}
\ee
where $x_\mu^{\mathrm{min}}$ is given in Eq.~(\ref{eq:xellmin}). 
\begin{figure}[]
    \centering
    \includegraphics[width=1.\linewidth]{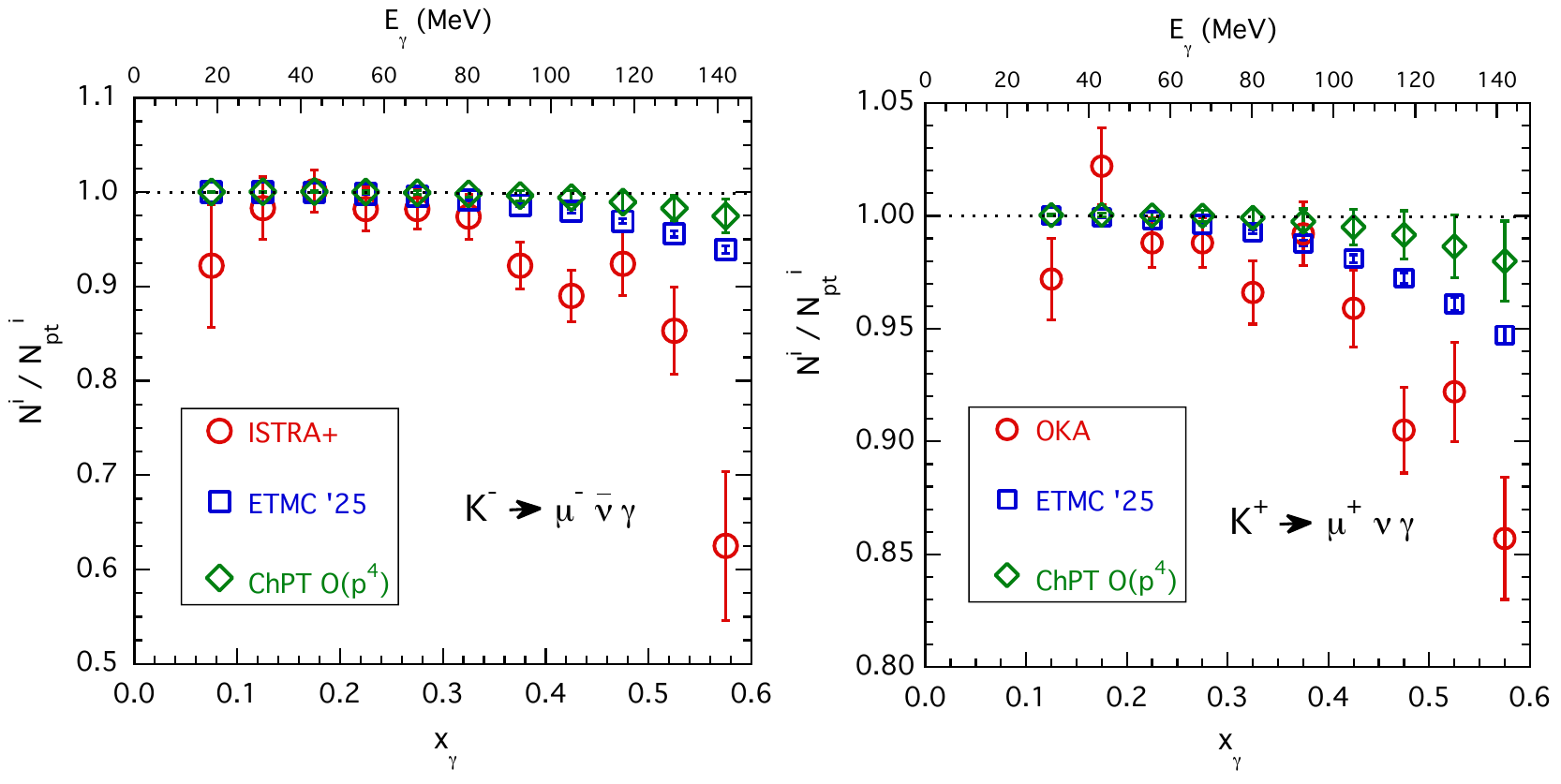}
    \caption{The ratio $N^{i}/N^{i}_{\mathrm{pt}}$ from ISTRA+ data (red circles), our lattice QCD 
results (blue squares), and the ChPT predictions at $O(p^4)$ (green diamonds).
\textit{Right}: The same ratio measured by OKA (red circles), obtained in this work (blue squares), and 
predicted by ChPT at $O(p^4)$ (green diamonds).
In both panels, the results are given for each ``strip'' in Tab.~\ref{tab:ISTRA+OKA}. 
}    \label{fig:ISTRA+OKA}
\end{figure}
A corresponding expression to Eq.~(\ref{Eq:ISTRA+OKAdr}) can be introduced for the point-like contribution $\Gamma^{\mathrm{pt}, i}_\mu$  by replacing the total double-differential decay rate in Eq.~(\ref{eq:d2Gamma_pt}) with its point-like component. ISTRA+~\cite{ISTRA:2010smy} and OKA~\cite{OKA:2019gav} have measured the following quantities:
\be
\frac{N^i}{N_{\mathrm{pt}}^i} = \frac{\Gamma^i_\mu}{\Gamma^{\mathrm{pt}, i}_\mu},
\label{Eq:ISTRA+OSKAobs}
\ee
using a Montecarlo estimate for the number of point-like events $N_{\mathrm{pt}}^i$. \\
A comparison between our results, the ChPT predictions at $O(p^4)$, and the experimental data is presented in Fig.~\ref{fig:ISTRA+OKA} and summarized in Tab.~\ref{tab:res_ISTRA+OKA}. For large values of $x_\gamma$ ($x_\gamma \gtrsim 0.45$), the plots reveal a significant tension between our lattice results and the ISTRA+ and OKA data, as well as between the ISTRA+ and OKA data and the ChPT predictions. In particular, both our lattice results and the ChPT predictions show a much flatter behaviour in $x_{\gamma}$ compared to the experimental data. 
To quantify this discrepancy, we compute the following reduced chi-squared variable for ISTRA+:  
\begin{equation}
\hat{\chi}^2_{\mathrm{ISTRA+}} = \frac{1}{N_s}  \sum_{i,j=1}^{N_s} \bigg( \frac{N_{\mathrm{exp}}^i}{N^i_{\mathrm{pt}}} - \frac{N_{\mathrm{th}}^i}{N^i_{\mathrm{pt}}} \bigg) C_{ij}^{-1} \bigg( \frac{N_{\mathrm{exp}}^j}{N^j_{\mathrm{pt}}} - \frac{N_{\mathrm{th}}^j}{N^j_{\mathrm{pt}}} \bigg),
\label{Eq:ISTRA+OKAchisq}
\end{equation}  
where $C$ is the correlation matrix, and $N_s$ the total number of strips ($N_{s}=11$ for ISTRA+).  $\hat{\chi}^2_{\mathrm{OKA}}$ is defined analogously. We find  
\begin{equation}
\hat{\chi}^2_{\mathrm{ISTRA+}} = 3.9, \qquad \hat{\chi}^2_{\mathrm{OKA}} = 3.6 \ ,
\label{Eq:ISTRA+OKAchisqres}
\end{equation}  
confirming the tension, already observed in our previous work~\cite{Frezzotti:2020bfa}, between lattice QCD predictions and ISTRA+ and OKA data. 
\begin{table}[]	 
\begin{minipage}[t]{0.475\linewidth}
{\footnotesize
\setlength{\tabcolsep}{7pt}
\begin{tabular}{cccc}
 \hline \hline  &  & &  \\ [-2. ex] 
strip & ~~$N^i_{\mathrm{exp}} / N^i_{\mathrm{pt}}$~~& ~~$N^i_{\mathrm{th}} / N^i_{\mathrm{pt}}$~~ & ~~~~ChPT~~~~\\ [1. ex] \hline &  & &   \\ [-2. ex] 
01 & $0.922~(65)$ & $1.0001~~~(1)$ & $1.0002~~~(1)$ \\ [1. ex]  &  &   & \\ [-2. ex]
02 & $0.983~(33)$ & $1.0001~~~(1)$ & $1.0004~~~(4)$ \\ [1. ex]  &  &   & \\ [-2. ex]
03 & $1.001~(22)$ & $0.9996~~~(2)$ & $1.0005~~~(8)$ \\ [1. ex]  &  &   & \\ [-2. ex]
04 & $0.982~(23)$ & $0.9981~~~(3)$ & $1.0002~~(14)$ \\ [1. ex]  &  &   & \\ [-2. ex]
05 & $0.982~(21)$ & $0.9953~~~(5)$ & $0.9994~~(23)$ \\ [1. ex]  &  &   & \\ [-2. ex]
06 & $0.974~(24)$ & $0.9913~~~(8)$ & $0.9981~~(37)$ \\ [1. ex]  &  &   & \\ [-2. ex]
07 & $0.922~(25)$ & $0.9860~~(12)$ & $0.9963~~(54)$ \\ [1. ex]  &  &   & \\ [-2. ex]
08 & $0.890~(27)$ & $0.9798~~(17)$ & $0.9942~~(78)$ \\ [1. ex]  &  &   & \\ [-2. ex]
09 & $0.924~(34)$ & $0.9691~~(23)$ & $0.989~(11)$ \\ [1. ex]  &  &   & \\ [-2. ex]
10 & $0.853~(46)$ & $0.9556~~(31)$ & $0.983~(14)$ \\ [1. ex]  &  &   & \\ [-2. ex]
11 & $0.625~(79)$ & $0.9390~~(41)$ & $0.975~(18)$ \\ [1. ex] \hline \hline
\end{tabular}
}
\end{minipage}
\hfill
\begin{minipage}[t]{0.475\linewidth}
{\footnotesize
\setlength{\tabcolsep}{7pt}
\begin{tabular}{cccc}
 \hline \hline  &  & &  \\ [-2. ex] 
strip & ~~$N_{\mathrm{exp}}^i / N_{\mathrm{pt}}^i$~~& ~~$N_{\mathrm{th}}^i / N_{\mathrm{pt}}^i$~~ & ~~~~ChPT~~~~\\ [1. ex] \hline &  & &   \\ [-2. ex] 
01 & $0.972~(18)$ & $1.0000~~~(1)$ & $1.0004~~~(3)$ \\ [1. ex]  &  &   & \\ [-2. ex]
02 & $1.022~(17)$ & $0.9994~~~(2)$ & $1.0004~~~(7)$ \\ [1. ex]  &  &   & \\ [-2. ex]
03 & $0.988~(11)$ & $0.9981~~~(3)$ & $1.0002~~(14)$ \\ [1. ex]  &  &   & \\ [-2. ex]
04 & $0.988~(11)$ & $0.9962~~~(5)$ & $1.0001~~(24)$ \\ [1. ex]  &  &   & \\ [-2. ex]
05 & $0.966~(14)$ & $0.9928~~~(8)$ & $0.9991~~(38)$ \\ [1. ex]  &  &   & \\ [-2. ex]
06 & $0.992~(14)$ & $0.9877~~(12)$ & $0.9975~~(56)$ \\ [1. ex]  &  &   & \\ [-2. ex]
07 & $0.959~(17)$ & $0.9810~~(17)$ & $0.9950~~(79)$ \\ [1. ex]  &  &   & \\ [-2. ex]
08 & $0.905~(19)$ & $0.97237~~(23)$ & $0.992~(11)$ \\ [1. ex]  &  &   & \\ [-2. ex]
09 & $0.922~(22)$ & $0.9609~~(31)$ & $0.986~(14)$ \\ [1. ex]  &  &   & \\ [-2. ex]
10 & $0.857~(27)$ & $0.9472~~(39)$ & $0.980~(18)$ \\ 
[1. ex] &  &   & \\ [-2. ex]
 & & & \\ [1. ex]
\hline \hline
\end{tabular}
}
\end{minipage}
\caption{Values of $N_{\mathrm{exp}}^i/N_{\mathrm{pt}}^i$ (see text for details) obtained by the ISTRA+ (left table) and OKA experiments (right table), compared to our theoretical predictions $N_{\mathrm{th}}^i/N_{\mathrm{pt}}^i$, for the kinematical ranges selected by the two experiments (see Tab.~\ref{tab:ISTRA+OKA}). The fourth columns correspond to the predictions of ChPT at order ${\cal{O}}(p^4)$, based on the vector and axial form factors given in Eq.\,(\ref{eq:ChPT_FF}).}
\label{tab:res_ISTRA+OKA}
\end{table}

\section{Conclusions}
\label{sec:conclusions}

In this work, we presented a lattice QCD study of radiative leptonic kaon decays at physical 
quark masses including, for the first time, quark-disconnected contributions. Our computation
uses gauge ensembles produced by the ETM Collaboration, featuring spatial volumes of up 
to $L \simeq 7.7\,\mathrm{fm}$ and lattice spacings in the range 
$a \in [0.08,\,0.058]\,\mathrm{fm}$. Compared to our previous work, we employed the 
3d method~\cite{Giusti:2023pot,Tuo:2021ewr} to compute the relevant form factors and have carried out thorough analyses of the 
systematic errors associated with isolating the initial-state kaon, as well as 
those arising from the finite temporal extent of the lattice. The improvement of the numerical setup, in combination with a high-statistics evaluation of the correlation functions, 
led to a reduction in the uncertainties on both the axial and vector form factors 
by nearly a factor of two relative to our earlier determination~\cite{Desiderio:2020oej}.

The dependence of the form factors on $x_\gamma$ (i.e., the photon energy in units of the kaon mass) can be accurately described using a linear behaviour. By performing a linear fit to our data as in Eq.~(\ref{Eq:F_vs_xg_linear}), we obtain  
\begin{align}
F_{A}^{0} &= 0.0428(34), 
\quad
F_{A}^{1} = -0.0116(40), 
\quad
\mathrm{corr}(F_{A}^{0}, F_{A}^{1}) = -0.37, \\[8pt]
F_{V}^{0} &= 0.1421(57),
\quad
F_{V}^{1} =  -0.0452(72),
\quad
\mathrm{corr}(F_{V}^{0}, F_{V}^{1}) = -0.77,
\end{align}
for the intercept $F_{A(V)}^0$ and the slope $F_{A(V)}^1$ of the axial (vector) form factor, respectively. These results  can be directly applied in phenomenological analyses and in estimating the acceptance of future experiments.
The sum and the difference of the vector and axial form factors at $x_\gamma=1$ are found to be
\begin{align}
    F^+(x_\gamma = 1)  & = 0.128(7) ~ , ~ \\[8pt]
    F^-(x_\gamma = 1)  & = 0.066(6) ~ . ~ \label{eq:fmus}
\end{align}
For comparison, the last PDG review\,\cite{ParticleDataGroup:2024cfk} quotes from the available experimental results
\begin{align}
    F^+(x_\gamma = 1)  & = 0.133(8)~~~~~ \qquad [\mbox{from } K \to e \nu \gamma ] ~ , ~ \\[8pt]
                       & = 0.165(7)(11) \qquad [\mbox{from } K \to \mu \nu \gamma ] ~ , ~ \\[8pt]
    F^-(x_\gamma = 1)  & = 0.153(33)~~~~ \qquad [\mbox{from } K \to \mu \nu \gamma ]~ . ~   \label{Eq:fmexp}
\end{align}

Our final results exhibit a mild tension of $2.6\,\sigma$ with the KLOE 
measurement of the CKM-independent ratio $R_{\gamma}$ in Eq.~(\ref{eq:Rgamma}) for $K^- \to e^- \bar{\nu}_e \gamma$, while they agree within $1\,\sigma$ with the more recent measurement 
by the E36 Collaboration at J-PARC. In the muonic decay channel $K^- \to \mu^- \bar{\nu}_\mu \gamma$, we confirm the tensions 
with ISTRA+ and OKA data (already noted in Ref.~\cite{Frezzotti:2020bfa}), which are particularly sensitive to the 
negative-helicity form-factor $F^{-}$. This results in a  $2.6~\sigma$  discrepancy  between our determination of $F^-(x_\gamma=1)$ in Eq.~(\ref{eq:fmus}) and the experimental estimate in Eq.~(\ref{Eq:fmexp}).
Additionally, a discrepancy of about 
two standard deviations is found when comparing our predictions with E787 data at large 
angles between the muon and the photon.

The correlation functions generated for this work will also be used to investigate the 
rare kaon decay $K^{-}\to \ell' \bar{\ell'}\,\ell^{-}\,\bar{\nu}_{\ell}$, an interesting 
probe of potential New Physics. A fully controlled determination of this decay rate 
requires, however, addressing the analytic continuation problem that emerges when the 
virtual-photon offshellness $q^{2}$ exceeds the two-pion threshold, $q^{2} > 4M_{\pi}^{2}$. Building 
on the approach proposed in Ref.~\cite{Frezzotti:2023nun} and employing the Hansen--Lupo--Tantalo (HLT) 
spectral-density reconstruction method~\cite{Hansen:2019idp} (already employed in our recent work on 
$B_{s}\to \mu^{+}\mu^{-}\gamma$~\cite{Frezzotti:2024kqk}), we aim to resolve the issue of the analytic 
continuation for $K^{-}\to \ell' \bar{\ell'}\,\ell^{-}\,\bar{\nu}_{\ell}$ decays. Our 
results will be presented in a forthcoming paper. 

\section{Acknowledgments}
\label{sec:akno}
We thank C. Tarantino for useful comments and all members of the ETMC for the most enjoyable collaboration. V.L., F.S., G.G., R.F., and N.T. are supported by the Italian Ministry
of University and Research (MUR) and the European
Union (EU) – Next Generation EU, Mission 4, Component 1, PRIN 2022, CUP F53D23001480006. 
F.S. is supported by ICSC – Centro Nazionale di Ricerca in High Performance Computing, Big Data and Quantum Computing, funded by European Union -NextGenerationEU and by Italian  Ministry of University and Research (MUR) projects FIS\_00001556 and PRIN\_2022N4W8WR. We acknowledge support from the LQCD123, ENP SFT, and SPIF Scientific Initiatives of
the Italian Nuclear Physics Institute (INFN). C.T.S is supported in part by STFC consolidated grant ST/X000583/1.

The open-source packages tmLQCD~\cite{Jansen:2009xp,Abdel-Rehim:2013wba,Deuzeman:2013xaa,Kostrzewa:2022hsv}, LEMON~\cite{Deuzeman:2011wz}, DD-$\alpha$AMG~\cite{Frommer:2013fsa,Alexandrou:2016izb,Bacchio:2017pcp,Alexandrou:2018wiv}, QPhiX~\cite{joo2016optimizing,Schrock:2015gik} and QUDA~\cite{Clark:2009wm,Babich:2011np,Clark:2016rdz} have been used in the ensemble generation.

We gratefully acknowledge CINECA for the provision of GPU time on Leonardo supercomputing facilities under the specific initiative INFN-LQCD123, and under project IscrB VITO-QCD and project IscrB SemBD. We gratefully acknowledge the Gauss Centre for Supercomputing e.V. (www.gauss-centre.eu) for funding this project by providing computing time on the GCS Supercomputers SuperMUC-NG at Leibniz Supercomputing Centre. We acknowledge the Texas Advanced Computing Center (TACC) at The University of Texas at Austin for providing HPC resources (Project ID PHY21001). The authors gratefully acknowledge PRACE for awarding access to HAWK at HLRS within the project with Id Acid 4886.

\bibliographystyle{JHEP}
\bibliography{biblio}

\end{document}